\documentclass[aps,prc,twocolumn,nofootinbib,superscriptaddress,groupedaddress]{revtex4}
\usepackage{dcolumn}   
\usepackage{bm}        
\usepackage{multirow}
\usepackage{xcolor}

\usepackage{amsmath}
\usepackage{amsfonts}
\usepackage{amssymb}
\usepackage{makeidx}
\usepackage{graphicx}
\usepackage{anysize}
\usepackage{hyperref}
\usepackage{slashed}

\usepackage{ulem} 

\newenvironment{mymathbox}
{\par\smallskip\centering\begin{lrbox}{0}%
\begin{minipage}[c]{0.8\textwidth}}
{\end{minipage}\end{lrbox}%
\framebox[0.9\textwidth]{\usebox{0}}%
\par\medskip
\ignorespacesafterend}
\newcommand{\bb}{\begin{mymathbox}}
\newcommand{\eb}{\end{mymathbox}}

\newcommand{\be}{\begin{equation}}
\newcommand{\ee}{\end{equation}}
\newcommand{\ba}{\begin{eqnarray}}
\newcommand{\ea}{\end{eqnarray}}

\newcommand{\nk}{{\bf      k}}
\newcommand{\np}{{\bf      p}}
\newcommand{\nq}{{\bf      q}}

\newcommand{\npsi}{{\bf \npsi}}

\newcommand{\de}{\text{d}}

\newcommand{\non}{\nonumber}

\newcommand{\bma}{\begin{pmatrix}}
\newcommand{\ema}{\end{pmatrix}}

\begin{document}

\title{ Neutrino energy reconstruction from semi-inclusive samples }

\author{R.~Gonz\'alez-Jim\'enez}
\affiliation{Grupo de F\'isica Nuclear, Departamento de Estructura de la Materia,~F\'isica T\'ermica y Electr\'onica and IPARCOS,~Facultad de Ciencias F\'isicas, Universidad Complutense de Madrid, CEI Moncloa, Madrid 28040, Spain\looseness=-4}
\author{M.~B.~Barbaro}
\affiliation{Dipartimento di Fisica, Universit\`{a} di Torino and INFN, Sezione di Torino, Via P. Giuria 1, 10125 Torino, Italy\looseness=-4}
\affiliation{Universit\'e Paris-Saclay, CNRS/IN2P3, IJCLab, 91405 Orsay, France\looseness=-4}
\author{J.~A.~Caballero}
\affiliation{Departamento de F\'{i}sica At\'omica, Molecular y Nuclear, Universidad de Sevilla, 41080 Sevilla, Spain\looseness=-4}
\affiliation{ Instituto de F\'{i}sica Te\'orica y Computacional Carlos I, Granada 18071, Spain\looseness=-4}
\author{T.~W.~Donnelly}
\affiliation{Center for Theoretical Physics, Laboratory for Nuclear Science and Department of Physics, Massachusetts Institute of Technology, Cambridge, Massachusetts 02139, USA\looseness=-4}
\author{N.~Jachowicz}
\affiliation{Department of Physics and Astronomy, Ghent University, Proeftuinstraat 86, B-9000 Gent, Belgium\looseness=-4}
\author{G.~D.~Megias}
\affiliation{Departamento de F\'{i}sica At\'omica, Molecular y Nuclear, Universidad de Sevilla, 41080 Sevilla, Spain\looseness=-4}
\affiliation{Research Center for Cosmic Neutrinos, Institute for Cosmic Ray Research. University of Tokyo, Kashiwa, Chiba 277-8582, Japan\looseness=-4}
\author{K.~Niewczas}
\affiliation{Department of Physics and Astronomy, Ghent University, Proeftuinstraat 86, B-9000 Gent, Belgium\looseness=-4}
\affiliation{Institute of Theoretical Physics, University of Wroc{\l}aw, Plac Maxa Borna 9, 50-204, Wroc{\l}aw, Poland\looseness=-4}
\author{A.~Nikolakopoulos}
\affiliation{Department of Physics and Astronomy, Ghent University, Proeftuinstraat 86, B-9000 Gent, Belgium\looseness=-4}
\author{J.~W.~Van~Orden}
\affiliation{Department of Physics, Old Dominion University, Norfolk, VA 23529, Jefferson Lab, 12000 Jefferson Avenue, Newport News, VA 23606, USA. \footnote{Notice: Authored by Jefferson Science Associates, LLC under U.S. DOE Contract No. DE-AC05-06OR23177.
		The U.S. Government retains a non-exclusive, paid-up, irrevocable, world-wide license to publish or reproduce this manuscript for U.S. Government purposes}
\looseness=-4}
\author{J.~M.~Ud\'ias}
\affiliation{Grupo de F\'isica Nuclear, Departamento de Estructura de la Materia,~F\'isica T\'ermica y Electr\'onica and IPARCOS,~Facultad de Ciencias F\'isicas, Universidad Complutense de Madrid, CEI Moncloa, Madrid 28040, Spain\looseness=-4}

\date{\today}

\begin{abstract}
We study neutrino-nucleus charged-current reactions on finite nuclei for the situation in which an outgoing muon and a proton are detected in coincidence, {\it i.e.,} we focus on semi-inclusive cross sections. We limit our attention to one-body current interactions (quasielastic scattering) and assess the impact of different nuclear effects in the determination of the neutrino energy. We identify the regions in phase space where the neutrino energy can be reconstructed relatively well, and study whether the cross section in those regions is significant. 
Our results indicate that it is possible to filter more than 50\% of all events according to the muon and proton kinematics, so that for the DUNE and T2K fluxes the neutrino energy can be determined with an uncertainty of less than 1\% and 3\%, respectively. 
Furthermore, we find that the reconstructed neutrino energy does not depend strongly on how one treats the final-state interactions and is not much affected by the description of the initial state. On the other hand, the estimations of the uncertainty on the neutrino energy show important sensitivity to the modeling of the initial state. 
\end{abstract}

\maketitle
 
\section{Introduction}

The T2K, MINERvA and LArTPC experiments have shown their capabilities to measure the final-state lepton ($\mu^\pm$ or $e^\pm$) and to identify one or more charged particles in coincidence~\cite{Argoneut14,Cai20,Lu18,Walton15,T2K18}. Future experiments, such as DUNE~\cite{DUNE16}, will incorporate an enhanced tracking capability for hadrons in the final state. 
Also, it is worth mentioning the SK-Gd project~\cite{Simpson19}, that improves the detection and identification capabilities of neutrons by adding Gd salts to the SuperKamiokande water tank. 
Compared to inclusive experiments, where only the final lepton is detected, the additional information about the hadrons in the final state, {\it viz.} semi-inclusive scattering, will improve the reconstruction of the incoming neutrino energy. 

In~\cite{Mosel14, Furmanski17, VanOrden19, Munteanu20} explorations of the possibilities arising from the extended knowledge of the final state, specifically focusing on events where there is simultaneous detection of the lepton and a nucleon, were presented. 
In reference~\cite{VanOrden19}, it was proposed to study the average neutrino energy corresponding to a given semi-inclusive event
\ba
    \langle E \rangle = \dfrac{ \int{ dE\, E\, \phi(E) \frac{ d^6\sigma (E)}{ d\Omega_l dk_l d\Omega_N dp_N } } }{ \int dE \phi(E)
 \frac{ d^6\sigma (E)}{ d\Omega_l dk_l d\Omega_N dp_N } } \,,\label{enu_ave}
\ea 
where $\frac{ d^6\sigma(E)}{ d\Omega_l dk_l d\Omega_N dp_N }$ is the six-fold differential cross section for a fixed neutrino energy $E$ and fixed muon and final nucleon kinematics. $\phi(E)$ is a given flux distribution, normalized as
$\int dE \phi(E)=1$.
Furthermore, the standard deviation for the average neutrino energy can be obtained from the first and second statistical moments
\ba 
    \Delta E = \sqrt{ \langle E^2 \rangle - \langle E \rangle^2}\,,\label{e_error}
\ea
where 
\ba
    \langle E^2 \rangle = \dfrac{ \int{ dE\, E^2\, \phi(E) \frac{ d^6\sigma (E)}{ d\Omega_l dk_l d\Omega_N dp_N } } }{ \int dE \phi(E) \frac{ d^6\sigma (E)}{ d\Omega_l dk_l d\Omega_N dp_N } }\,.
\ea

Thus, provided that expressions for the flux and cross section are known, and given that the 4-momenta of the lepton and proton in the final-state are both measured, the average neutrino energy will be defined up to $\langle E \rangle \pm \Delta E$. The impact of different assumptions for the nuclear models involved on the neutrino energies and uncertainties were studied in~\cite{VanOrden19} . To summarize the conclusions 
in said reference, we note that
  \begin{enumerate}
      \item The reconstructed neutrino energy $\langle E\rangle$ depends only moderately on the nuclear model introduced in Eq.~(\ref{enu_ave}). 
      \item The corresponding uncertainty of the reconstructed energy does depend on the nuclear model, but it may be relatively low for a large fraction of the events. 
   \end{enumerate}
The basis for these observations is the fact that the neutrino energy gets essentially blurred by the missing-energy of the nuclear system, that is, the energy required to knock out the observed nucleon, while the nuclear recoil is generally very small. 
In light nuclei, such as $^{12}$C or $^{16}$O, for many events the main contribution will come from the nucleons in the $p$-shell(s), for which the missing-energy is a rather well known quantity.

In this paper we extend the previous analysis presented in~\cite{VanOrden19} by examining the whole phase space and scrutinize in detail the potential for model-independent neutrino-energy determinations. We consider semi-inclusive reactions involving an incident neutrino followed by detection of a charged lepton and a nucleon in the final state together with no produced pions; that is we focus on events of the type CC1$\mu$1$p$0$\pi$, having chosen in the present work to emphasize muons and protons in the final state. As discussed later, this selection of events does not mean that one and only one nucleon is assumed to be present in the final state, only that for sure at least one is present. Indeed, depending on the kinematics chosen there must be other nucleons beyond the one actually detected. The study presented in \cite{VanOrden19} is extended here to include various models that treat the issue of hadronic final-state interactions. We discuss a typical situation, that is, we make specific choices for the measured 4-momenta, in order to orient the reader to the basic characteristics of semi-inclusive reactions before going on to analyze a broad region of the full phase space.

We first introduce the semi-inclusive kinematics and cross section in general terms (Sect.~\ref{sec:CS-Kin}), and then particularize for the quasielastic (QE) interaction (Sect.~\ref{sec:NMQE}). 
In Sect.~\ref{sec:SEAT}, the semi-inclusive cross section is studied for a fixed set of kinematics. Full phase space results are shown in Sect.~\ref{sec:FPSR}. In Sects.~\ref{subsec:FSI} and~\ref{subsec:SFs}, we assess the effect of final-state interactions and the description of the initial state on the neutrino energy determination. In Sect.~\ref{subsec:BestKin}, we show the regions of phase space where the neutrino energy is reconstructed with the lowest error. Finally, we draw our conclusions in Sect.~\ref{sec:Conclusions}.

\section{Kinematics and cross section}\label{sec:CS-Kin}

For the discussion that follows we will assume that the final-state lepton (here a muon is assumed) with 4-momentum $(E_l, \nk_l)$ and a nucleon (here a proton is assumed) with 4-momentum $(E_N, \np_N)$ are detected in coincidence. No other particles are assumed to be detected, although, depending on the specific kinematics assumed, they must be present (see below). We work in the laboratory frame where the target nucleus is at rest, the incoming neutrino momentum is along $\hat{z}$, and the lepton kinematical variables are contained in the $\hat{x}-\hat{z}$ plane. The angle between the incident neutrino and the outgoing lepton is $\theta_l$, while in the chosen coordinate system the polar and azimuthal angles that specify the direction of the outgoing nucleon are $\theta_N$ and $\phi_N$, respectively. The magnitude of the nucleon's 3-momentum is given by $p_N=|\np_N|$. 
Apart from the detected nucleon, the hadronic final state contains an undetected hadronic system having missing 4-momentum $(E_B, {\bf p}_B)$, namely, a total energy of $E_B$ and a missing 3-momentum ${\bf p}_B \equiv {\bf p}_m$. If one denotes by ${\bf q}$ the 3-momentum transferred from the leptons to the hadronic system one has 
\ba
{\bf p}_m = {\bf q} - {\bf p}_N\,. 
\ea
The undetected hadronic system has invariant mass $M_B$ ($M_B^0$ at threshold with $M_B \ge M_B^0$) and total energy 
\ba
E_B =T_B+M_B=\sqrt{ (M_B)^2 + {p_m}^2 }\,,\label{Emiss1}
\ea
which defines the kinetic energy of the unobserved final-state system, $T_B$. From Eq.~(\ref{Emiss1}) one has
\ba
E_B = E - E_l - T_N - (M­_A^0-m_N)\,,\label{Em1}
\ea 
where $M_A^0$ is the target ground-state mass and $T_N=E_N-m_N$ is the kinetic energy of the detected nucleon. This leads to an expression for the so-called missing-energy,
\ba
E_m &=& (M_B-M_B^0)+E_s\non\\ 
    &=& E - E_l - T_N - T_B\,,\label{eq:E_m}
\ea
where $E_s=M_B^0+m_N-M_A^0$ is the separation energy and the (typically very small) recoil kinetic energy difference has been neglected. Clearly, if one knew the missing-energy $E_m$ then the incident neutrino energy $E$ would also be known. 
The magnitude of this missing-momentum $p_m$ is given by
\ba
    p_m &=& \bigl[k^2+{k_l}^2+p_N^2-2kk_l\cos\theta_l-2kp_N\cos\theta_N\non\\
    &+& 2k_lp_N(\cos\theta_l\cos\theta_N+\sin\theta_l\sin\theta_N\cos\phi_N)\bigr]^\frac{1}{2}.\non\\ \label{eq:p_missing}
\ea

Depending on the specific kinematics, {\it i.e.,} the value of the missing-energy, the residual system may be the daughter nucleus in its ground state (this defines the threshold for the semi-inclusive reaction to become possible); or it may be in a discrete excited state (these are states in the residual nucleus that lie below the threshold where a second nucleon can be ejected), and, while they de-excite by $\gamma$-decay, that process is slow on the nuclear timescale and thus these states may be treated effectively as stationary states. Then, at a well-defined threshold a second nucleon must be emitted (this is not optional: there are no nuclear states involving one nucleon and a residual bound nucleus above this point); and so on with more and more particles in the final state in addition to the one special nucleon that is assumed to be detected. At even larger missing energy (roughly 140 MeV) pion production becomes possible (still with the lepton and one nucleon assumed to be detected) and beyond where more particles may be present in the undetected part of the final-state system.

We consider neutrino energy distributions from DUNE and T2K fluxes. 
The flux-averaged semi-inclusive cross section for this process is given by:
\ba
    \langle \dfrac{d^{6}\sigma}{\de k_l\de\Omega_l \de p_N\de\Omega_N} \rangle  = \label{semiincl-cs}
    \int{ dE \phi(E) \dfrac{d^{6}\sigma(E)}{\de k_l\de\Omega_l \de p_N\de\Omega_N} }\,.\non\\
\ea
With the neutrino energy and missing-energy related through energy-momentum conservation, it is convenient to change the integration variable from $E$ to $E_m$, thereby writing the expresion for the semi-inclusive cross section in a more familiar way, found for instance in inclusive and exclusive electron scattering. The semi-inclusive cross section is then given by:
\ba
    \langle \dfrac{d^{6}\sigma}{\de k_l\de\Omega_l \de p_N\de\Omega_N} \rangle 
    = \int\de E_m \, \phi(E)\non\\ 
    \times {\cal F}\frac{k_l^2p_N^2 M^*_B}{(2\pi)^5E_Bf_{rec}} \, \ell_{\mu\nu}H^{\mu\nu}\, ,\label{d6sig0}
\ea 
with 
\ba
    f_{rec} = \left| 1 - \dfrac{E-\hat\nk_\nu\cdot(\np_N+\nk_l)}{E_B} \right|\,
\ea
the nuclear recoil factor, where $\hat\nk_\nu$ is a unit vector along the neutrino beam line. 
${\cal F}$ is given by  
\ba
    {\cal F} = \left(\frac{G_F}{\sqrt2}\right)^2\cos^2\theta_c\,,
\ea 
where $G_F$ is the Fermi constant and $\theta_c$ is the Cabibbo mixing angle.
The charged-current lepton tensor is 
\ba
    \ell_{\mu\nu} &=& \frac{2}{EE_l} \left( K_{i,\mu} K_{f,\nu} + K_{i,\nu} K_{f,\mu} - g_{\mu\nu}K_i\cdot K_f\right. \non\\
    &-&ih \left. \epsilon_{\mu\nu\alpha\beta}K_i^\alpha K_f^\beta \right)\,,
\ea
with $h=+1$ for antineutrinos and $-1$ for neutrinos. The terms $K_i$ and $K_f$ represent the four-momenta of the initial and final leptons involved in the process. 
Up to this point the discussion is general, {\it i.e.,} it does not depend on the nuclear model or the reaction channel. All the complexity regarding the hadronic part of the interaction is in the hadron tensor $H^{\mu\nu}$, which is discussed in Sect.~\ref{sec:NMQE}. 

Equation~(\ref{Em1}) tells us that the neutrino energy can be reconstructed very well from a semi-inclusive sample of events when it is dominated by a narrow and well-known missing-energy region. This should be the case for QE scattering, where the neutrino scatters elastically from a bound nucleon such that the missing-energy is of the order of the binding energy of the nucleon. 
In the following, we focus on the QE reaction and we describe the content of the different nuclear models employed in this work.

\subsection{Nuclear models for quasielastic scattering}\label{sec:NMQE}

We focus on neutrino-induced charge-current QE scattering, with one boson exchanged between lepton and one-body hadron currents, within the impulse approximation. Thus we do not consider meson-exchange currents (MEC) nor the processes where real pions may be produced in the final state.
In this work we restrict our attention to $^{16}$O, although analyses along these lines can easily be performed for other nuclei~\cite{Megias18}. 
 
We will describe the initial state as a set of relativistic mean-field (RMF) wave
functions that correspond to different shells labeled by the relativistic quantum number $\kappa$. The hadron tensor for a given shell is:
\ba
    H_\kappa^{\mu\nu} &=& \rho_\kappa(E_m)\non\\
     &\times& \sum_{m_j, s_N}[J_{\kappa,m_j,s_N}^\mu(Q,P_N)]^*J_{\kappa,m_j,s_N}^\nu(Q,P_N)\,,\non
\ea
where $Q=K_i-K_f$, $\rho_\kappa(E_m)$ is the missing-energy density and
$J_{\kappa,m_j,s_N}^\mu$ is the hadron current, in momentum space defined as:
\begin{flalign}
    &J_{\kappa,m_j,s_N}^\mu(P_N,Q) && \\
    &=\int \de\np\overline{\Psi}^{s_N}(\np_N,\nq+\np)\ {\cal O}^\mu(Q)\ \Psi_\kappa^{m_j}(\np)\,,&&\non 
\end{flalign}
$m_j$ is the third component of the total angular momentum $j$ of the bound nucleon and $s_N$ is the spin projection of the final nucleon. $\Psi$-functions are relativistic independent-particle wave functions describing the bound and scattered nucleon and ${\cal O}^\mu$ is the usual boson-nucleon-nucleon operator, for which we use its CC2 form (see~\cite{Udias93,Udias95,Martinez06} for details).

In the framework of a pure shell model, one simply has $\rho_\kappa(E_m)=\delta(E_m-E_m^\kappa)$, where $E_m^\kappa$ is the energy eigenvalue for a given shell. 
This missing-energy distribution of the shell model, however, is only a first approximation to the one in real nuclei, that for the valence shells has been measured in quite a number of electron scattering experiments. Generally speaking, effects beyond mean field, such as short- and long-range correlations, modify the actual missing-energy distribution predicted by the shell model. 

In this work, as already done in~\cite{VanOrden19}, we also take as a reference the spectral function formalism. The spectral function incorporates the probability of finding the nucleon in the initial state with certain energy and momentum. It includes the depletion of the occupation of the shell-model states, and the appearance of nucleons at deeper (namely, higher) missing-energy, in both cases due to correlations, both long- and short-ranged. For the purpose of this work, we will consider the Rome spectral function~\cite{Benhar94,Benhar05} as a fair representation of the missing-energy and missing-momentum distribution of the nucleons in the target nuclei, as measured in electron scattering experiments.  
Other spectral function calculations are available in the literature~\cite{Amir97, Ivanov19}; however, since all of these have been constrained to some extent to reproduce the $(e,e'p)$ electron scattering cross sections, their results will not differ much from the ones here.

The spectral function is easy to incorporate in a fully factorized, plane-wave calculation (as in~\cite{VanOrden19,Franco-Patino20}), with the exclusive 6-differential cross section given by
\ba
    \dfrac{d^{6}\sigma}{\de k_l\de\Omega_l \de p_N\de\Omega_N} = K S(E_m,p_m) \sigma_{\nu N}\,.
    \label{eq:fact}
\ea
$K$ is a function containing kinematic factors, $S(E_m,p_m)$ is the initial-state spectral function and $\sigma_{\nu N}$ is the charged-current elastic neutrino-nucleon cross section for an off-shell nucleon with initial momentum $p_m$ (see Refs.~\cite{Benhar05,VanOrden19,Franco-Patino20} for details). 

\begin{figure}[htbp]
\centering  
(a)\includegraphics[width=.3\textwidth,angle=270]{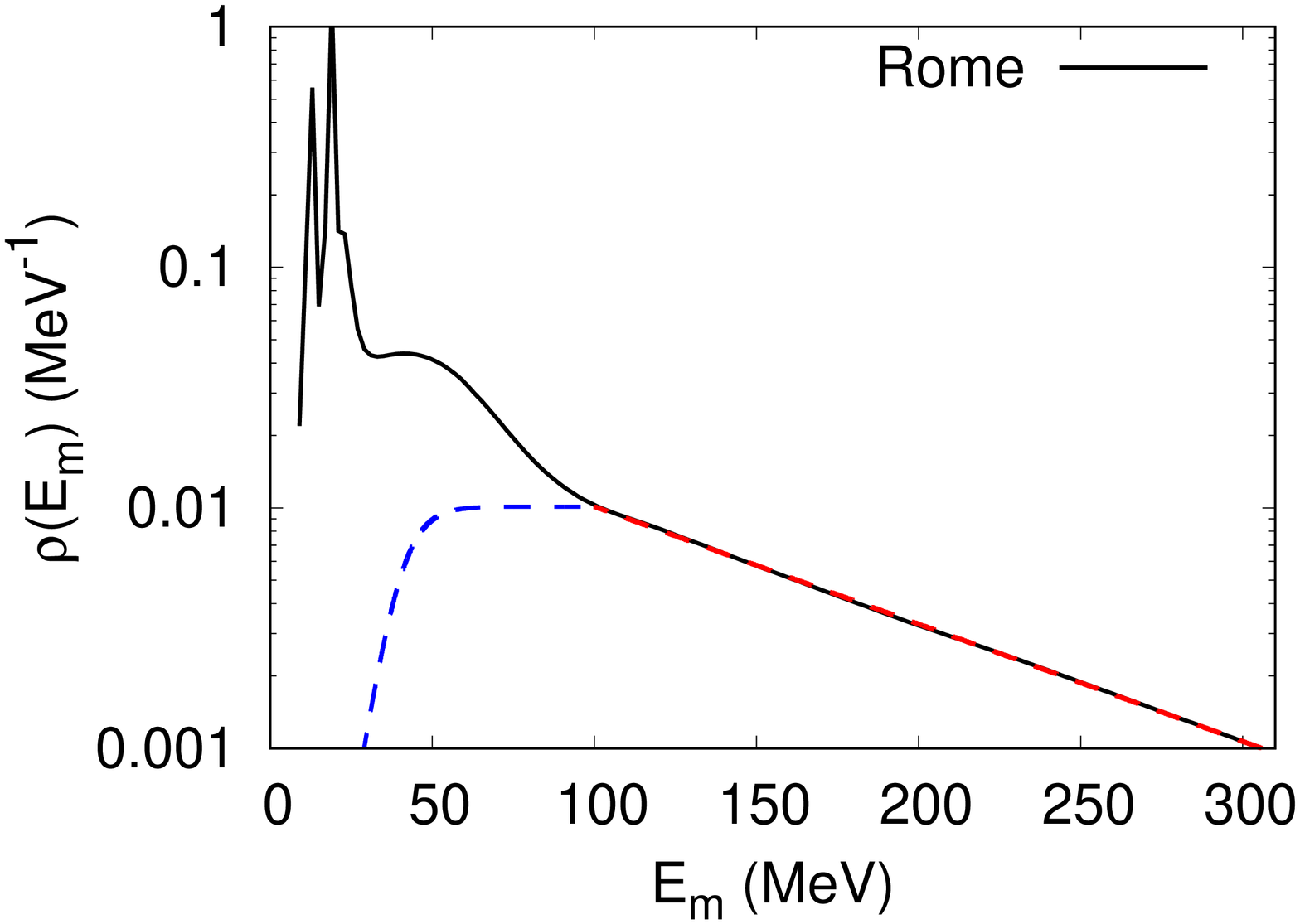}
(b)\includegraphics[width=.3\textwidth,angle=270]{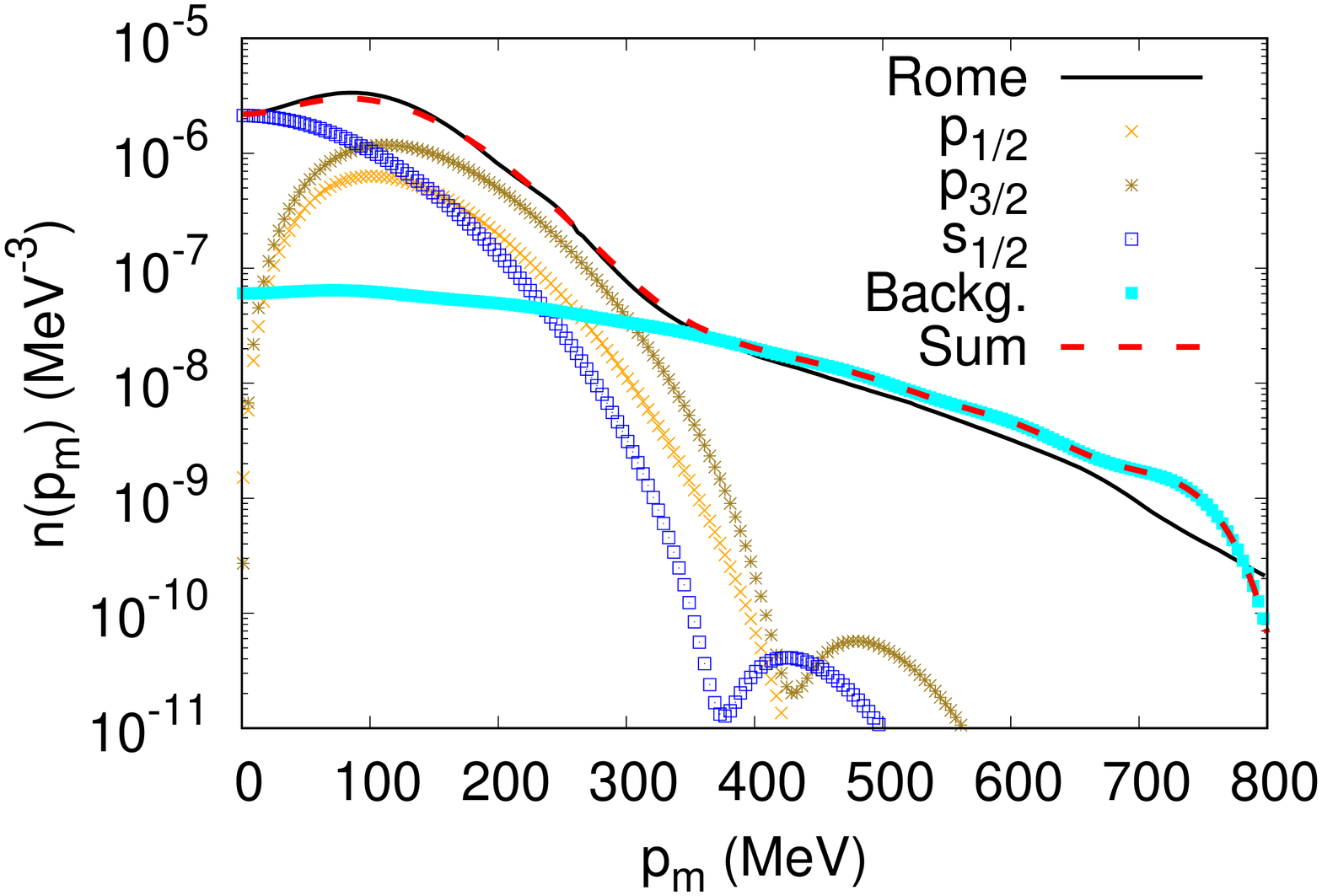}
\caption{(Color online) (a) The Rome spectral function (integrated over $\np_m$) as a function of the missing-energy. Our parameterization of the background is represented by the dashed line. (b) Momentum distributions from the Rome spectral function and from our representation.}\label{fig:rho&n}
\end{figure}

In this work, however, we use a representation of the spectral function amenable to our relativistic distorted-wave and unfactorized calculations, that is sufficient to achieve the goals of this work. 
Thus, we divide the missing-energy phase space into several regions. In the lowest-energy region, below the two-nucleon threshold, we will have the $p$-shell states, with an energy dependence not given by $\delta$ functions but rather by the energy distribution seen in the spectral function [Fig.~\ref{fig:rho&n}(a)]. We identify each shell with a different region of missing-energy following the analysis presented in~\cite{VanOrden19}. The high $E_m$ and $p_m$ part of the spectral function due to correlations, is accounted for by introducing an $s$ wave~\cite{Colle14,Ryckebusch15}, broad in momentum space (narrow in $r$-space, approximately $0.85$ fm), that is fitted to reproduce the momentum distribution of the spectral function in this region. The momentum distribution obtained in this representation, compared to the one of the Rome spectral function is shown in Fig.~\ref{fig:rho&n}(b).
The specific regions and occupation numbers are summarized in Table~\ref{Table}. Above the two-nucleon knockout threshold, the independent-particle shells and the background coexist. To account for it, we have parameterized the missing-energy profile of the background in the region $25 < E_m < 100$ MeV [dashed-blue line in Fig.~\ref{fig:rho&n}(a)]. In the region $E_m>100$ MeV we assume that there is only background, which is well described by an exponential fall-off. The explicit expressions for these functions are given in Appendix~\ref{backg-function}. 

\begin{figure}[htbp]
\centering  
(a)\includegraphics[width=.3\textwidth,angle=270]{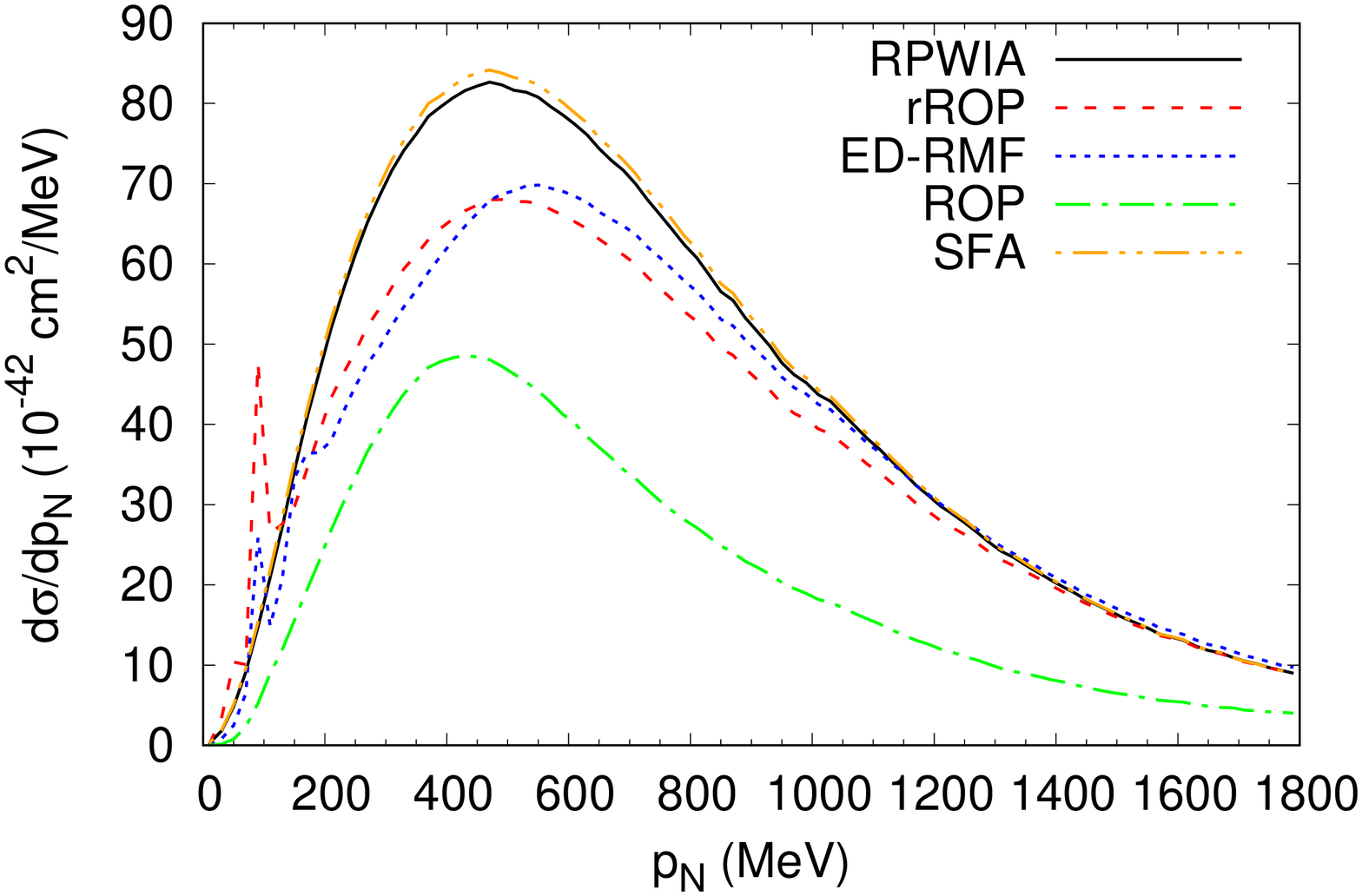}
(b)\includegraphics[width=.3\textwidth,angle=270]{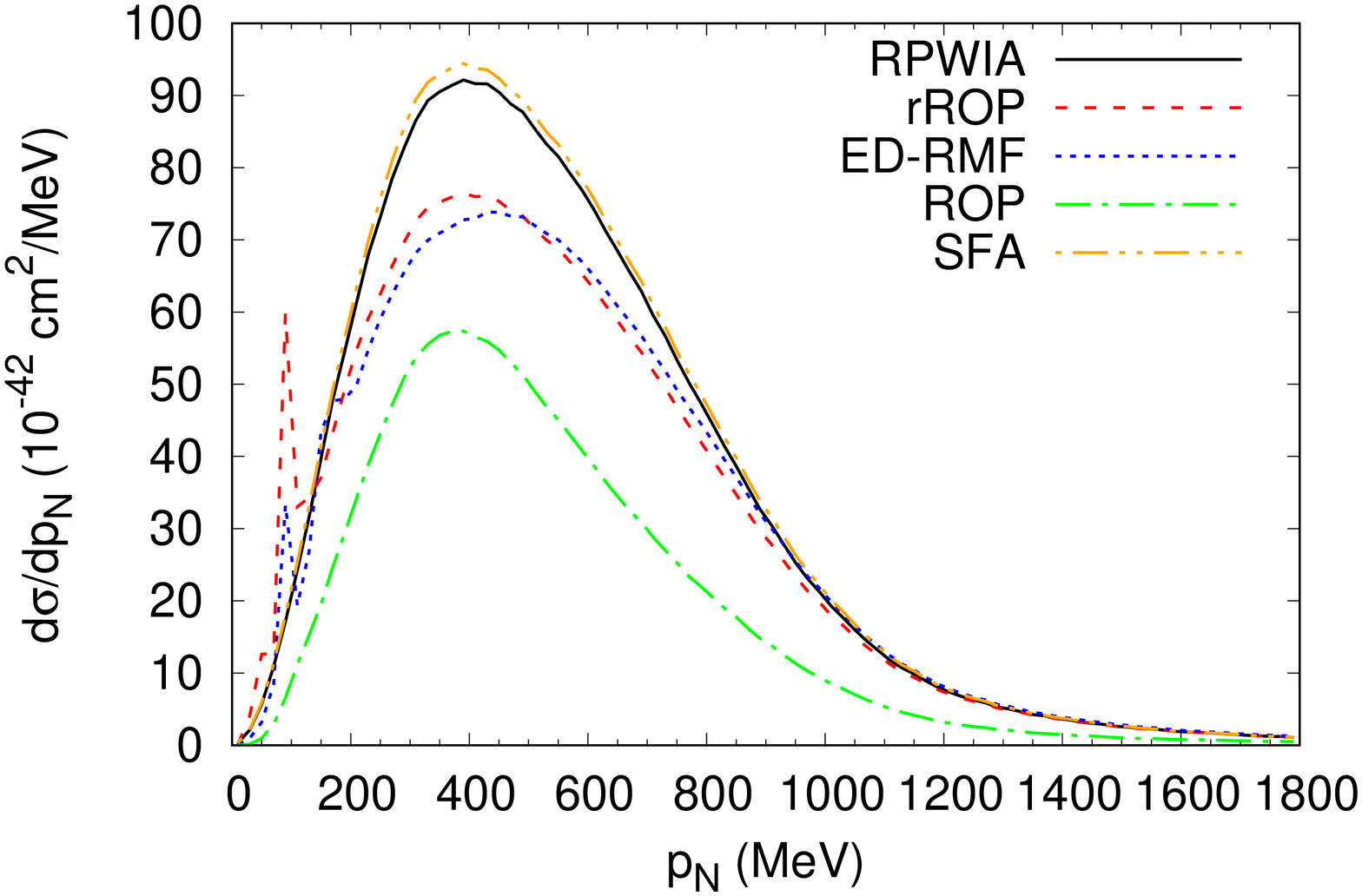}
\caption{(Color online) Single-differential cross sections for the DUNE (a) and T2K (b) fluxes with the models discussed in the text.}\label{fig:sdc}
\end{figure}

The representation of the initial state we use in our modeling essentially contains the same (albeit somewhat simplified) missing-energy and -momentum structure of the Rome spectral function. 
Indeed, despite the fact that our calculations are unfactorized (due to the relativistic effects~\cite{Caballero98a,Gardner94} and eventually FSI~\cite{Udias95}), our cross section results, when FSI are neglected, are within few percent of the ones obtained with the fully factorized calculation based on the spectral function approach of the Rome model (see, {\it e.g.,} Fig.~\ref{fig:sdc}), showing that ingredients preventing factorization and negative-energy components have a small effect on the cross sections computed here~\cite{Ivanov13}. This means that our results are representative of what MC event generator based on the spectral function+factorized calculations, even considering FSI, would produce.

\begin{table}
\renewcommand{\tabcolsep}{3mm}
\centering
\begin{tabular}{ c | c | c }
 $E_m$ (MeV) & Shells & $^{16}$O \\  
 \hline
 \hline
 $0-16.5$ & p$_{1/2}$ & 1.51  \\
 \hline
 $16.5-25$ & p$_{3/2}$ & 3.47 \\
 \hline
 $25-100$  & s$_{1/2}$ + backg. & 2.22 \\
           & s$_{1/2}$  & 1.62 \\
           & backg.     & 0.60 \\
 \hline
 $100-300$ & backg.     & 0.80 
\end{tabular}
\caption{Correspondence between missing-energy regions and shells in oxygen. In the last column are the occupation numbers.}
\label{Table}
\end{table}

To summarize, in our rationale (impulse approximation) the spectral function is a reasonably realistic representation of the initial state (energy and  momentum) of the nucleon which will contribute, for the reaction at hand, to a final state in which we will see at least one knocked-out nucleon. We want to emphasize that in this way, we incorporate the experimental constraints provided by electron scattering experiments on missing-energy and -momentum distribution in the initial nucleon, certainly much better than within any Fermi gas approach or a pure shell model. 

In the following we discuss the description of the final state that we incorporate in our calculations. 
In a way, the final state is to a large extent determined by the experimental signal that is to be described:
{\it is there only a proton and no other hadrons in the final state? }
Or, at the other extreme, {\it does the experimental signal contain every event for which at least one proton is seen? }
Generally, the actual experimental situation will be a combination of these two extreme, simplified, cases. We will look for some representation of these situations, in order to study the effects of the different definitions of the final state on the reliability of the determination of the neutrino energy.

\begin{enumerate}
\item {\bf Real relativistic optical potential (rROP) or energy-dependent relativistic mean-field (ED-RMF).} 
In this case the final-state nucleon is a solution of the Dirac equation with a real potential, and the absence of an imaginary part in the potential means that no flux is lost. 
We will use both the rROP and the ED-RMF, the difference between them being in the relativistic mean-field potential seen by the final nucleon~\cite{Gonzalez-Jimenez20}. 
In the rROP case, we use the real part of the (energy-dependent A-independent oxygen) optical potential EDAI-O~\cite{Cooper93}, while the ED-RMF is the RMF potential (the same as for the bound state) but multiplied by a phenomenological function that weakens the potential for increasing nucleon momenta~\cite{Gonzalez-Jimenez19}. The nucleon wave functions within the ED-RMF model are eigenstates of the same Hamiltonian, therefore, orthogonality between initial and final state is satisfied, {\it i.e.,} Pauli blocking is consistently incorporated. This orthogonality is not as good in the rROP model; therefore, one should be cautious when the momentum of the nucleon is smaller than approximately $p_N<300$ MeV. For $p_N$ larger than around $400$ MeV, the overlap between the initial and final states is negligible and hence, orthogonality is not an issue. 
These approaches, in the pure shell-model case, have been shown to succesfully describe inclusive scattering data for both neutrinos and electrons~\cite{Gonzalez-Jimenez19,Gonzalez-Jimenez20}.
In the case of semi-inclusive scattering, as considered here, these models would provide an estimate of the situation in which the hadronic final-state signature consists of {\it at least one proton}. There may be other nucleons that arise from correlations in the initial state or other hadrons, such as nucleons or pions, produced during the interaction of the knocked-out nucleon(s) with the residual system. The scenarios with two or more nucleons in the final configurations necessarily arise from kinematics in which the missing-energy, $E_m$, is above the two-nucleon knockout threshold.
 \item {\bf Elastic-only channel in the FSI, represented by a complex relativistic optical potential (ROP).}
The whole ROP which is fitted to reproduce elastic proton-nucleus scattering, and that contains real and imaginary parts, is employed in this case. Hence, this calculation allows us to estimate the probability that the (primary) nucleon knocked out during the interaction with the boson propagates through the residual system with elastic scattering only. This primary nucleon does not knock out other nucleons or create new hadrons in its way out, nor does it lose energy in any way apart from elastic recoils. 
The angle of the nucleon can change, though. 

This situation can be considered equivalent to running the cascade models retaining only elastic interactions. However, the calculation presented here is of course a fully quantum mechanical one. 
The loss of flux implied by the imaginary part of the potential would lead to a strong underestimation of the inclusive cross section, in which the outgoing nucleon remains undetected.  
Thus, the ROP estimation would be more in line with an experimental signature of having one proton detected, and no other hadron. 
Additional hadrons, however, could appear due to correlations in the initial state and subsequent FSI of the secondary nucleon. 
The `one proton and one only' signature might be enforced in the calculation by keeping the missing-energy below the two-nucleon emission threshold, for the initial state, and the elastic (full optical potential) condition in the final state.

These ingredients, but with the pure shell model, have been widely applied to analyze exclusive $(e,e'p)$ data on different target nuclei~\cite{Udias93,Udias95,Udias01} for which there is certainty that there is only one proton in the final state. 
The theoretical prediction is scaled to the data by the spectroscopic factor and the agreement with data is outstanding. As in this work we use the spectral function representation, the scale factor is already included in the theory. Were we to use these ingredients to analyze the exclusive data, we would find very good agreement for the $p_{1/2}$-shell. For the $p_{3/2}$-shell, one has to consider that with the binning of the spectral function used here, the small states located at missing-energies around the main $p_{3/2}$ state are all summed up. When all these states are taken into account, the agreement is good, at least for low-to-moderate values of $E_m$. 
\item {\bf Relativistic plane-wave impulse approximation (RPWIA):} The final nucleon is described by a relativistic plane wave, so final-state interactions and Pauli blocking effects are neglected. Although RPWIA is an oversimplified description of the process that is not suitable for some experimental situations, it is a very common first-order theoretical estimate that can help in improving our understanding of the dynamical properties involved in semi-inclusive processes. We include it here as a reference.
\item In addition to the previous models, and as a further reference, we present here our calculations in the factorized {\bf spectral function approach (SFA)} described by Eq.~(\ref{eq:fact}). The SFA is completely or partially implemented in some of the Monte Carlo event generators used in neutrino experiments~\cite{T2K20,NuWro-web} and thus it is useful for reference.  
\end{enumerate}

In Fig.~\ref{fig:sdc} we compare the predictions of the models described above for the single-differential cross sections as a function of the final nucleon momentum  for the DUNE and T2K fluxes. As expected, our representation of the spectral function calculation in RPWIA and the factorized SFA results are essentially identical and yield the largest cross sections, due to the absence of Pauli blocking and of the distortion, that shifts the strength to regions kinematically suppressed~\cite{Nikolakopoulos19,Gonzalez-Jimenez19}. The ratios of the peak values of the rROP results to those of the RPWIA in the two panels in the figure are 0.84 (panel a) and 0.82 (panel b). The rROP and ED-RMF results are very similar both in shape and magnitude. The ROP result has a very similar shape to that of the rROP, albeit smaller in magnitude as a consequence of the restriction to elastic-only propagation for the knocked-out proton: the ratios at the peaks are 0.71 (panel a) and 0.75 (panel b). 
These results are consistent with those found in previous studies, {\it e.g.}~\cite{Martinez06}.

The main idea in this work is to compare the accuracy in the energy determination performed based on cross sections and kinematics (missing-energy and -momentum) given by these models, which either correspond approximately to possible experimental `samples' or `signatures' or to common ingredients of MC event generators. We will first do so for a fixed final-state kinematics (Sect.~\ref{sec:SEAT}) and then extend the analysis to the full phase space (Sect.~\ref{sec:FPSR}).

\section{Selected events and trajectories}
\label{sec:SEAT}

To get a deeper understanding of the procedures used to reconstruct the neutrino energy and the definition of its error, we analyze in detail the semi-inclusive cross section for a fixed set of kinematics. After this discussion, which should help in understanding the basic characteristics of semi-inclusive reactions of the type considered in the present study as well as those involving electron scattering, specifically, $(e,e'p)$ reactions, in the following section we present an analysis that extends over the full phase space. 

\begin{figure}[htbp]
\centering  
\includegraphics[width=.34\textwidth,angle=270]{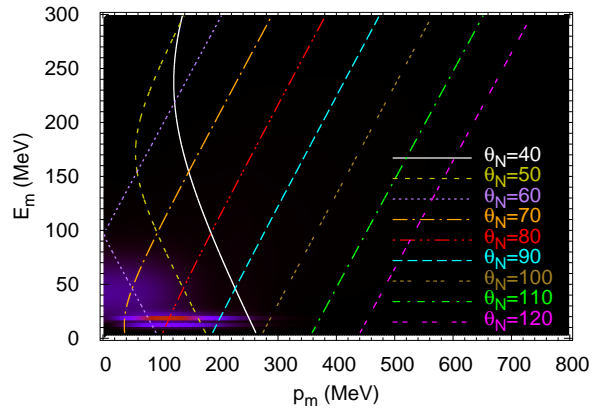}
\caption{(Color online) The $E_m-p_m$ trajectories are shown for selected ``typical'' kinematics: $E_l=3800$ MeV, $\theta_l=7$ deg, $T_N=140$ MeV, $\phi_N=180$ deg. Each line corresponds to a different value of the proton scattering angle $\theta_N$ (in  degrees). Here, we plot the Rome spectral function as a background to allow one to easily identify the different regions of the spectral function that are crossed by the trajectories. }\label{fig:trajectories}
\end{figure}

From Eqs.~(\ref{eq:E_m}) and (\ref{eq:p_missing}, it is clear that for fixed values of the observable parameters the values of $E_m$ and $p_m$ are determined for each value of $E$. Thus, for fixed values of the observables, the integral over $E$ in Eq.~(\ref{semiincl-cs}) follows a curve or trajectory in the $p_m$-$E_m$ plane. In Fig.~\ref{fig:trajectories} trajectories are shown for selected kinematics for the detected particles, namely, $E_l=3800$ MeV, $\theta_l=7$ deg, $T_N=140$ MeV, $\phi_N=180$ deg. This represents a ``typical'' situation and clearly many others could have been chosen to illustrate the basic behavior to be expected. What is varied here is the polar angle for the detected proton, $\theta_N$. As stated above, as one goes along a given trajectory the neutrino energy $E$ varies. It starts at the lower boundary which defines the threshold for the semi-inclusive reaction to occur and grows as the missing-energy increases. The semi-inclusive cross section thereby produced for a particular neutrino flux is then obtained as a line integral along the specific trajectory. In effect, each event where a muon and proton are detected in coincidence corresponds to a specific trajectory. Only were one to have a mono-energetic neutrino beam would a point in the $E_m-p_m$ plane be selected; however, with broad-band beams a weighted line integral is required. That said, one still sees a striking pattern to the behavior one should expect when performing such line integrals. The strength in the Rome spectral function, which should provide a good starting point for the characteristics to be expected in semi-inclusive reactions, is extremely localized. One sees the largest concentration of strength where the $p$-shells are located (at around $E_m = 20$ MeV) with less where the broad $s$-shell is located (at around $E_m = 50$ MeV); at still larger values of $E_m$ (and $p_m$) the Rome group spectral function does have some strength, although it is spread over a wide region in the $E_m-p_m$ plane and is too small to be seen in this representation. Furthermore, we note that pion production cannot occur until one reaches $E_m \sim m_\pi$ and that it is not appreciable until $E_m \sim m_\Delta - m_N \sim 300$ MeV. Of course, in the present work we are considering only events that have no pions.

We do have some knowledge about this generic behavior of the distribution of strength from inclusive electron scattering, $(e,e')$. Inclusive scattering corresponds to performing integrals over specific regions in the $E_m-p_m$ plane~\cite{Day90, Moreno14, Amaro20}. A very similar pattern is expected in that case and what is found for $^{16}$O is that somewhat over 50\% of the inclusive cross section stems from the $p$-shells, about 25\% comes from the $s$-shell region and the rest comes from a broad region at higher missing-energy. This is borne out in semi-inclusive $(e,e'p)$ electron scattering studies, although only a few data exist in that case. Thus, we expect this general picture to be the case for CC$\nu$ reactions. Indeed, in the semi-inclusive case that forms the focus of the present work, we expect that the line integrals discussed above are at their largest when the trajectories cross the peaks in the $p$-shell region together with somewhat reduced strength coming from the $s$-shell region and a negligible amount arising from the higher-$E_m$ region. Given that this is the case, we can then also expect that the optical model based approach discussed above is a reasonable one, whereas, were the high-$E_m$ region to be important this would not be obvious. 

In passing we note that, while other choices of two variables to replace $E_m$ and $p_m$ can of course be made, the generic behavior seen here strongly suggests that the present choice is a good one and that other choices may not reflect the highly localized nature of the nuclear response.

\begin{figure*}[htbp]
\centering  
(a)\includegraphics[width=.33\textwidth,angle=270]{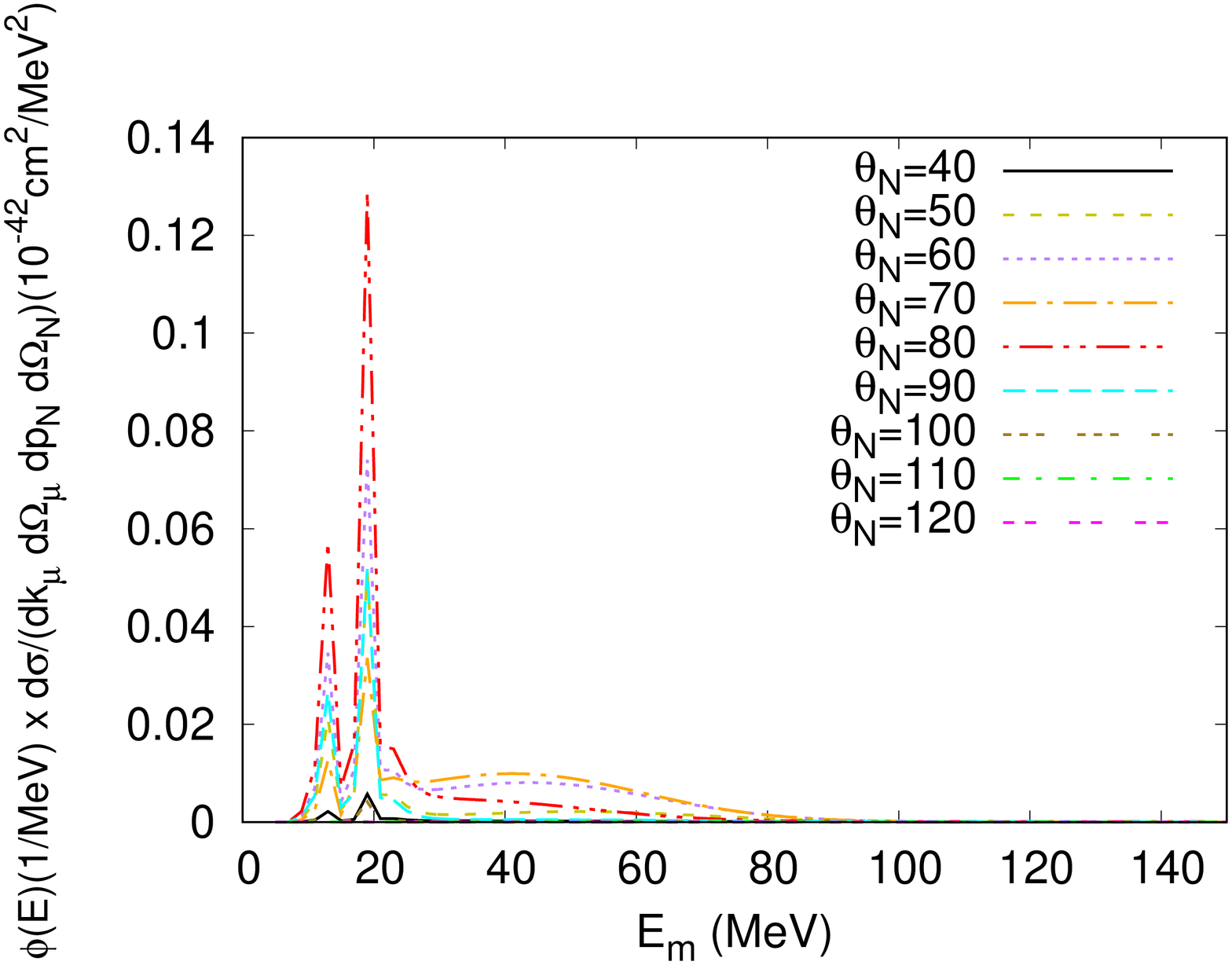}
(b)\includegraphics[width=.31\textwidth,angle=270]{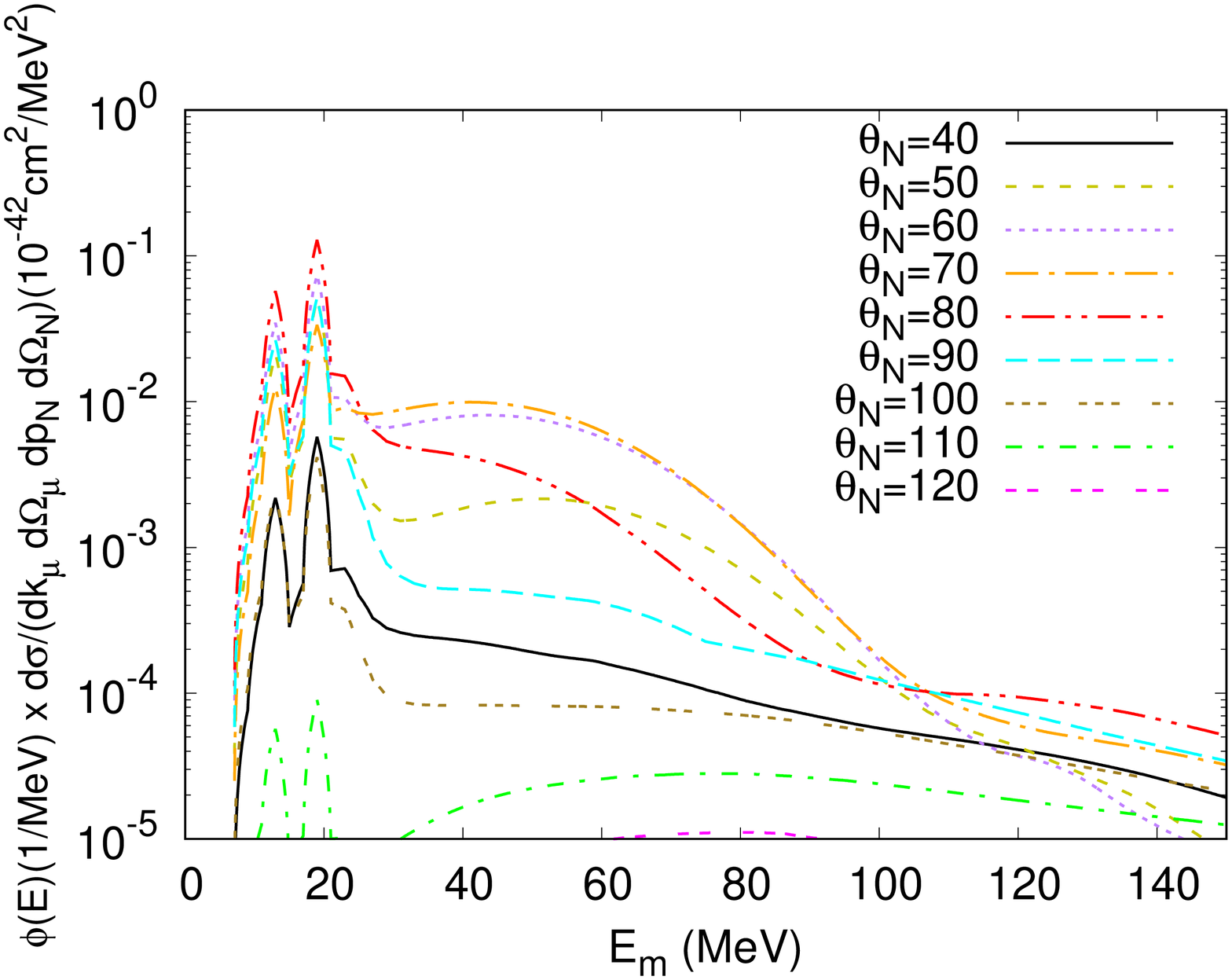}
\caption{(Color online) The six-fold differential cross section computed with the DUNE flux and the SFA model is shown as a function of the missing-energy $E_m$ on linear (a) and semi-log (b) scales. Muon and proton variables are fixed to: $E_l=3800$ MeV, $\theta_l=7$ deg, $T_N=140$ MeV, $\phi_N=180$ deg, as in Fig.~\ref{fig:trajectories}. }\label{fig:6fold}
\end{figure*}

Let us next see what the various models yield for the weighted cross sections. In Fig.~\ref{fig:6fold} we represent the integrand of Eq.~(\ref{semiincl-cs}) along the trajectories shown in Fig.~\ref{fig:trajectories}, {\it i.e.,} the six-fold differential cross section for fixed muon and proton kinematics as a function of the missing-energy. 
By varying the angle $\theta_N$, the cross section changes its magnitude and its shape but, in general, we observe a profile that resembles the $\rho(E_m)$ function used in our model [Fig.~\ref{fig:rho&n}(a)], {\it viz.,} two prominent peaks corresponding to the $p$-shells, a wide belly for the $s$-shell and a background that extends up to high missing-energies. Clearly, as expected, the $p$-shell strength is largest, the $s$-shell strength is smaller and the high-$E_m$ strength is completely negligible, being down by several orders of magnitude. By examining these results in the light of the trajectories shown in Fig.~\ref{fig:trajectories} we see that the general behavior we expect to occur is borne out. For example, the trajectories for $\theta_N = 60$ and 80 degrees both pass through the $p$-shell region near its peak. However, one trajectory intersects the $s$-shell region more than the other one does and this results in relatively different amounts from the $s$-shell compared with the $p$-shell. Or, if one has events that correspond to large values of $\theta_N$ with the chosen kinematics introduced above, then the cross sections are very small, as they should be, since neither the $p$- nor $s$-shell regions are crossed.

\begin{table*}[htbp]
\centering
\begin{tabular}{ c | c | c | c | c }
 $\theta_N$ (deg) & $\langle E\rangle$ (MeV) & $\Delta E/\langle E\rangle$ (\%) & c.s. ($10^{-42}$cm$^2/$MeV$^2$) & $p$-shell weight ($r$)\\  
 \hline
 $40$  & 3976 & 0.80 & 0.036 & 0.60 \\
 $50$  & 3970 & 0.53 & 0.30 & 0.65 \\
 $60$  & 3974 & 0.51 & 0.65 & 0.48 \\
 $70$  & 3980 & 0.51 & 0.56 & 0.29 \\
 $80$  & 3965 & 0.44 & 0.68 & 0.76 \\
 $90$  & 3964 & 0.56 & 0.24 & 0.86 \\
 $100$ & 3981 & 1.2  & 0.026 & 0.66 \\
 $110$ & 4043 & 1.6  & 0.0038 & 0.11 \\
 $120$ & 4061 & 1.5 & 0.0014 & 0.0025 
 \end{tabular}
\caption{ The kinematics chosen here are the following: $E_l=3800$ MeV, $\theta_l=7$ deg, $T_N=140$ MeV, and $\phi_N=180$ deg and different $\theta_N$ (first column). We show the mean neutrino energy $\langle E\rangle$, its relative one-sigma error $\Delta E/\langle E\rangle$, the cross section (c.s.), and the weight of the $p$-shells [$r$, defined in Eq.~(\ref{weight-r})]. We have used the DUNE flux and the SFA model. 
}
\label{Table2}
\end{table*}

Thus, we are interested in events that fulfill two conditions:
\begin{enumerate}
 \item The neutrino energy needs to be reconstructed rather well. For this to happen, most of the strength should be concentrated in a small $E_m$ region.  
 \item The cross section is large and, hence, the probability of finding events around such kinematics is high. 
\end{enumerate}
Given the structure of the spectral function, it seems very likely that for events in which the $p$-shells dominate these requirements will be fulfilled. Another reason that makes the `$p$-shell dominated events' particularly interesting is that in $^{16}$O the $p$-shells are below the two-nucleon emission threshold. Hence, this is the part of the spectral function that can be determined experimentally from exclusive $(e,e'p)$ experiments and, therefore, is well constrained.

By looking at Fig.~\ref{fig:trajectories}, the trajectories of greatest interest are those in the region around $80<\theta_N<100$ deg; however, for larger $\theta_N$ the size of the cross section starts to decrease rapidly. To quantify and better assess these results, for each of the curves shown in Fig.~\ref{fig:trajectories}, we present in Table~\ref{Table2} the values of: $\langle E\rangle$, $\Delta E/\langle E\rangle$, the cross section (integrated over $E_m$), and the weight of the $p$-shell region ($0 < E_m < 25$ MeV)~\footnote{We compute the weight of the $p$-shells with respect to the full cross section as: 
\ba
    r = \dfrac{ \int^{25}_0{ dE_m\, \phi(E) \frac{ d^6\sigma (E)}{ d\Omega_l dk_l d\Omega_N dp_N } } }{ \int^{300}_0{ dE_m\, \phi(E) 
\frac{ d^6\sigma (E)}{ d\Omega_l dk_l d\Omega_N dp_N } } }\,,\label{weight-r}
\ea 
with $E_m$ in MeV.}. For the particular kinematics studied here, the error in the reconstructed energy is small in all cases. 
Also, one does not see a simple correlation between $p$-shell dominance and small error. For example, for $\theta_N=40$ deg the error is larger than for $\theta_N=60$ deg, while the weight of the $p$-shells is smaller in the latter case. However, we should not draw general conclusions from the study of just one particular choice of kinematics. Thus, in the next section we address these and other questions in a more systematic way by analyzing the whole phase space.

\section{Full phase-space results}
\label{sec:FPSR}
 
In this section we extend the previous analysis (restricted to very particular kinematics) to the whole phase space, using the DUNE and T2K neutrino fluxes, peaked around $E\sim2.5$ GeV and $E\sim0.6$ GeV, respectively. Also, we study the dependence of the outcomes upon the different models described in Sect.~\ref{sec:FPSR}.
With that purpose in mind, we populate the whole phase space with a few millions of events distributed following the six-fold differential semi-inclusive cross section [Eq.~(\ref{semiincl-cs})] given by each of the models mentioned above, and for each event we compute the average neutrino energy ($\langle E\rangle$) and its error ($\Delta E$) according to Eqs.~(\ref{enu_ave}) and (\ref{e_error}).

We stress that the analysis presented here does not take into account non-QE interactions that can contribute to the 1 muon-1 proton sample. Furthermore, it does not account for detector efficiency and resolution. For these reasons, the error in the reconstructed neutrino energy reported here should be understood as a lowest intrinsic bound. Also, the reconstructed energy will likely change when non-QE processes are explicitly considered in the cross section. 

\begin{figure*}[ht]
\centering  
(a)\includegraphics[width=.32\textwidth,angle=270]{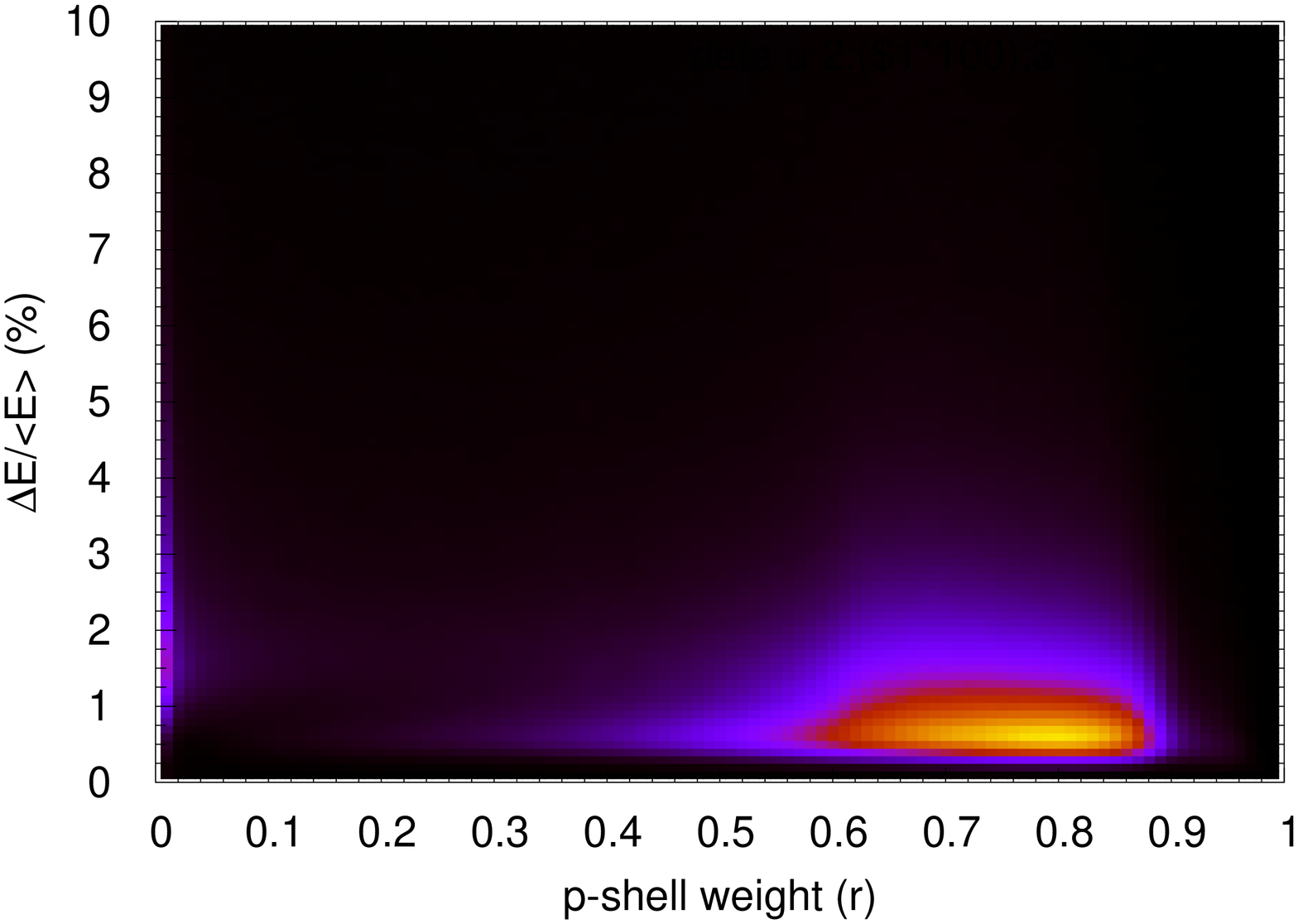}
(b)\includegraphics[width=.32\textwidth,angle=270]{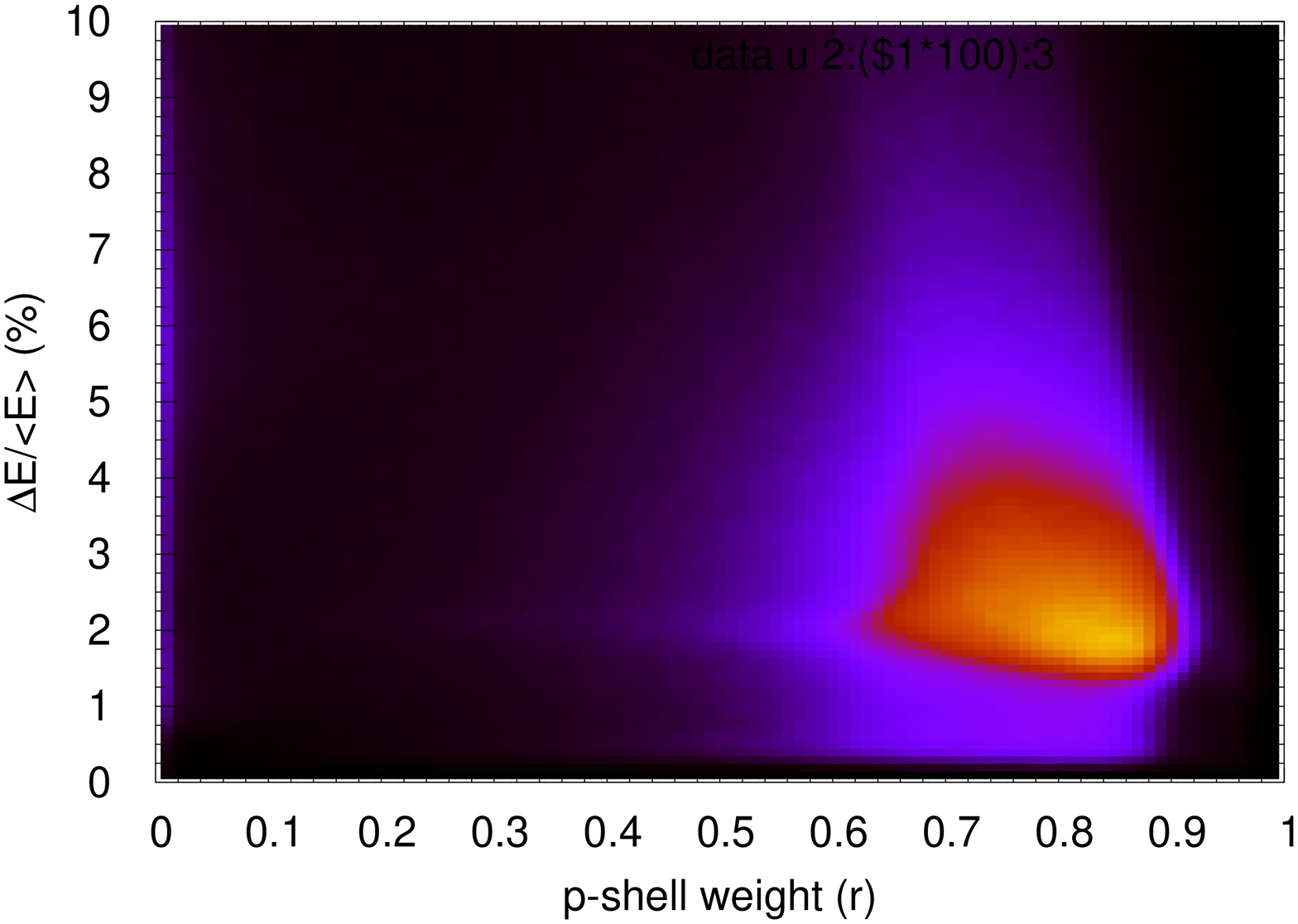}
\caption{(Color online) 2D histogram with the number of events in bins of the relative error (\%) and $p$-shell weight. The DUNE and T2K fluxes were employed in panels (a) and (b), respectively. The calculations correspond to the rROP model, but similar results are found with the other approaches. Brighter areas correspond to bins with more events.}\label{fig:Error_vs_ratio}
\end{figure*}

It is important to realize that the uncertainty in the neutrino energy for these semi-inclusive experiments is basically given 
by the relevant range of missing-energy.
We can see that for QE processes (only nuclear excitations and no pions produced), from about $E_m > 200$ MeV the cross section has fallen by several orders of magnitude, and hence the neutrino energy is effectively restricted to a limited region of $E_m$.
Thus, in absolute terms, DUNE and T2K events have similar errors in the neutrino energy, while in relative terms, the errors for the DUNE events are smaller, owing to the larger neutrino energies in the DUNE flux. 
This is clearly shown in Fig.~\ref{fig:Error_vs_ratio} that presents a 2D histogram with the number of events in bins of $\frac{\Delta E}{<E>}$ and $p$-shell weight ($r$). We see that the bulk of the events (intense yellow-orange regions) concentrates in a small region corresponding to $0.6<r<0.9$ and $\frac{\Delta E}{<E>}<1\%$ ($1<\Delta E<3\%$) for the DUNE (T2K) flux. 
This means that for most of these `small-error events', it is likely that the detected proton was ejected from a $p$-shell.

\begin{figure*}[htbp]
\centering  
(a)\includegraphics[width=.32\textwidth,angle=270]{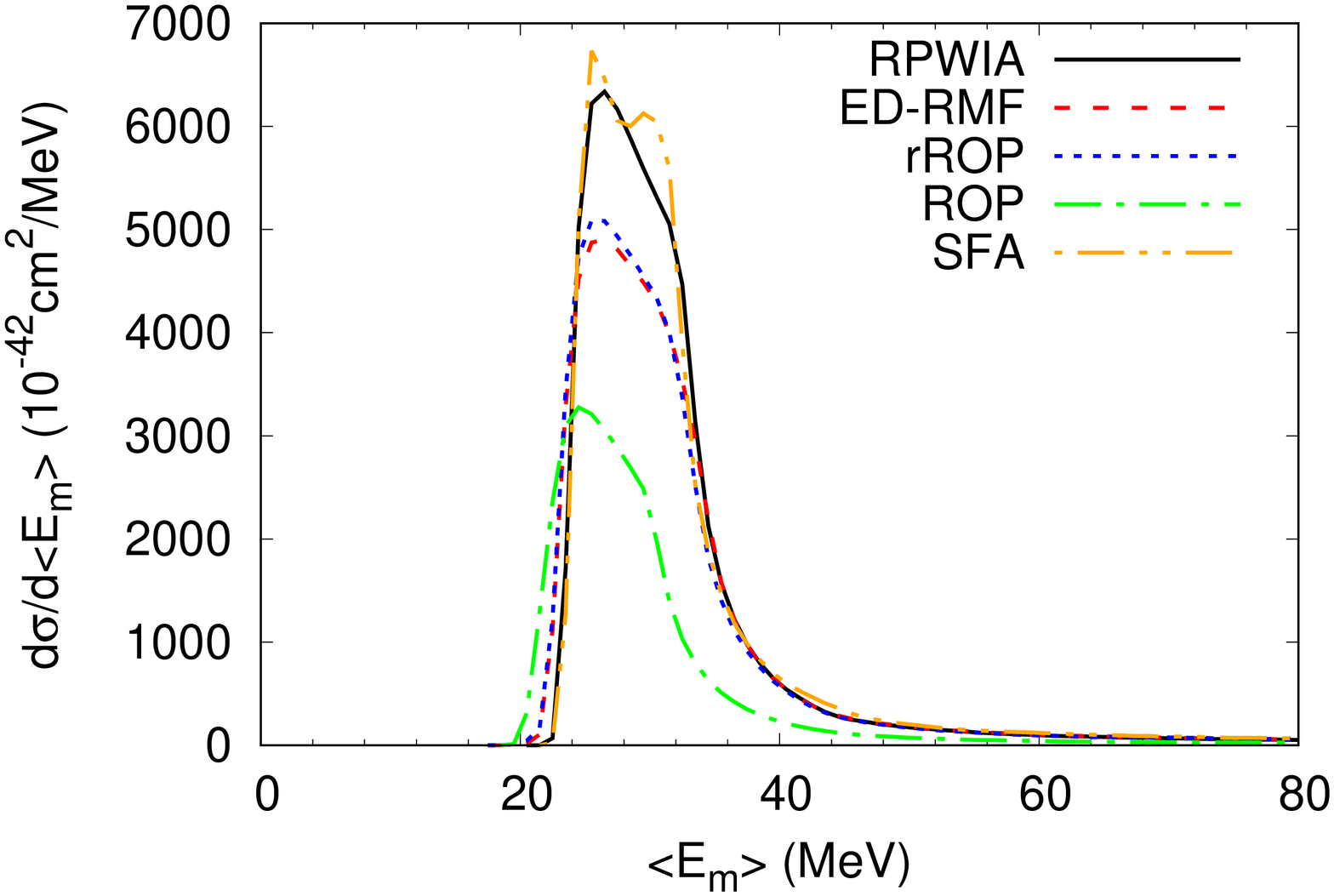}
(b)\includegraphics[width=.32\textwidth,angle=270]{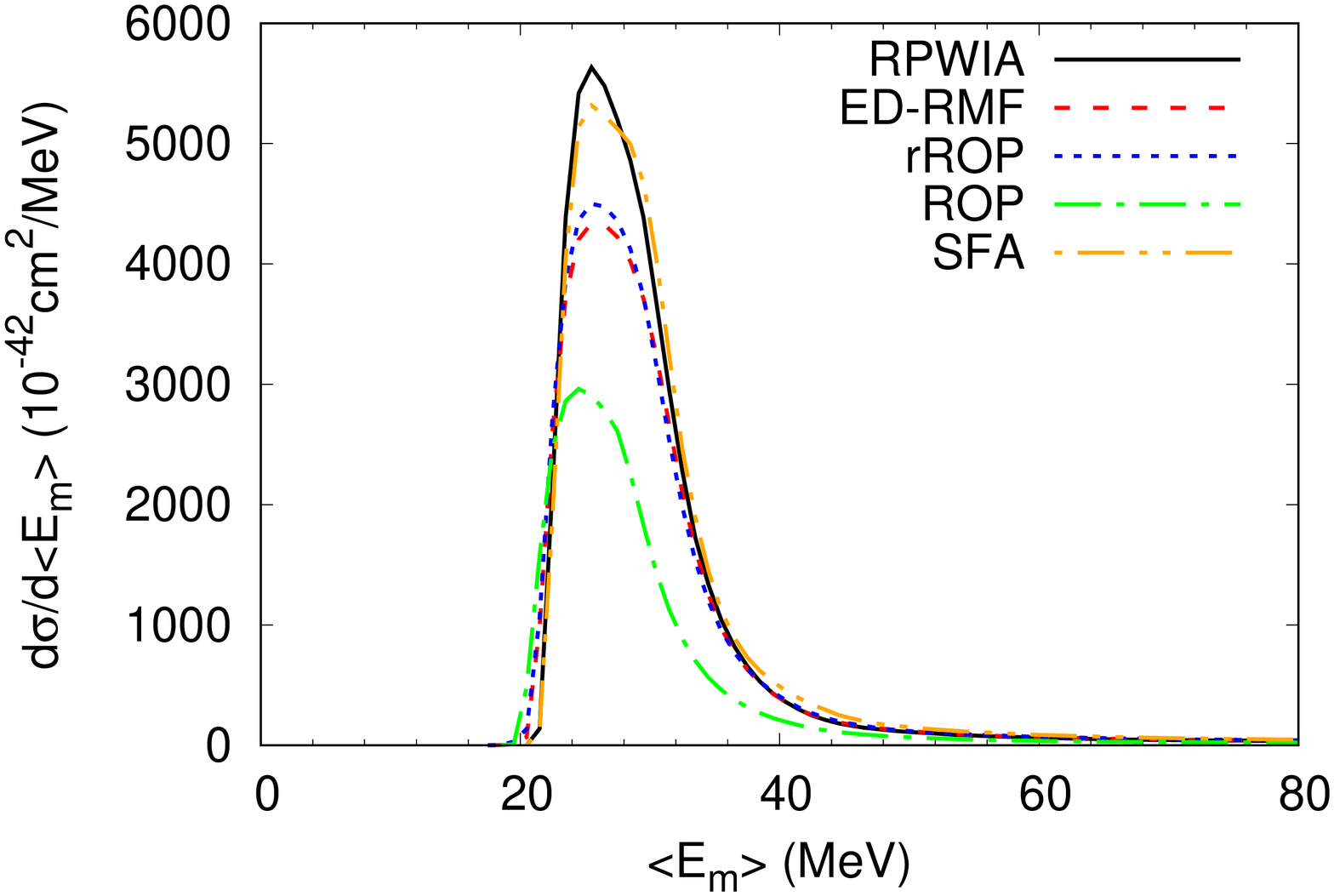}
\vspace{-0.5cm}
\caption{(Color online) Single-differential cross section as a function of the reconstructed missing-energy defined in Eq.~(\ref{<Em>}), for the DUNE (a) and T2K (b) fluxes.}\label{fig:Em_nevents}
\end{figure*}

As noted earlier, if one neglects the nuclear recoil, the missing-energy is trivially defined from the reconstructed neutrino. Accordingly, we define the reconstructed missing-energy as:
\ba
    \langle E_m\rangle = \langle E\rangle - E_l - T_N\,.\label{<Em>}
\ea
In Fig.~\ref{fig:Em_nevents} we show the single-differential cross section as a function of $\langle E_m\rangle$ for the five models and the DUNE and T2K neutrino fluxes. All models show similar shapes and significant differences in the size of the cross section, in line with the results discussed in Fig.~\ref{fig:sdc}. It is interesting that for all cases one observes a clear dominance of the $\langle E_m\rangle<40$ MeV region, corresponding to the $p$-shell region.

\begin{figure*}[htbp]
\centering  
(a)\includegraphics[width=.32\textwidth,angle=270]{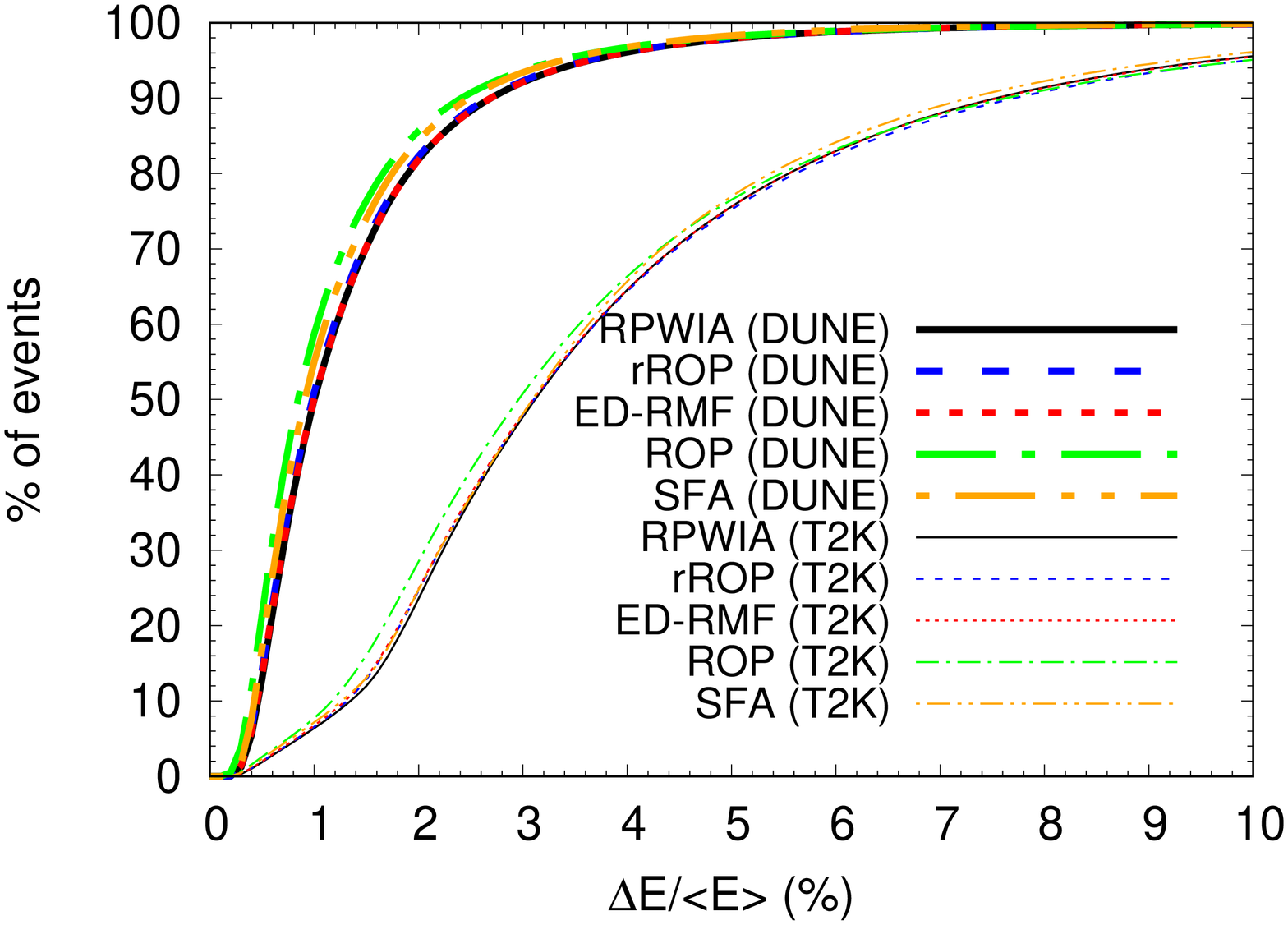}
(b)\includegraphics[width=.32\textwidth,angle=270]{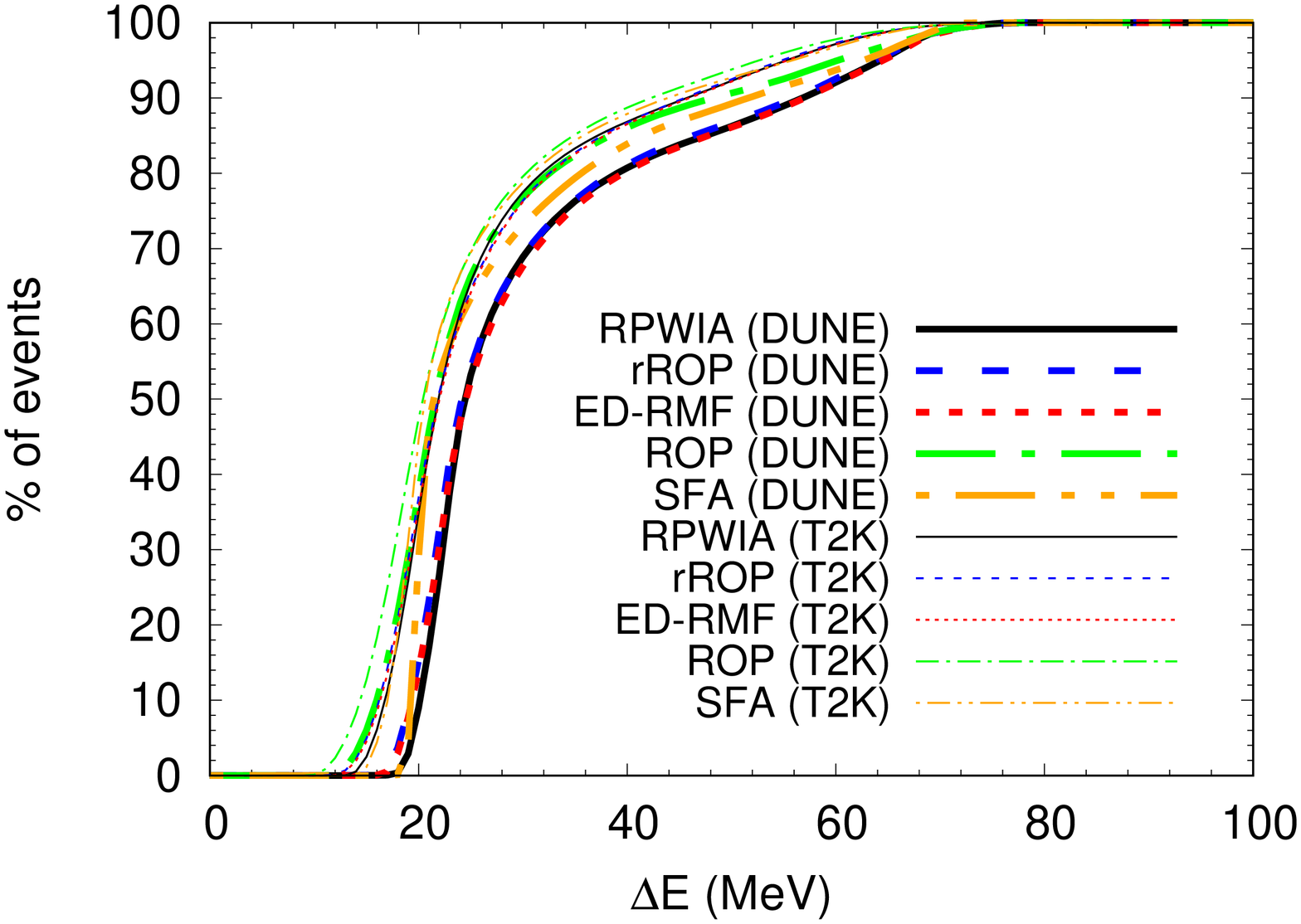}
\caption{(Color online) (a) Cumulative distributions as a function of the relative error (in percentage) in the reconstructed neutrino energy [error given by Eq.~(\ref{e_error})]. The thinner (thicker) lines are the results for the DUNE (T2K) flux.
(b) As for panel (a), except that the cumulative distributions are represented as a function of the absolute errors (in MeV).}\label{fig:cumulative_intrinsic}
\end{figure*}

In Fig.~\ref{fig:cumulative_intrinsic}(a) we show the cumulative distributions as a function of the relative error in the reconstructed neutrino energy. 
We observe that around 50\% of the events have an error lower than 1\% for the DUNE flux (thicker lines). For the T2K flux (thinner lines), the relative errors are somewhat larger and for around 47\% of the events $\frac{\Delta E}{\langle E\rangle}<3\%$.
Although for all models we see a similar trend, it seems that the `elastic only' ROP estimation has a slightly better capability for 
good reconstruction of the neutrino energy. In this case, we find that around 60\% (51\%) of the events have a
neutrino energy uncertainty below 1\% (3\%) for the DUNE (T2K) flux. This is somewhat expected, as this calculation corresponds to a signal very much enriched in `just one proton events'. The trade-off is, of course, that fewer events will qualify in the first place.

It is also interesting to look at these cumulative distributions as a function of the absolute errors. These are presented in Fig.~\ref{fig:cumulative_intrinsic}(b). For the two fluxes, the majority of the expected events (around 80\%) has an error between 15 and 40 MeV, consistent with a majority of events coming from the $p$-shells. 

To end this section, we comment that the estimations we made here from our theoretical prescription not only allow one to predict how many events there may be in an experiment leading to a given uncertainty, but they also allow one to identify where in the phase space these events lie, as we show in Sect.~\ref{subsec:BestKin}. But first, we will study the dependence of these predictions on the ingredients of the models.

\subsection{Dependence of the reconstructed energy and its error on the final-state interactions}\label{subsec:FSI}

To study how the reconstructed neutrino energy $\langle E\rangle$ depends on the model, we compute for each event (that is, for each set of muon and proton kinematics) a systematic error that quantifies the deviation of the predictions between a given model and a reference one. We choose the rROP model as reference, although the conclusions are independent of this choice. Thus, for each event we compute:
\ba
    \Delta E_\text{FSI,i} = \frac{1}{2}\left|\langle E\rangle_\text{rROP}-\langle E\rangle_{i}\right|\,,\label{e_error_FSI}
\ea
where the index $i$ refers to RPWIA, ROP, ED-RMF or factorized SFA.
Notice that in this section, all the results share the same description of the initial state, except for the case of SFA, for which it is just slightly different due to factorization, the absence of negative-energy components, and the representation of the spectral function used in our calculation. Hence, by comparing the results of these models we are actually evaluating the impact of FSI (and Pauli blocking effects) on the quality of the neutrino energy reconstruction; this is the reason for the index FSI in the previous equation.

\begin{figure*}[htbp]
\centering  
(a)\includegraphics[width=.32\textwidth,angle=270]{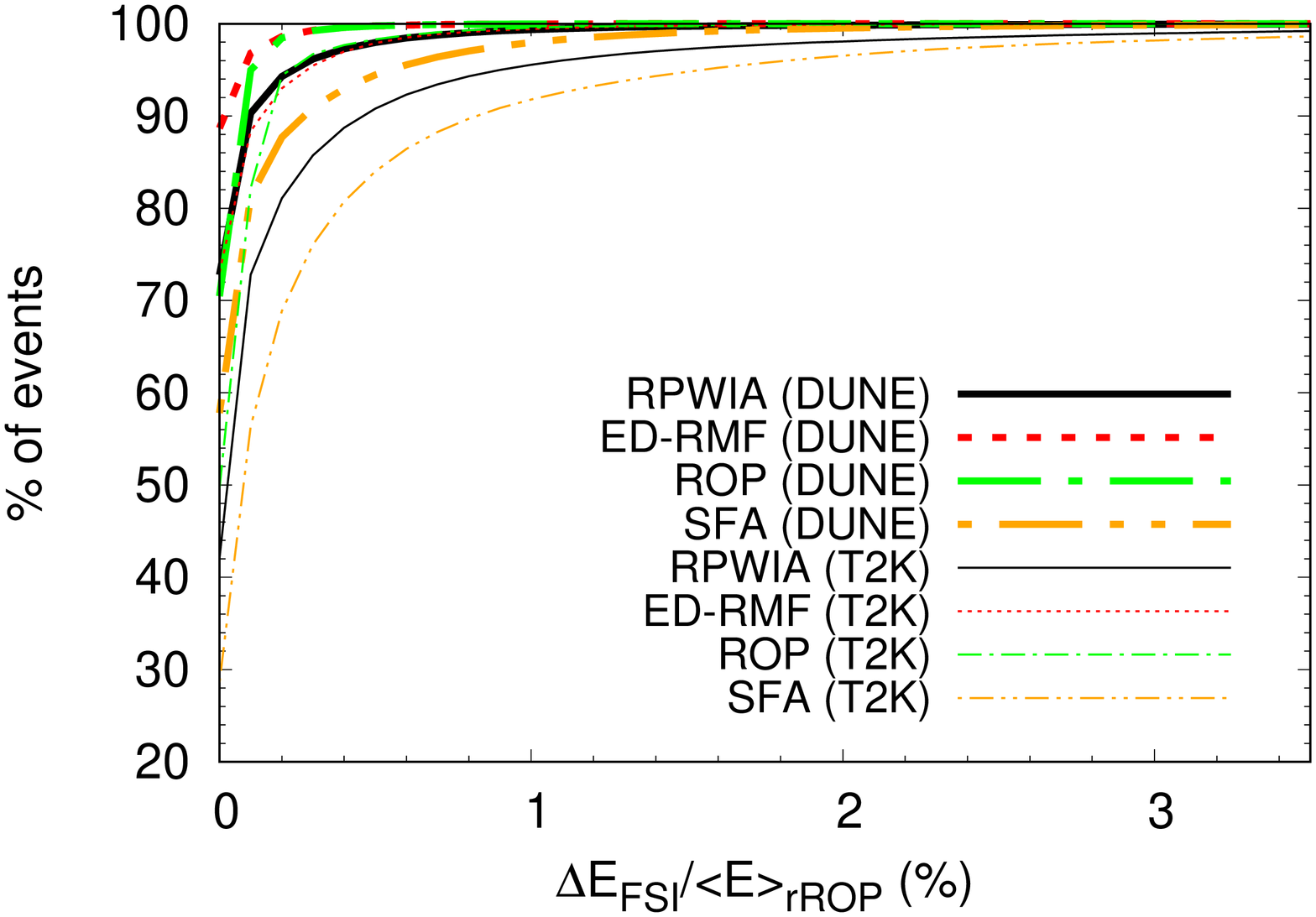}
(b)\includegraphics[width=.32\textwidth,angle=270]{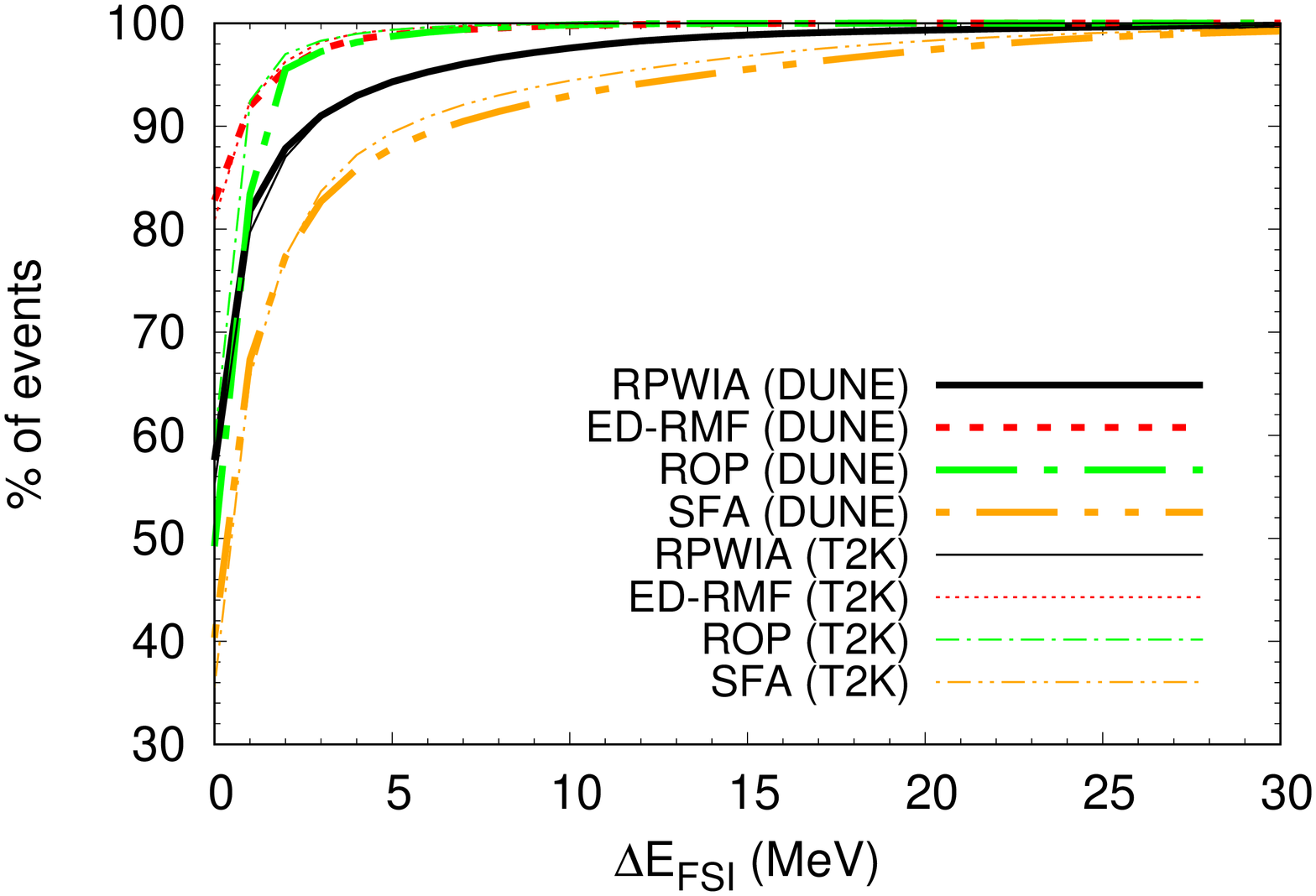}
\caption{(Color online) (a) Cumulative distributions as a function of the error in the reconstructed neutrino energy due to the description of FSI [error given by Eq.~(\ref{e_error_FSI})]. 
The thinner (thicker) lines are the results for the DUNE (T2K) flux.
(b) As for panel (a), except that the cumulative distributions are represented as a function of the absolute errors (in MeV).}\label{fig:cumulative_FSI}
\end{figure*}

The cumulative distributions as a function of $\Delta E_\text{FSI}$ are shown in Fig.~\ref{fig:cumulative_FSI}. 
These results clearly show that the FSI uncertainty, evaluated via Eq.~(\ref{e_error_FSI}), is very small. $\Delta E_\text{ED-RMF}$ and $\Delta E_\text{FSI,ROP}$ are, as expected, very small since the cross section shapes in these models are similar to the reference one, rROP. The largest $\Delta E_\text{FSI}$ is found for the SFA model. This is due to the differences in the initial state, the factorization assumption, and to the lack of FSI and Pauli blocking. In any case, for nearly 98\% (90\%) of the events, the FSI error remains below 1\% for the DUNE (T2K) fluxes. 
This important result tells us that given a flux and missing-energy distribution, for such semi-inclusive samples, the reconstructed neutrino energy would show a relatively small dependence on FSI.
This does not mean, however, that the treatment of FSI is unimportant, as the magnitude of the cross section between different models is very large.

\subsection{Dependence of the reconstructed energy and its error on the initial state}\label{subsec:SFs}

In what follows we study the uncertainty in the neutrino energy linked to the particular description of the initial state. As said before, we could consider the spectral function to provide a fair description of the situation in which the neutrino knocks out a proton, yielding a final state in which there is {\it at least} a proton. The experimental sample, however, may depart from this definition to some degree.
To better understand how the experimental situation affects the region of the spectral function that needs to be taken into
account, let us consider a particular scenario of QE scattering with a high missing-energy, {\it e.g.} $100$ MeV. In this case the residual system has an excitation energy far larger than a bound nucleus and thus necessarily additional nucleons will be ejected. If these nucleons are detected and removed from the experimental signal definition, this region of missing-energy does not contribute fully. On the other hand, these additional nucleons may be undetectable or counted as part of the experimental signal, in which case this region of missing-energy does contribute.

\begin{figure}[htbp]
\centering  
\includegraphics[width=.32\textwidth,angle=270]{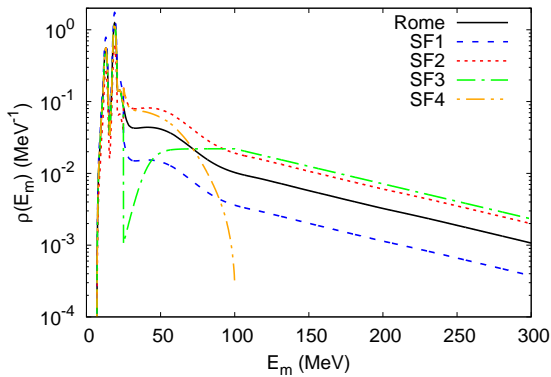}
\caption{(Color online) Different missing-energy profiles employed to analyze the impact of the description of the initial state in the 
reconstructed neutrino energy and its error.}\label{fig:SFs}
\end{figure}

To study more in detail the impact that the description of the initial state may have in the reconstructed neutrino energy and its error, we compare the results of the rROP and our representation of the spectral function, chosen again as our reference model, with several calculations for which we have built different versions of the spectral functions. These versions correspond to variations of the reference spectral function obtained by artificially varying the occupation of the shells and of the background, while keeping the total strength constant (8 neutrons in the target nucleus).
Each of these models should be more suitable for describing a specific experimental selection of the final state.
In particular, we have used the $\rho(E_m)$ functions shown in Fig.~\ref{fig:SFs} and summarized in what follows: 
\begin{itemize}
 \item SF1: The occupation number of the $p$-shells is increased from its original value of 4.98 neutrons (Table~\ref{Table}) to 6.92, consequently, the deeper $E_m$ regions decrease their relative weight. This would correspond to an experimental sample enriched in events with one and only one proton, as opposed to the `at least one proton' signature.
 \item SF2: The occupation number of the $p$-shells is decreased to 2.30 neutrons and, consequently, the deeper $E_m$ regions increase their relative weight. This is effectively the opposite to the previous situation.
\end{itemize} 

Since the $p$-shells are well constrained experimentally, it seems reasonable to keep them as they are (Table~\ref{Table}). Thus, we produce two more variations, SF3 and SF4, where we change the occupation numbers of the high missing-energy background and the $s$-shell in the following way, representing extreme opposite limiting cases:
\begin{itemize}
 \item SF3: We set the $s$-shell to zero and increase the background to keep the total number of nucleons.
 \item SF4: We set the background to zero and increase the $s$-shell to keep the total number of nucleons.
\end{itemize} 

\begin{figure*}[htbp]
\centering  
(a)\includegraphics[width=.32\textwidth,angle=270]{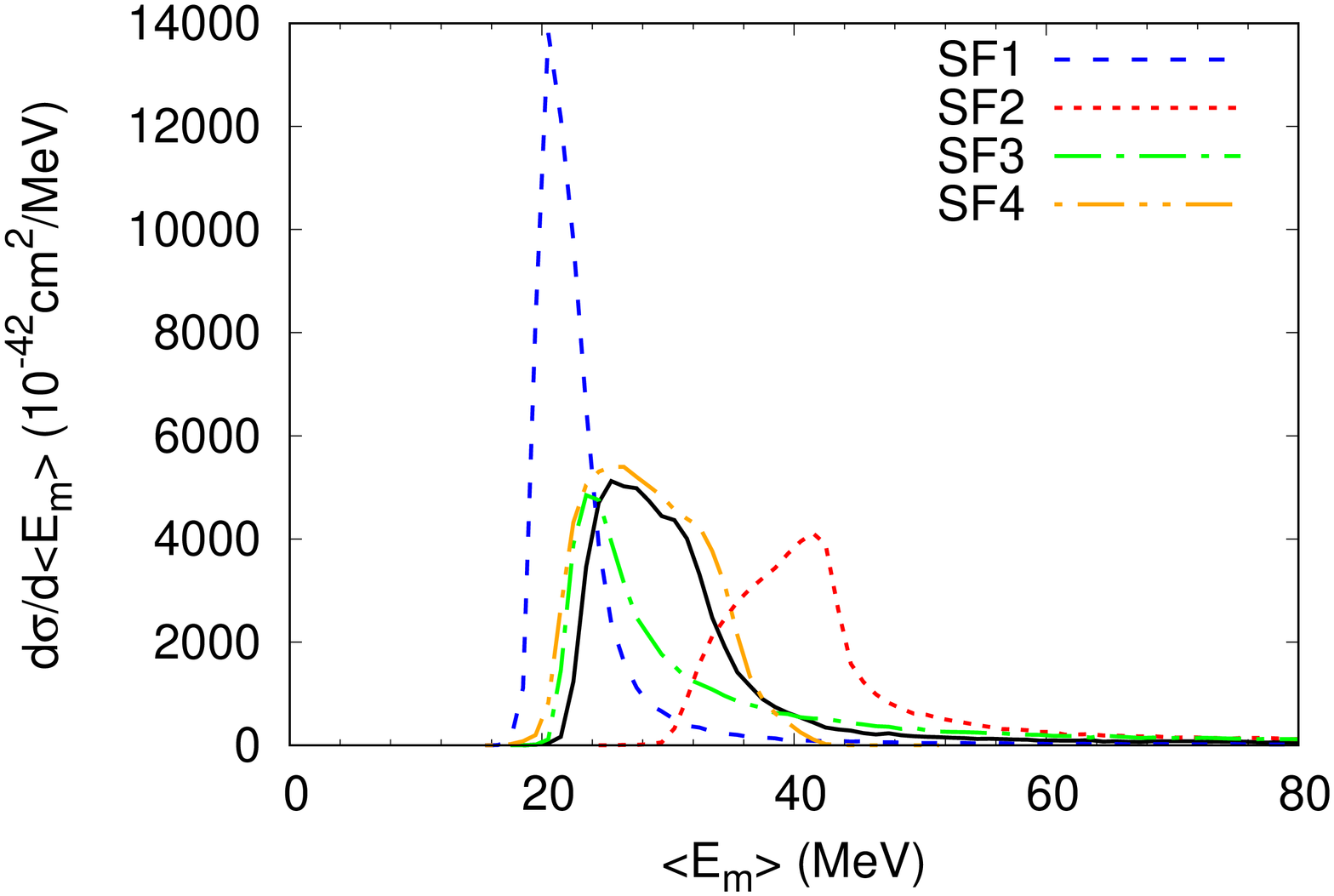}
\vspace{-0.3cm}
(b)\includegraphics[width=.32\textwidth,angle=270]{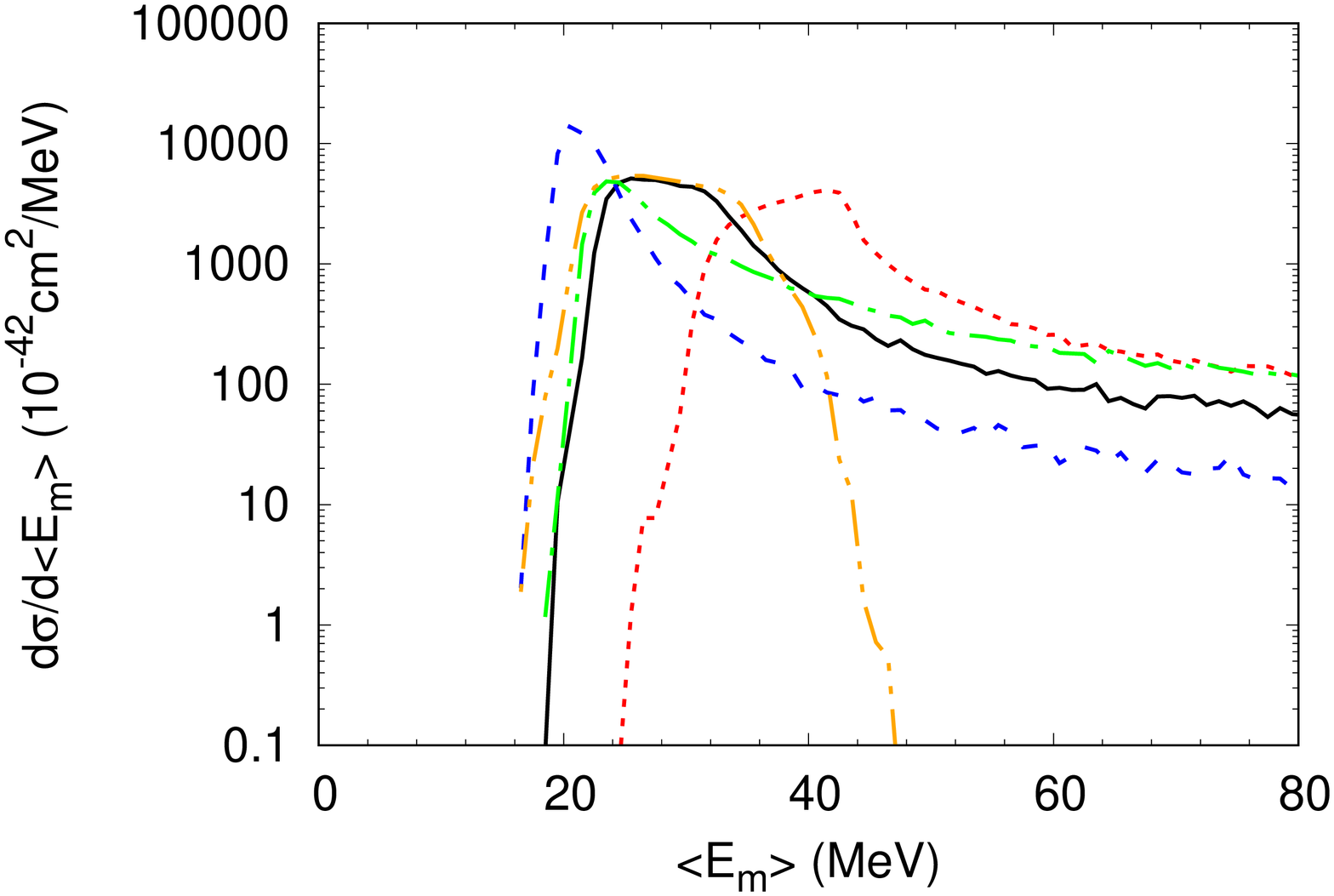}\\
(c)\includegraphics[width=.32\textwidth,angle=270]{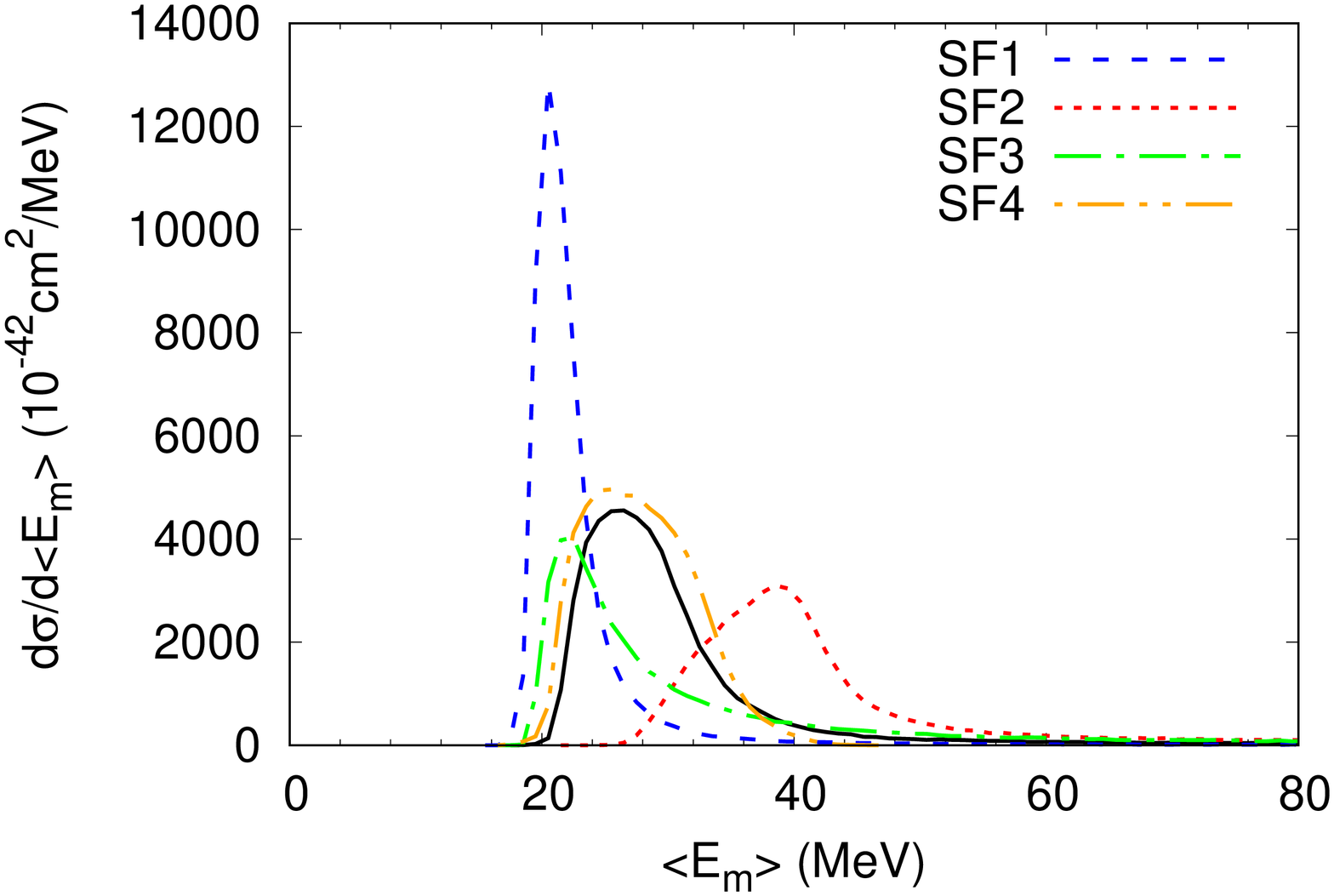}
(d)\includegraphics[width=.32\textwidth,angle=270]{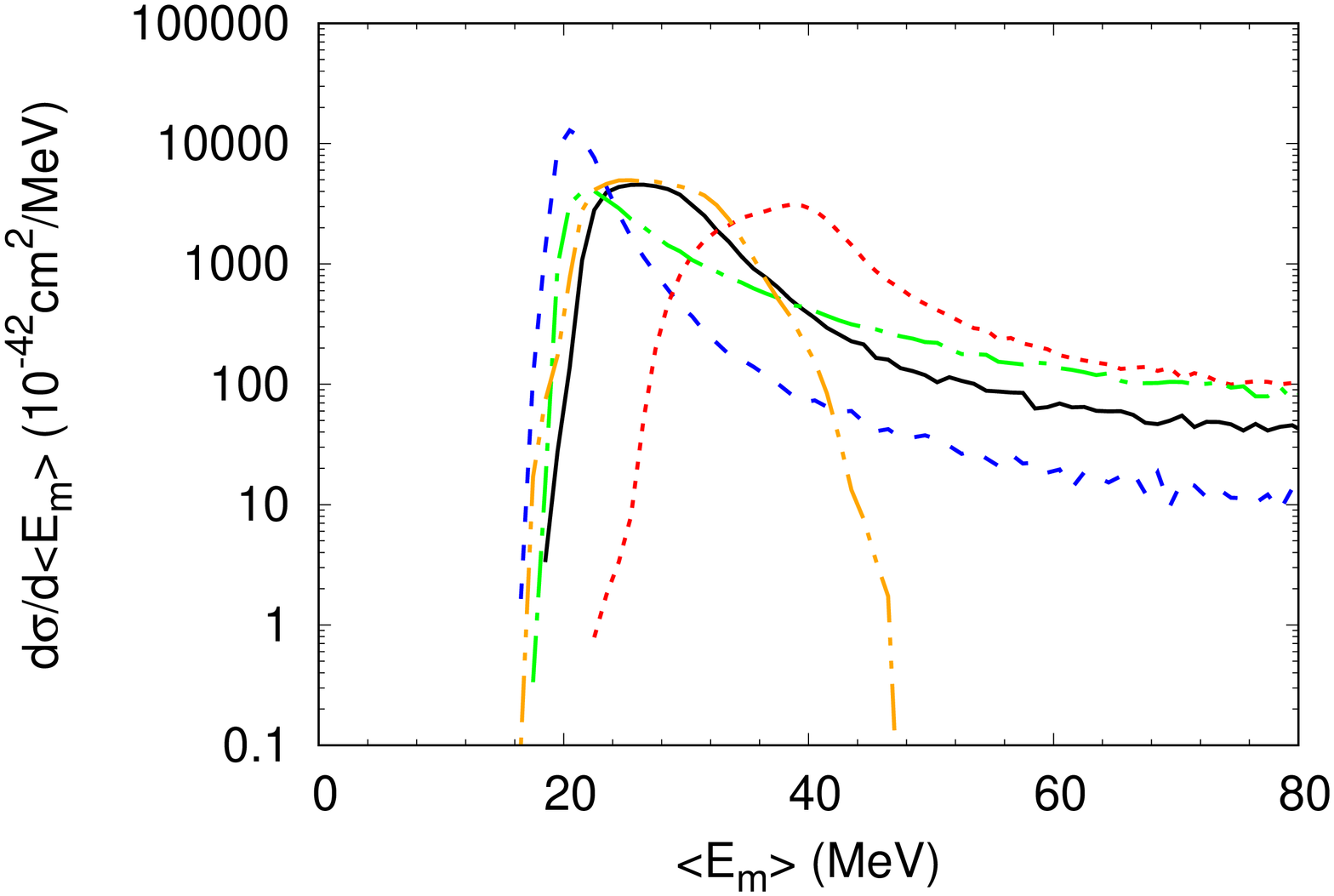}
\vspace{-0.5cm}
\caption{(Color online) (a) Single-differential cross section as a function of the reconstructed missing-energy for the DUNE flux, computed with the rROP model and the different missing-energy profiles shown in Fig.~\ref{fig:SFs}. The black-solid line correspond to the reference case, {\it i.e.,} the rROP model with the Rome missing-energy profile. (b) The same as panel (a), except using a semi-logarithmic scale. 
The results in panels (c) and (d) are for the T2K flux. }\label{fig:Em_nevents_SFs}
\end{figure*}

In Fig.~\ref{fig:Em_nevents_SFs} we show the single-differential cross section as a function of $\langle E_m\rangle$ -- defined in Eq.~\eqref{<Em>} -- for the different missing-energy profiles of Fig.~\ref{fig:SFs}. For the SF1 we observe a sharp and high peak at around 20 MeV, which is due to the increase of the occupation number of the $p$-shell. 
The effect of decreasing the occupation number of the $p$-shells (SF2) is to shift the distribution to the right, as expected since the relative weight of the deeper $E_m$ region is higher.
For the SF3 model, the distribution is narrower than in the reference case in the region $E_m<40$ MeV, while its magnitude increases from about $E_m>40$ MeV.
For the SF4 model, the distribution is just slightly wider than in the reference case in the region $E_m<40$ MeV, and practically zero above that.
Similar results are found for the DUNE and T2K fluxes. 

\begin{figure*}[htbp]
\centering  
(a)\includegraphics[width=.32\textwidth,angle=270]{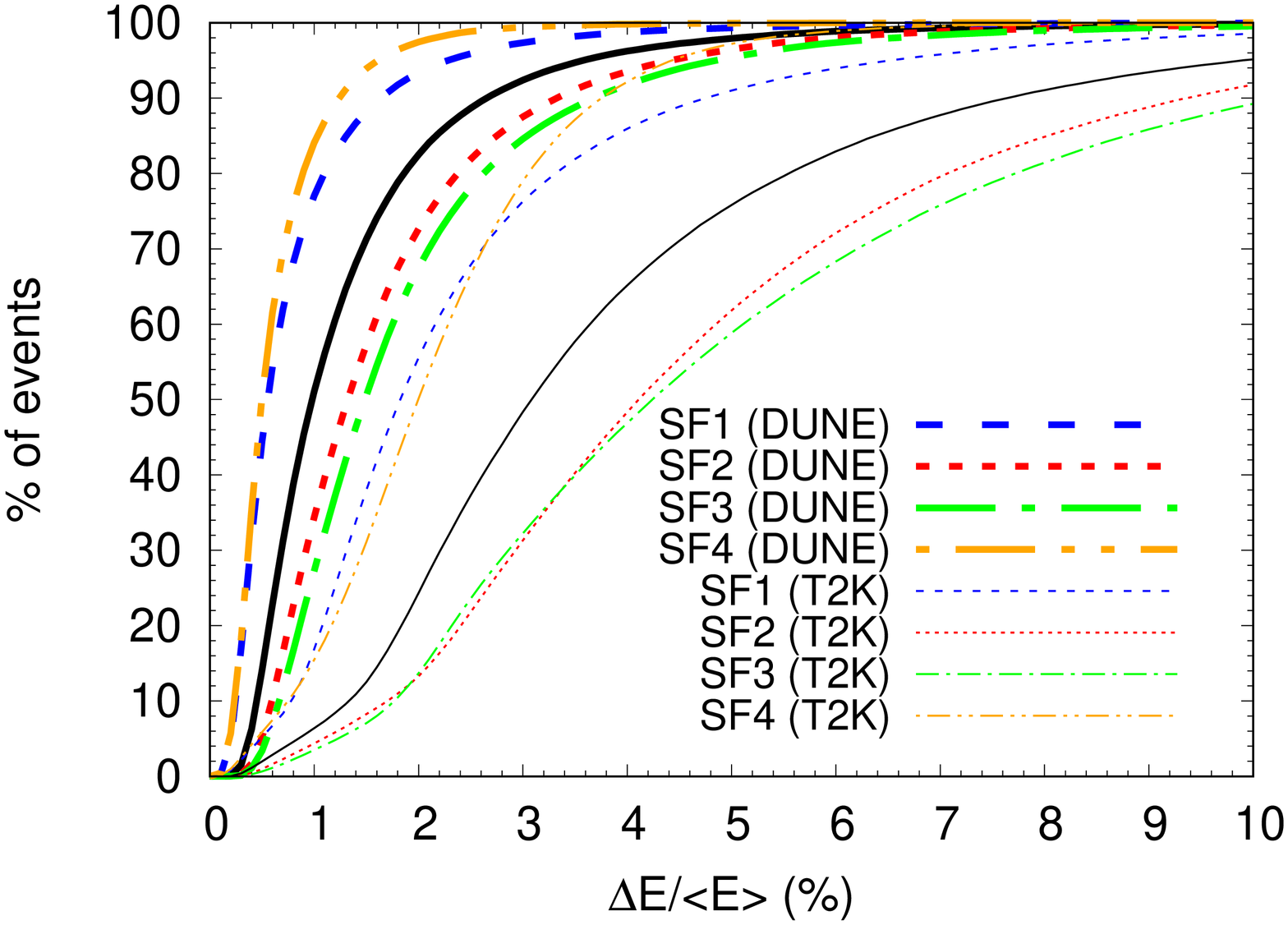}
(b)\includegraphics[width=.32\textwidth,angle=270]{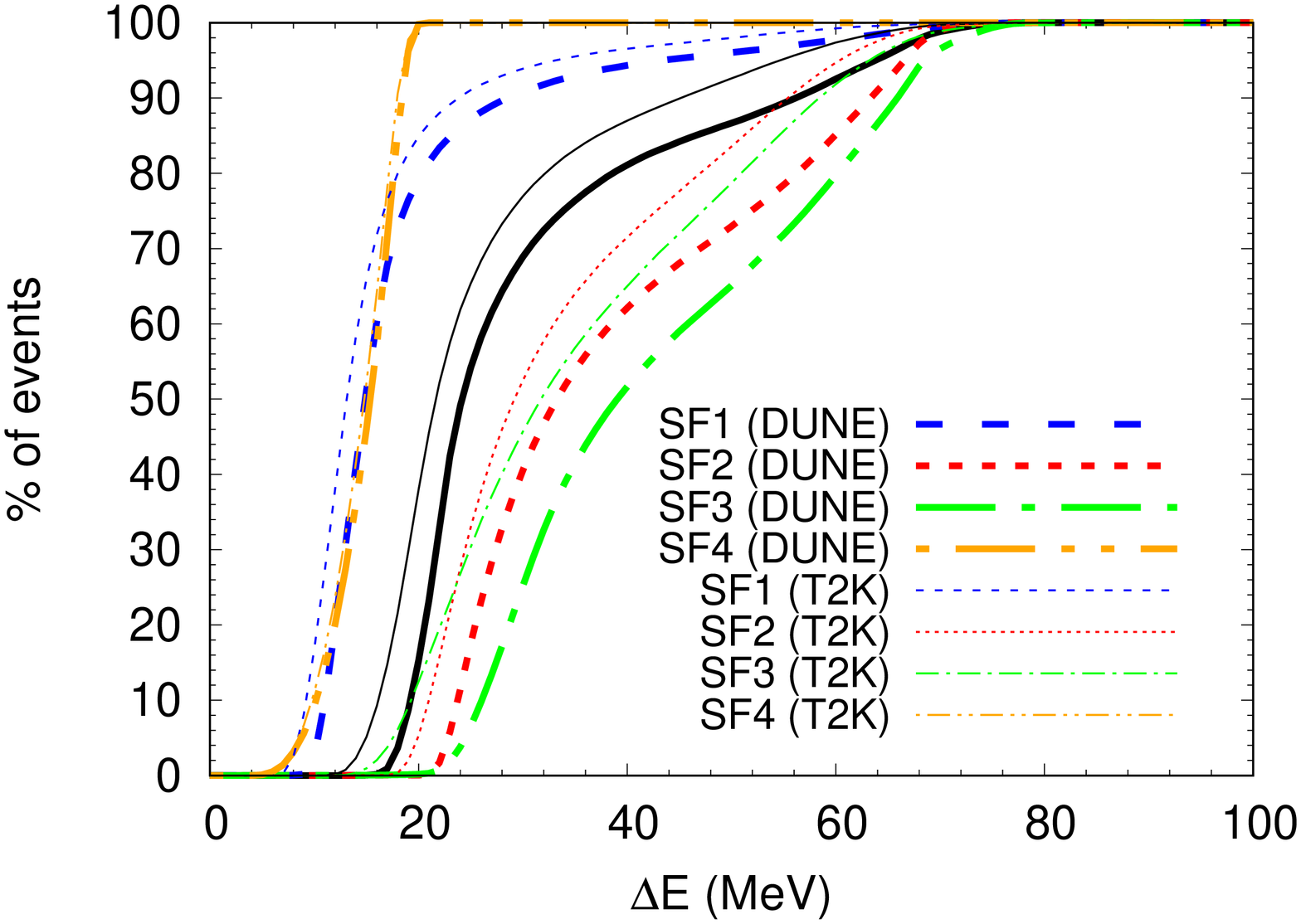}
(c)\includegraphics[width=.32\textwidth,angle=270]{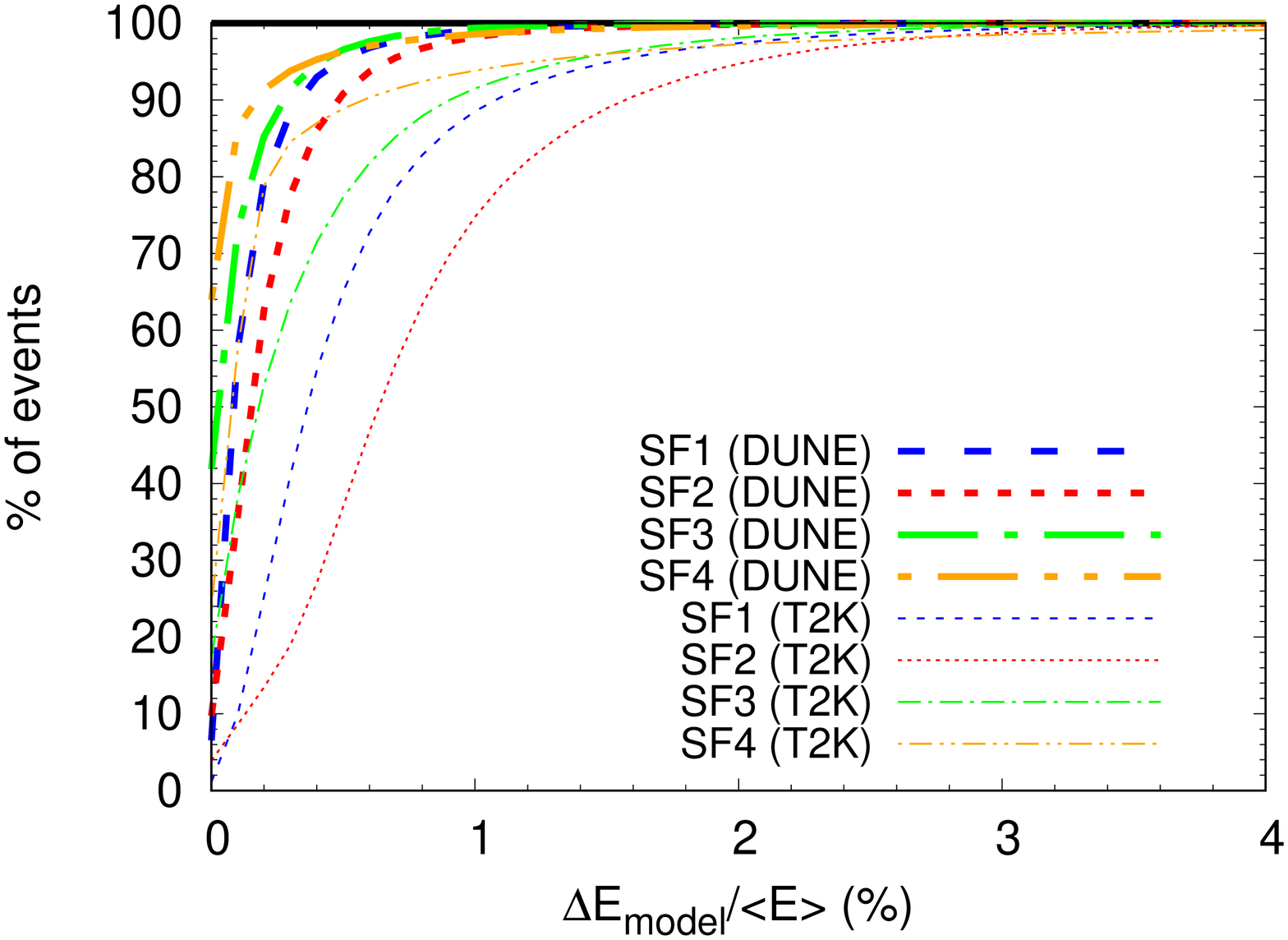}
(d)\includegraphics[width=.32\textwidth,angle=270]{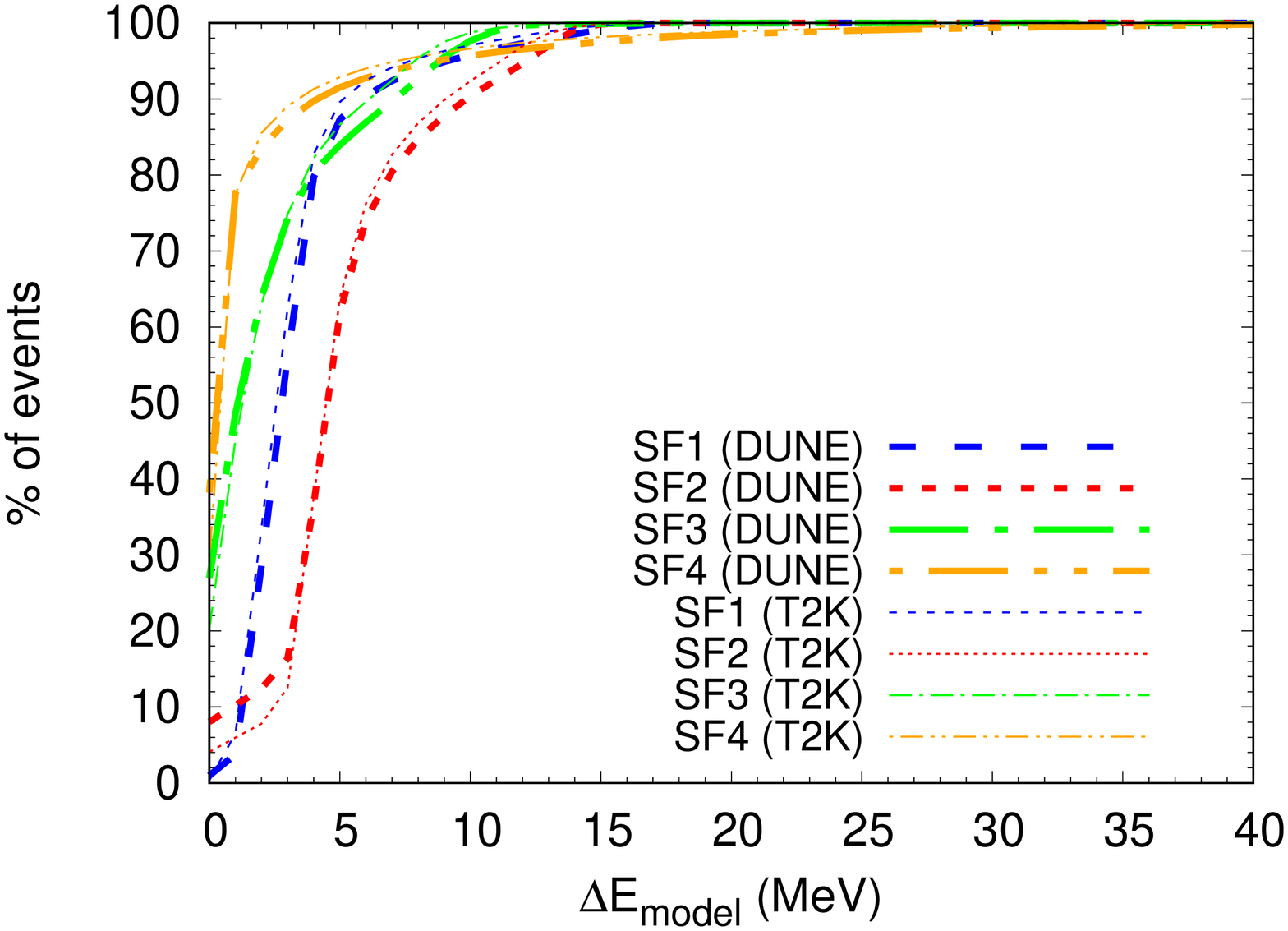}
\caption{(Color online) (a) Cumulative distributions as a function of the relative error (in percentage) in the reconstructed neutrino energy [error given by Eq.~(\ref{e_error})]. 
(b) As for panel (a), except here the cumulative distributions are represented as a function of the absolute errors (in MeV).
(c) Cumulative distributions as a function of the error in the reconstructed neutrino energy due to the description of initial state [error given by Eq.~(\ref{e_error_SF})]. (d) As for panel (c), except here the cumulative distributions are represented as a function of the absolute errors (in MeV).
In all panels, the thinner (thicker) lines are the results for the DUNE (T2K) flux.}\label{fig:cumulative_SFs}
\end{figure*}

The cumulative distributions of events with a given neutrino energy uncertainty, for these spectral functions are shown 
in Fig.~\ref{fig:cumulative_SFs}. Here we show the cumulative distributions as a function of the `intrinsic error' given by 
Eq.~(\ref{e_error}) (upper panels) and the model error (lower panels). The latter represents half of the distance to the reference case, {\it i.e.:}
\ba
    \Delta E_{\text{model,}i} = \frac{1}{2}\left|\langle E\rangle_\text{rROP}-\langle E\rangle_{\text{SF}i}\right|\,,\label{e_error_SF}
\ea
where the index $i=$1, 2, 3 or 4.
As was the case for the final-state variation, looking at the upper panels, we conclude that despite the extreme differences among the ingredients of the initial state, they all lead to very similar values for the reconstructed neutrino energy, {\it viz.} at the few percent level. That is, were we using our `standard' model to reconstruct the energy of every event, those values of energy will differ very little if the actual experimental behavior was given by the SF1 to SF4 variations in the spectral function. 

On the other hand, from the lower panels we conclude that the estimate for the uncertainty in the neutrino energy might be more dependent on the shape of the spectral function assumed. The most remarkable cases occurs for the model with an increased occupation of the $p$-shells (SF1) and the one without background (SF4), they exhibit the least uncertainty in the reconstructed energy, as expected. This figure can help us estimate the reduction in the neutrino energy uncertainty that an experimental sample enriched in the `just one proton signal' may achieve.

\subsection{Best kinematics for energy reconstruction}\label{subsec:BestKin}

Once we have described the uncertainties in the reconstructed energy associated with the description of the initial and final nucleon states, we identify the phase-space regions where the neutrino energy can best be reconstructed.
In the left panels in Figs.~{\ref{fig:El_thetal}-\ref{fig:thetaN_phiN}, we show the double-differential cross section as a function of the final lepton energy and polar angle. In the right-hand panels, for the same combination of variables, we show the average relative error $\Delta E/\langle E\rangle$ per bin, with $\Delta E$ the intrinsic error given by eq.~(\ref{e_error}). 
For example, if we were to select events in which the relative error is lower than 1\%, Fig.~\ref{fig:El_thetal}(b) tells us that it would be very likely to find those events in the darkest region that corresponds to the bins in which the average $\Delta E/\langle E\rangle$ is lower than 1\%. On the contrary, it will be very unlikely to find an event with $\Delta E/\langle E\rangle<1\%$ for $\theta_l>\pi/4$. 
The top (bottom) panels correspond to the DUNE (T2K) flux. The results were performed with the rROP model, although a similar behavior is found with the other models. 
 
In Fig.~\ref{fig:thetaN_phiN} we show the results versus the polar and azimuthal angles that define the kinematics of the ejected (and measured) nucleon. As observed, for both DUNE (upper panels) and TK2 (bottom panels) fluxes, but particularly for DUNE, most of the strength concentrates in a small region at around $\phi_N=180$ deg and $\theta_N=45$ deg. It is important to point out that this is also the region where one expects to find events whose neutrino energy can best be reconstructed.
 
We study events in the $p_N-E_l$ plane in Fig.~\ref{fig:pN_El}. For both fluxes, the strength spreads over a wide region in $p_N$ that goes from $150$ MeV up to $1000$ MeV. In terms of $E_l$ the strength is located in the regions $500<E_l<5000$ MeV and $100<E_l<1000$ for DUNE and T2K, respectively. The results in panels (b) and (d) suggest that $p_N$ does not play a decisive role when it comes to choosing small-error events. In spite of that, one should be careful when $p_N$ is lower than approximately $300$ MeV (or equivalently in the small energy transfer region) because in that region nuclear effects like Pauli blocking, distortion and long-range correlations~\cite{Pandey16, Nikolakopoulos19, Gonzalez-Jimenez19} play a more relevant role, and therefore, models based on plane-wave impulse approximation should be used with care.

Finally, in Fig.~\ref{fig:pN_thetaN} we show the results in the $p_N-\theta_N$ plane. The location of the strength (left panels) and of small-error regions (right panels) is consistent with that observed in the previous figures. It is clear again that the nucleon polar angle $\theta_N$ plays a more relevant role than $p_N$ when it comes to selecting small-error events.

It is important to recall here that, even though these regions of `good events' have been identified with the reference model, 
events filtered based upon these findings, but generated from any of the other variations of FSI or initial state ingredients, will still be of relatively low uncertainty.

\section{Conclusions}\label{sec:Conclusions}

In this work we have analyzed in detail the case of charged-current muon neutrino interactions with oxygen for the T2K and DUNE neutrino fluxes. 
Our interest has been focused on the neutrino energy reconstruction process, and we have shown that a successful reconstruction (at the few percent level) of the neutrino energy can be obtained from samples consisting in the muon and at least one proton detected. This is mainly due to the large contribution from the $p$-shells to the cross section. We show that this outcome is largely unaffected by the treatment of hadronic final-state interactions. Hence, provided a realistic distribution of missing-energy and -momentum in the initial state is available, the neutrino energy can be reconstructed quite well.

Our study allows one to estimate not only the error for each given muon and proton kinematics, but also how the uncertainty in the neutrino energy will evolve with cuts in the experimental sample, in the kinematics or in the number of hadrons in the final state. These results for the error in the reconstructed energy should be understood as a lower bound, since additional contributions not included in this analysis, {\it e.g.} from the efficiency and resolution of the detectors and from non-quasielastic contributions to the sample, potentially can play a significant role. For example, while real pion production is excluded here and to some extent can be vetoed experimentally (remember we are considering only CC1$\mu$1$p$0$\pi$ events), pions do play a role as virtual particles. They enter as particles exchanged between nucleons, both one-pion exchange and two-pion exchange perhaps via an effective scalar meson; however, these effects are already included through the use of the effective interactions we employ. Pions can also enter as parts of two-body currents as has been studied for inclusive scattering~\cite{VanOrden81, DePace04, Amaro02, Martini09, Nieves11, VanCuyck17, Amaro20}. These contributions are yet to be fully implemented for semi-inclusive reactions. 

We expect that the findings of this study will not differ very much from the ones based on a factorized spectral function approach plus a cascade model for the propagation of the hadrons in the nuclear medium. Since our calculations are based on models that describe the scattering process within a fully relativistic and quantum mechanical framework, incorporating Pauli blocking and binding energies in a realistic way, we believe that these results should be of great interest for the whole neutrino interaction community. 

In this work we have restricted our attention to the case of oxygen, taking into account DUNE and T2K neutrino fluxes. A next step will be to extend the present study to some other nuclear systems. In particular, we will explore the cases of carbon and argon. The former, although more similar to oxygen, is of great interest for T2K and HyperKamiokande experiments. A detailed analysis in the region of low values of the kinematical variables and its comparison with the present oxygen results will be very valuable. The analysis of the significantly more complex case of argon will be of crucial interest for the DUNE collaboration. 

\begin{acknowledgments}
 This work was supported by the Madrid Government under the Multiannual Agreement with Complutense University in the line Program to Stimulate Research for Young Doctors in the context of the V PRICIT (Regional Programme of Research and Technological Innovation), Project PR65/19-22430 (R.G.-J.);
 by the Project BARM-RILO-18-02 of University of Turin and from INFN, National Project NUCSYS (M.B.B.);
 by the Spanish Ministerio de Economia y Competitividad and ERDF (European Regional Development Fund) under contract FIS2017-88410-P, by the Junta de Andalucia (grants No. FQM160, SOMM17/6105/UGR) and by the University of Tokyo ICRR's Inter-University Research Program FY2020, Ref. No. A07 (J.A.C. and G.D.M.); 
 by the Research Foundation Flanders (FWO-Flanders), and by the Special Research Fund, Ghent University (N.J., K.N. and A.N.);
 by the NCN Preludium Grant No. 2020/37/N/ST2/01751 and also by the Polish Ministry of Science and Higher Education, Grant No. DIR WK/2017/05 (K.N.);
 by the European Union's Horizon 2020 research and innovation programme under the Marie Sklodowska-Curie grant agreement No. 839481 (G.D.M.); 
 by Jefferson Science Associates, LLC under U.S. DOE Contract DE-AC05-06OR23177 (J.W.V.O.); 
 and by the Office of Nuclear Physics of the US Department of Energy under Grant Contract DE-FG02-94ER40818 (T.W.D). 
 The authors also thank Omar Benhar for kindly providing the model for the spectral function of ${}^{16}$O employed in this study.
 
 The computations of this work were performed in Brigit, the HPC server of the Universidad Complutense de Madrid. 
\end{acknowledgments}

\appendix 

\section{Background missing-energy profile}\label{backg-function}

The missing-energy profile of the background component of the spectral function shown in Fig.~\ref{fig:rho&n}(a) is parameterized as follows: 
\ba
  F(E_m) = a\exp(-b\ E_m)\,,
\ea
if $E_m > 100$ MeV, and 
\ba
  F(E_m)  =  \dfrac{a\exp( -100\, b)}{\exp[-(E_m-c)/w] + 1 }\,,
\ea
if $25 < E_m < 100$  MeV.
The parameters are $a=0.03113$ MeV$^{-1}$ and $b=0.0112371$ MeV$^{-1}$ for the exponential, and $c = 40$ MeV and $w = 5$ MeV for the Fermi function. 
The function is normalized so that it gives the proper occupation numbers shown in Table~\ref{Table}, {\it i.e.,} $\int_{25}^{100} dE_m F(E_m) = 0.6$ and $\int_{100}^{300} dE_m F(E_m) = 0.8$, with $E_m$ in MeV.

\bibliographystyle{apsrev4-1}
\bibliography{bibliography}

\begin{thebibliography}{42}%
\makeatletter
\providecommand \@ifxundefined [1]{%
 \@ifx{#1\undefined}
}%
\providecommand \@ifnum [1]{%
 \ifnum #1\expandafter \@firstoftwo
 \else \expandafter \@secondoftwo
 \fi
}%
\providecommand \@ifx [1]{%
 \ifx #1\expandafter \@firstoftwo
 \else \expandafter \@secondoftwo
 \fi
}%
\providecommand \natexlab [1]{#1}%
\providecommand \enquote  [1]{``#1''}%
\providecommand \bibnamefont  [1]{#1}%
\providecommand \bibfnamefont [1]{#1}%
\providecommand \citenamefont [1]{#1}%
\providecommand \href@noop [0]{\@secondoftwo}%
\providecommand \href [0]{\begingroup \@sanitize@url \@href}%
\providecommand \@href[1]{\@@startlink{#1}\@@href}%
\providecommand \@@href[1]{\endgroup#1\@@endlink}%
\providecommand \@sanitize@url [0]{\catcode `\\12\catcode `\$12\catcode
  `\&12\catcode `\#12\catcode `\^12\catcode `\_12\catcode `\%12\relax}%
\providecommand \@@startlink[1]{}%
\providecommand \@@endlink[0]{}%
\providecommand \url  [0]{\begingroup\@sanitize@url \@url }%
\providecommand \@url [1]{\endgroup\@href {#1}{\urlprefix }}%
\providecommand \urlprefix  [0]{URL }%
\providecommand \Eprint [0]{\href }%
\providecommand \doibase [0]{http://dx.doi.org/}%
\providecommand \selectlanguage [0]{\@gobble}%
\providecommand \bibinfo  [0]{\@secondoftwo}%
\providecommand \bibfield  [0]{\@secondoftwo}%
\providecommand \translation [1]{[#1]}%
\providecommand \BibitemOpen [0]{}%
\providecommand \bibitemStop [0]{}%
\providecommand \bibitemNoStop [0]{.\EOS\space}%
\providecommand \EOS [0]{\spacefactor3000\relax}%
\providecommand \BibitemShut  [1]{\csname bibitem#1\endcsname}%
\let\auto@bib@innerbib\@empty
\bibitem [{\citenamefont {Acciarri}\ \emph {et~al.}(2014)\citenamefont
  {Acciarri} \emph {et~al.}}]{Argoneut14}%
  \BibitemOpen
  \bibfield  {author} {\bibinfo {author} {\bibfnamefont {R.}~\bibnamefont
  {Acciarri}} \emph {et~al.},\ }\href {\doibase 10.1103/PhysRevD.90.012008}
  {\bibfield  {journal} {\bibinfo  {journal} {Phys. Rev. D}\ }\textbf {\bibinfo
  {volume} {90}},\ \bibinfo {pages} {012008} (\bibinfo {year}
  {2014})}\BibitemShut {NoStop}%
\bibitem [{\citenamefont {Cai}\ \emph {et~al.}(2020)\citenamefont {Cai} \emph
  {et~al.}}]{Cai20}%
  \BibitemOpen
  \bibfield  {author} {\bibinfo {author} {\bibfnamefont {T.}~\bibnamefont
  {Cai}} \emph {et~al.} (\bibinfo {collaboration} {The MINER\ensuremath{\nu}A
  Collaboration}),\ }\href {\doibase 10.1103/PhysRevD.101.092001} {\bibfield
  {journal} {\bibinfo  {journal} {Phys. Rev. D}\ }\textbf {\bibinfo {volume}
  {101}},\ \bibinfo {pages} {092001} (\bibinfo {year} {2020})}\BibitemShut
  {NoStop}%
\bibitem [{\citenamefont {Lu}\ \emph {et~al.}(2018)\citenamefont {Lu} \emph
  {et~al.}}]{Lu18}%
  \BibitemOpen
  \bibfield  {author} {\bibinfo {author} {\bibfnamefont {X.-G.}\ \bibnamefont
  {Lu}} \emph {et~al.} (\bibinfo {collaboration} {MINERvA Collaboration}),\
  }\href {\doibase 10.1103/PhysRevLett.121.022504} {\bibfield  {journal}
  {\bibinfo  {journal} {Phys. Rev. Lett.}\ }\textbf {\bibinfo {volume} {121}},\
  \bibinfo {pages} {022504} (\bibinfo {year} {2018})}\BibitemShut {NoStop}%
\bibitem [{\citenamefont {Walton}\ \emph {et~al.}(2015)\citenamefont {Walton}
  \emph {et~al.}}]{Walton15}%
  \BibitemOpen
  \bibfield  {author} {\bibinfo {author} {\bibfnamefont {T.}~\bibnamefont
  {Walton}} \emph {et~al.} (\bibinfo {collaboration} {MINERvA Collaboration}),\
  }\href {\doibase 10.1103/PhysRevD.91.071301} {\bibfield  {journal} {\bibinfo
  {journal} {Phys. Rev. D}\ }\textbf {\bibinfo {volume} {91}},\ \bibinfo
  {pages} {071301} (\bibinfo {year} {2015})}\BibitemShut {NoStop}%
\bibitem [{\citenamefont {Abe}\ \emph {et~al.}(2018)\citenamefont {Abe} \emph
  {et~al.}}]{T2K18}%
  \BibitemOpen
  \bibfield  {author} {\bibinfo {author} {\bibfnamefont {K.}~\bibnamefont
  {Abe}} \emph {et~al.} (\bibinfo {collaboration} {The T2K Collaboration}),\
  }\href {\doibase 10.1103/PhysRevD.98.032003} {\bibfield  {journal} {\bibinfo
  {journal} {Phys. Rev. D}\ }\textbf {\bibinfo {volume} {98}},\ \bibinfo
  {pages} {032003} (\bibinfo {year} {2018})}\BibitemShut {NoStop}%
\bibitem [{\citenamefont {Acciarri}\ \emph {et~al.}(2016)\citenamefont
  {Acciarri} \emph {et~al.}}]{DUNE16}%
  \BibitemOpen
  \bibfield  {author} {\bibinfo {author} {\bibfnamefont {R.}~\bibnamefont
  {Acciarri}} \emph {et~al.} (\bibinfo {collaboration} {DUNE}),\ }\href@noop {}
  {\  (\bibinfo {year} {2016})},\ \Eprint {http://arxiv.org/abs/1601.05471}
  {arXiv:1601.05471 [physics.ins-det]} \BibitemShut {NoStop}%
\bibitem [{\citenamefont {Simpson}\ \emph {et~al.}(2019)\citenamefont {Simpson}
  \emph {et~al.}}]{Simpson19}%
  \BibitemOpen
  \bibfield  {author} {\bibinfo {author} {\bibfnamefont {C.}~\bibnamefont
  {Simpson}} \emph {et~al.},\ }\href {\doibase 10.3847/1538-4357/ab4883}
  {\bibfield  {journal} {\bibinfo  {journal} {The Astrophysical Journal}\
  }\textbf {\bibinfo {volume} {885}},\ \bibinfo {pages} {133} (\bibinfo {year}
  {2019})}\BibitemShut {NoStop}%
\bibitem [{\citenamefont {Mosel}\ \emph {et~al.}(2014)\citenamefont {Mosel},
  \citenamefont {Lalakulich},\ and\ \citenamefont {Gallmeister}}]{Mosel14}%
  \BibitemOpen
  \bibfield  {author} {\bibinfo {author} {\bibfnamefont {U.}~\bibnamefont
  {Mosel}}, \bibinfo {author} {\bibfnamefont {O.}~\bibnamefont {Lalakulich}}, \
  and\ \bibinfo {author} {\bibfnamefont {K.}~\bibnamefont {Gallmeister}},\
  }\href {\doibase 10.1103/PhysRevLett.112.151802} {\bibfield  {journal}
  {\bibinfo  {journal} {Phys. Rev. Lett.}\ }\textbf {\bibinfo {volume} {112}},\
  \bibinfo {pages} {151802} (\bibinfo {year} {2014})}\BibitemShut {NoStop}%
\bibitem [{\citenamefont {Furmanski}\ and\ \citenamefont
  {Sobczyk}(2017)}]{Furmanski17}%
  \BibitemOpen
  \bibfield  {author} {\bibinfo {author} {\bibfnamefont {A.~P.}\ \bibnamefont
  {Furmanski}}\ and\ \bibinfo {author} {\bibfnamefont {J.~T.}\ \bibnamefont
  {Sobczyk}},\ }\href {\doibase 10.1103/PhysRevC.95.065501} {\bibfield
  {journal} {\bibinfo  {journal} {Phys. Rev. C}\ }\textbf {\bibinfo {volume}
  {95}},\ \bibinfo {pages} {065501} (\bibinfo {year} {2017})}\BibitemShut
  {NoStop}%
\bibitem [{\citenamefont {Van~Orden}\ and\ \citenamefont
  {Donnelly}(2019)}]{VanOrden19}%
  \BibitemOpen
  \bibfield  {author} {\bibinfo {author} {\bibfnamefont {J.~W.}\ \bibnamefont
  {Van~Orden}}\ and\ \bibinfo {author} {\bibfnamefont {T.~W.}\ \bibnamefont
  {Donnelly}},\ }\href {\doibase 10.1103/PhysRevC.100.044620} {\bibfield
  {journal} {\bibinfo  {journal} {Phys. Rev. C}\ }\textbf {\bibinfo {volume}
  {100}},\ \bibinfo {pages} {044620} (\bibinfo {year} {2019})}\BibitemShut
  {NoStop}%
\bibitem [{\citenamefont {Munteanu}\ \emph {et~al.}(2020)\citenamefont
  {Munteanu}, \citenamefont {Suvorov}, \citenamefont {Dolan}, \citenamefont
  {Sgalaberna}, \citenamefont {Bolognesi}, \citenamefont {Manly}, \citenamefont
  {Yang}, \citenamefont {Giganti}, \citenamefont {Iwamoto},\ and\ \citenamefont
  {Jes\'us-Valls}}]{Munteanu20}%
  \BibitemOpen
  \bibfield  {author} {\bibinfo {author} {\bibfnamefont {L.}~\bibnamefont
  {Munteanu}}, \bibinfo {author} {\bibfnamefont {S.}~\bibnamefont {Suvorov}},
  \bibinfo {author} {\bibfnamefont {S.}~\bibnamefont {Dolan}}, \bibinfo
  {author} {\bibfnamefont {D.}~\bibnamefont {Sgalaberna}}, \bibinfo {author}
  {\bibfnamefont {S.}~\bibnamefont {Bolognesi}}, \bibinfo {author}
  {\bibfnamefont {S.}~\bibnamefont {Manly}}, \bibinfo {author} {\bibfnamefont
  {G.}~\bibnamefont {Yang}}, \bibinfo {author} {\bibfnamefont {C.}~\bibnamefont
  {Giganti}}, \bibinfo {author} {\bibfnamefont {K.}~\bibnamefont {Iwamoto}}, \
  and\ \bibinfo {author} {\bibfnamefont {C.}~\bibnamefont {Jes\'us-Valls}},\
  }\href {\doibase 10.1103/PhysRevD.101.092003} {\bibfield  {journal} {\bibinfo
   {journal} {Phys. Rev. D}\ }\textbf {\bibinfo {volume} {101}},\ \bibinfo
  {pages} {092003} (\bibinfo {year} {2020})}\BibitemShut {NoStop}%
\bibitem [{\citenamefont {Megias}\ \emph {et~al.}(2018)\citenamefont {Megias},
  \citenamefont {Barbaro}, \citenamefont {Caballero}, \citenamefont {Amaro},
  \citenamefont {Donnelly}, \citenamefont {Simo},\ and\ \citenamefont
  {Orden}}]{Megias18}%
  \BibitemOpen
  \bibfield  {author} {\bibinfo {author} {\bibfnamefont {G.~D.}\ \bibnamefont
  {Megias}}, \bibinfo {author} {\bibfnamefont {M.~B.}\ \bibnamefont {Barbaro}},
  \bibinfo {author} {\bibfnamefont {J.~A.}\ \bibnamefont {Caballero}}, \bibinfo
  {author} {\bibfnamefont {J.~E.}\ \bibnamefont {Amaro}}, \bibinfo {author}
  {\bibfnamefont {T.~W.}\ \bibnamefont {Donnelly}}, \bibinfo {author}
  {\bibfnamefont {I.~R.}\ \bibnamefont {Simo}}, \ and\ \bibinfo {author}
  {\bibfnamefont {J.~W.~V.}\ \bibnamefont {Orden}},\ }\href {\doibase
  10.1088/1361-6471/aaf3ae} {\bibfield  {journal} {\bibinfo  {journal} {J.
  Phys. G: Nucl. Part. Phys.}\ }\textbf {\bibinfo {volume} {46}},\ \bibinfo
  {pages} {015104} (\bibinfo {year} {2018})}\BibitemShut {NoStop}%
\bibitem [{\citenamefont {Ud\'{\i}as}\ \emph {et~al.}(1993)\citenamefont
  {Ud\'{\i}as}, \citenamefont {Sarriguren}, \citenamefont {Moya~de Guerra},
  \citenamefont {Garrido},\ and\ \citenamefont {Caballero}}]{Udias93}%
  \BibitemOpen
  \bibfield  {author} {\bibinfo {author} {\bibfnamefont {J.~M.}\ \bibnamefont
  {Ud\'{\i}as}}, \bibinfo {author} {\bibfnamefont {P.}~\bibnamefont
  {Sarriguren}}, \bibinfo {author} {\bibfnamefont {E.}~\bibnamefont {Moya~de
  Guerra}}, \bibinfo {author} {\bibfnamefont {E.}~\bibnamefont {Garrido}}, \
  and\ \bibinfo {author} {\bibfnamefont {J.~A.}\ \bibnamefont {Caballero}},\
  }\href {\doibase 10.1103/PhysRevC.48.2731} {\bibfield  {journal} {\bibinfo
  {journal} {Phys. Rev. C}\ }\textbf {\bibinfo {volume} {48}},\ \bibinfo
  {pages} {2731} (\bibinfo {year} {1993})}\BibitemShut {NoStop}%
\bibitem [{\citenamefont {Ud\'{\i}as}\ \emph {et~al.}(1995)\citenamefont
  {Ud\'{\i}as}, \citenamefont {Sarriguren}, \citenamefont {Moya~de Guerra},
  \citenamefont {Garrido},\ and\ \citenamefont {Caballero}}]{Udias95}%
  \BibitemOpen
  \bibfield  {author} {\bibinfo {author} {\bibfnamefont {J.~M.}\ \bibnamefont
  {Ud\'{\i}as}}, \bibinfo {author} {\bibfnamefont {P.}~\bibnamefont
  {Sarriguren}}, \bibinfo {author} {\bibfnamefont {E.}~\bibnamefont {Moya~de
  Guerra}}, \bibinfo {author} {\bibfnamefont {E.}~\bibnamefont {Garrido}}, \
  and\ \bibinfo {author} {\bibfnamefont {J.~A.}\ \bibnamefont {Caballero}},\
  }\href {\doibase 10.1103/PhysRevC.51.3246} {\bibfield  {journal} {\bibinfo
  {journal} {Phys. Rev. C}\ }\textbf {\bibinfo {volume} {51}},\ \bibinfo
  {pages} {3246} (\bibinfo {year} {1995})}\BibitemShut {NoStop}%
\bibitem [{\citenamefont {Mart\'{\i}nez}\ \emph {et~al.}(2006)\citenamefont
  {Mart\'{\i}nez}, \citenamefont {Lava}, \citenamefont {Jachowicz},
  \citenamefont {Ryckebusch}, \citenamefont {Vantournhout},\ and\ \citenamefont
  {Ud\'{\i}as}}]{Martinez06}%
  \BibitemOpen
  \bibfield  {author} {\bibinfo {author} {\bibfnamefont {M.~C.}\ \bibnamefont
  {Mart\'{\i}nez}}, \bibinfo {author} {\bibfnamefont {P.}~\bibnamefont {Lava}},
  \bibinfo {author} {\bibfnamefont {N.}~\bibnamefont {Jachowicz}}, \bibinfo
  {author} {\bibfnamefont {J.}~\bibnamefont {Ryckebusch}}, \bibinfo {author}
  {\bibfnamefont {K.}~\bibnamefont {Vantournhout}}, \ and\ \bibinfo {author}
  {\bibfnamefont {J.~M.}\ \bibnamefont {Ud\'{\i}as}},\ }\href {\doibase
  10.1103/PhysRevC.73.024607} {\bibfield  {journal} {\bibinfo  {journal} {Phys.
  Rev. C}\ }\textbf {\bibinfo {volume} {73}},\ \bibinfo {pages} {024607}
  (\bibinfo {year} {2006})}\BibitemShut {NoStop}%
\bibitem [{\citenamefont {Benhar}\ \emph {et~al.}(1994)\citenamefont {Benhar},
  \citenamefont {Fabrocini}, \citenamefont {Fantoni},\ and\ \citenamefont
  {Sick}}]{Benhar94}%
  \BibitemOpen
  \bibfield  {author} {\bibinfo {author} {\bibfnamefont {O.}~\bibnamefont
  {Benhar}}, \bibinfo {author} {\bibfnamefont {A.}~\bibnamefont {Fabrocini}},
  \bibinfo {author} {\bibfnamefont {S.}~\bibnamefont {Fantoni}}, \ and\
  \bibinfo {author} {\bibfnamefont {I.}~\bibnamefont {Sick}},\ }\href {\doibase
  https://doi.org/10.1016/0375-9474(94)90920-2} {\bibfield  {journal} {\bibinfo
   {journal} {Nuclear Physics A}\ }\textbf {\bibinfo {volume} {579}},\ \bibinfo
  {pages} {493 } (\bibinfo {year} {1994})}\BibitemShut {NoStop}%
\bibitem [{\citenamefont {Benhar}\ \emph {et~al.}(2005)\citenamefont {Benhar},
  \citenamefont {Farina}, \citenamefont {Nakamura}, \citenamefont {Sakuda},\
  and\ \citenamefont {Seki}}]{Benhar05}%
  \BibitemOpen
  \bibfield  {author} {\bibinfo {author} {\bibfnamefont {O.}~\bibnamefont
  {Benhar}}, \bibinfo {author} {\bibfnamefont {N.}~\bibnamefont {Farina}},
  \bibinfo {author} {\bibfnamefont {H.}~\bibnamefont {Nakamura}}, \bibinfo
  {author} {\bibfnamefont {M.}~\bibnamefont {Sakuda}}, \ and\ \bibinfo {author}
  {\bibfnamefont {R.}~\bibnamefont {Seki}},\ }\href {\doibase
  10.1103/PhysRevD.72.053005} {\bibfield  {journal} {\bibinfo  {journal} {Phys.
  Rev. D}\ }\textbf {\bibinfo {volume} {72}},\ \bibinfo {pages} {053005}
  (\bibinfo {year} {2005})}\BibitemShut {NoStop}%
\bibitem [{\citenamefont {Amir-Azimi-Nili}\ \emph {et~al.}(1997)\citenamefont
  {Amir-Azimi-Nili}, \citenamefont {Udias}, \citenamefont {Müther},
  \citenamefont {Skouras},\ and\ \citenamefont {Polls}}]{Amir97}%
  \BibitemOpen
  \bibfield  {author} {\bibinfo {author} {\bibfnamefont {K.}~\bibnamefont
  {Amir-Azimi-Nili}}, \bibinfo {author} {\bibfnamefont {J.}~\bibnamefont
  {Udias}}, \bibinfo {author} {\bibfnamefont {H.}~\bibnamefont {Müther}},
  \bibinfo {author} {\bibfnamefont {L.}~\bibnamefont {Skouras}}, \ and\
  \bibinfo {author} {\bibfnamefont {A.}~\bibnamefont {Polls}},\ }\href
  {\doibase https://doi.org/10.1016/S0375-9474(97)00595-2} {\bibfield
  {journal} {\bibinfo  {journal} {Nuclear Physics A}\ }\textbf {\bibinfo
  {volume} {625}},\ \bibinfo {pages} {633 } (\bibinfo {year}
  {1997})}\BibitemShut {NoStop}%
\bibitem [{\citenamefont {Ivanov}\ \emph {et~al.}(2019)\citenamefont {Ivanov},
  \citenamefont {Antonov}, \citenamefont {Megias}, \citenamefont {Caballero},
  \citenamefont {Barbaro}, \citenamefont {Amaro}, \citenamefont {Ruiz~Simo},
  \citenamefont {Donnelly},\ and\ \citenamefont {Ud\'{\i}as}}]{Ivanov19}%
  \BibitemOpen
  \bibfield  {author} {\bibinfo {author} {\bibfnamefont {M.~V.}\ \bibnamefont
  {Ivanov}}, \bibinfo {author} {\bibfnamefont {A.~N.}\ \bibnamefont {Antonov}},
  \bibinfo {author} {\bibfnamefont {G.~D.}\ \bibnamefont {Megias}}, \bibinfo
  {author} {\bibfnamefont {J.~A.}\ \bibnamefont {Caballero}}, \bibinfo {author}
  {\bibfnamefont {M.~B.}\ \bibnamefont {Barbaro}}, \bibinfo {author}
  {\bibfnamefont {J.~E.}\ \bibnamefont {Amaro}}, \bibinfo {author}
  {\bibfnamefont {I.}~\bibnamefont {Ruiz~Simo}}, \bibinfo {author}
  {\bibfnamefont {T.~W.}\ \bibnamefont {Donnelly}}, \ and\ \bibinfo {author}
  {\bibfnamefont {J.~M.}\ \bibnamefont {Ud\'{\i}as}},\ }\href {\doibase
  10.1103/PhysRevC.99.014610} {\bibfield  {journal} {\bibinfo  {journal} {Phys.
  Rev. C}\ }\textbf {\bibinfo {volume} {99}},\ \bibinfo {pages} {014610}
  (\bibinfo {year} {2019})}\BibitemShut {NoStop}%
\bibitem [{\citenamefont {Franco-Patino}\ \emph {et~al.}(2020)\citenamefont
  {Franco-Patino}, \citenamefont {Gonzalez-Rosa}, \citenamefont {Caballero},\
  and\ \citenamefont {Barbaro}}]{Franco-Patino20}%
  \BibitemOpen
  \bibfield  {author} {\bibinfo {author} {\bibfnamefont {J.~M.}\ \bibnamefont
  {Franco-Patino}}, \bibinfo {author} {\bibfnamefont {J.}~\bibnamefont
  {Gonzalez-Rosa}}, \bibinfo {author} {\bibfnamefont {J.~A.}\ \bibnamefont
  {Caballero}}, \ and\ \bibinfo {author} {\bibfnamefont {M.~B.}\ \bibnamefont
  {Barbaro}},\ }\href {\doibase 10.1103/PhysRevC.102.064626} {\bibfield
  {journal} {\bibinfo  {journal} {Phys. Rev. C}\ }\textbf {\bibinfo {volume}
  {102}},\ \bibinfo {pages} {064626} (\bibinfo {year} {2020})}\BibitemShut
  {NoStop}%
\bibitem [{\citenamefont {Colle}\ \emph {et~al.}(2014)\citenamefont {Colle},
  \citenamefont {Cosyn}, \citenamefont {Ryckebusch},\ and\ \citenamefont
  {Vanhalst}}]{Colle14}%
  \BibitemOpen
  \bibfield  {author} {\bibinfo {author} {\bibfnamefont {C.}~\bibnamefont
  {Colle}}, \bibinfo {author} {\bibfnamefont {W.}~\bibnamefont {Cosyn}},
  \bibinfo {author} {\bibfnamefont {J.}~\bibnamefont {Ryckebusch}}, \ and\
  \bibinfo {author} {\bibfnamefont {M.}~\bibnamefont {Vanhalst}},\ }\href
  {\doibase 10.1103/PhysRevC.89.024603} {\bibfield  {journal} {\bibinfo
  {journal} {Phys. Rev. C}\ }\textbf {\bibinfo {volume} {89}},\ \bibinfo
  {pages} {024603} (\bibinfo {year} {2014})}\BibitemShut {NoStop}%
\bibitem [{\citenamefont {Ryckebusch}\ \emph {et~al.}(2015)\citenamefont
  {Ryckebusch}, \citenamefont {Vanhalst},\ and\ \citenamefont
  {Cosyn}}]{Ryckebusch15}%
  \BibitemOpen
  \bibfield  {author} {\bibinfo {author} {\bibfnamefont {J.}~\bibnamefont
  {Ryckebusch}}, \bibinfo {author} {\bibfnamefont {M.}~\bibnamefont
  {Vanhalst}}, \ and\ \bibinfo {author} {\bibfnamefont {W.}~\bibnamefont
  {Cosyn}},\ }\href {\doibase 10.1088/0954-3899/42/5/055104} {\bibfield
  {journal} {\bibinfo  {journal} {J. Phys. G: Nucl. Part. Phys.}\ }\textbf
  {\bibinfo {volume} {42}},\ \bibinfo {pages} {055104} (\bibinfo {year}
  {2015})}\BibitemShut {NoStop}%
\bibitem [{\citenamefont {Caballero}\ \emph {et~al.}(1998)\citenamefont
  {Caballero}, \citenamefont {Donnelly}, \citenamefont {{Moya de Guerra}},\
  and\ \citenamefont {Udías}}]{Caballero98a}%
  \BibitemOpen
  \bibfield  {author} {\bibinfo {author} {\bibfnamefont {J.}~\bibnamefont
  {Caballero}}, \bibinfo {author} {\bibfnamefont {T.}~\bibnamefont {Donnelly}},
  \bibinfo {author} {\bibfnamefont {E.}~\bibnamefont {{Moya de Guerra}}}, \
  and\ \bibinfo {author} {\bibfnamefont {J.}~\bibnamefont {Udías}},\ }\href
  {\doibase https://doi.org/10.1016/S0375-9474(97)00817-8} {\bibfield
  {journal} {\bibinfo  {journal} {Nuclear Physics A}\ }\textbf {\bibinfo
  {volume} {632}},\ \bibinfo {pages} {323 } (\bibinfo {year}
  {1998})}\BibitemShut {NoStop}%
\bibitem [{\citenamefont {Gardner}\ and\ \citenamefont
  {Piekarewicz}(1994)}]{Gardner94}%
  \BibitemOpen
  \bibfield  {author} {\bibinfo {author} {\bibfnamefont {S.}~\bibnamefont
  {Gardner}}\ and\ \bibinfo {author} {\bibfnamefont {J.}~\bibnamefont
  {Piekarewicz}},\ }\href {\doibase 10.1103/PhysRevC.50.2822} {\bibfield
  {journal} {\bibinfo  {journal} {Phys. Rev. C}\ }\textbf {\bibinfo {volume}
  {50}},\ \bibinfo {pages} {2822} (\bibinfo {year} {1994})}\BibitemShut
  {NoStop}%
\bibitem [{\citenamefont {Ivanov}\ \emph {et~al.}(2013)\citenamefont {Ivanov},
  \citenamefont {Gonz\'alez-Jim\'enez}, \citenamefont {Caballero},
  \citenamefont {Barbaro}, \citenamefont {Donnelly},\ and\ \citenamefont
  {Ud\'ias}}]{Ivanov13}%
  \BibitemOpen
  \bibfield  {author} {\bibinfo {author} {\bibfnamefont {M.}~\bibnamefont
  {Ivanov}}, \bibinfo {author} {\bibfnamefont {R.}~\bibnamefont
  {Gonz\'alez-Jim\'enez}}, \bibinfo {author} {\bibfnamefont {J.}~\bibnamefont
  {Caballero}}, \bibinfo {author} {\bibfnamefont {M.}~\bibnamefont {Barbaro}},
  \bibinfo {author} {\bibfnamefont {T.}~\bibnamefont {Donnelly}}, \ and\
  \bibinfo {author} {\bibfnamefont {J.}~\bibnamefont {Ud\'ias}},\ }\href
  {\doibase https://doi.org/10.1016/j.physletb.2013.10.001} {\bibfield
  {journal} {\bibinfo  {journal} {Physics Letters B}\ }\textbf {\bibinfo
  {volume} {727}},\ \bibinfo {pages} {265 } (\bibinfo {year}
  {2013})}\BibitemShut {NoStop}%
\bibitem [{\citenamefont {Gonz\'alez-Jim\'enez}\ \emph
  {et~al.}(2020)\citenamefont {Gonz\'alez-Jim\'enez}, \citenamefont {Barbaro},
  \citenamefont {Caballero}, \citenamefont {Donnelly}, \citenamefont
  {Jachowicz}, \citenamefont {Megias}, \citenamefont {Niewczas}, \citenamefont
  {Nikolakopoulos},\ and\ \citenamefont {Ud\'{\i}as}}]{Gonzalez-Jimenez20}%
  \BibitemOpen
  \bibfield  {author} {\bibinfo {author} {\bibfnamefont {R.}~\bibnamefont
  {Gonz\'alez-Jim\'enez}}, \bibinfo {author} {\bibfnamefont {M.~B.}\
  \bibnamefont {Barbaro}}, \bibinfo {author} {\bibfnamefont {J.~A.}\
  \bibnamefont {Caballero}}, \bibinfo {author} {\bibfnamefont {T.~W.}\
  \bibnamefont {Donnelly}}, \bibinfo {author} {\bibfnamefont {N.}~\bibnamefont
  {Jachowicz}}, \bibinfo {author} {\bibfnamefont {G.~D.}\ \bibnamefont
  {Megias}}, \bibinfo {author} {\bibfnamefont {K.}~\bibnamefont {Niewczas}},
  \bibinfo {author} {\bibfnamefont {A.}~\bibnamefont {Nikolakopoulos}}, \ and\
  \bibinfo {author} {\bibfnamefont {J.~M.}\ \bibnamefont {Ud\'{\i}as}},\ }\href
  {\doibase 10.1103/PhysRevC.101.015503} {\bibfield  {journal} {\bibinfo
  {journal} {Phys. Rev. C}\ }\textbf {\bibinfo {volume} {101}},\ \bibinfo
  {pages} {015503} (\bibinfo {year} {2020})}\BibitemShut {NoStop}%
\bibitem [{\citenamefont {Cooper}\ \emph {et~al.}(1993)\citenamefont {Cooper},
  \citenamefont {Hama}, \citenamefont {Clark},\ and\ \citenamefont
  {Mercer}}]{Cooper93}%
  \BibitemOpen
  \bibfield  {author} {\bibinfo {author} {\bibfnamefont {E.~D.}\ \bibnamefont
  {Cooper}}, \bibinfo {author} {\bibfnamefont {S.}~\bibnamefont {Hama}},
  \bibinfo {author} {\bibfnamefont {B.~C.}\ \bibnamefont {Clark}}, \ and\
  \bibinfo {author} {\bibfnamefont {R.~L.}\ \bibnamefont {Mercer}},\ }\href
  {\doibase 10.1103/PhysRevC.47.297} {\bibfield  {journal} {\bibinfo  {journal}
  {Phys. Rev. C}\ }\textbf {\bibinfo {volume} {47}},\ \bibinfo {pages} {297}
  (\bibinfo {year} {1993})}\BibitemShut {NoStop}%
\bibitem [{\citenamefont {Gonz\'alez-Jim\'enez}\ \emph
  {et~al.}(2019)\citenamefont {Gonz\'alez-Jim\'enez}, \citenamefont
  {Nikolakopoulos}, \citenamefont {Jachowicz},\ and\ \citenamefont
  {Ud\'{\i}as}}]{Gonzalez-Jimenez19}%
  \BibitemOpen
  \bibfield  {author} {\bibinfo {author} {\bibfnamefont {R.}~\bibnamefont
  {Gonz\'alez-Jim\'enez}}, \bibinfo {author} {\bibfnamefont {A.}~\bibnamefont
  {Nikolakopoulos}}, \bibinfo {author} {\bibfnamefont {N.}~\bibnamefont
  {Jachowicz}}, \ and\ \bibinfo {author} {\bibfnamefont {J.~M.}\ \bibnamefont
  {Ud\'{\i}as}},\ }\href {\doibase 10.1103/PhysRevC.100.045501} {\bibfield
  {journal} {\bibinfo  {journal} {Phys. Rev. C}\ }\textbf {\bibinfo {volume}
  {100}},\ \bibinfo {pages} {045501} (\bibinfo {year} {2019})}\BibitemShut
  {NoStop}%
\bibitem [{\citenamefont {Ud\'{\i}as}\ \emph {et~al.}(2001)\citenamefont
  {Ud\'{\i}as}, \citenamefont {Caballero}, \citenamefont {de~Guerra},
  \citenamefont {Vignote},\ and\ \citenamefont {Escuderos}}]{Udias01}%
  \BibitemOpen
  \bibfield  {author} {\bibinfo {author} {\bibfnamefont {J.~M.}\ \bibnamefont
  {Ud\'{\i}as}}, \bibinfo {author} {\bibfnamefont {J.~A.}\ \bibnamefont
  {Caballero}}, \bibinfo {author} {\bibfnamefont {E.~M.}\ \bibnamefont
  {de~Guerra}}, \bibinfo {author} {\bibfnamefont {J.~R.}\ \bibnamefont
  {Vignote}}, \ and\ \bibinfo {author} {\bibfnamefont {A.}~\bibnamefont
  {Escuderos}},\ }\href {\doibase 10.1103/PhysRevC.64.024614} {\bibfield
  {journal} {\bibinfo  {journal} {Phys. Rev. C}\ }\textbf {\bibinfo {volume}
  {64}},\ \bibinfo {pages} {024614} (\bibinfo {year} {2001})}\BibitemShut
  {NoStop}%
\bibitem [{\citenamefont {Abe}\ \emph {et~al.}(2020)\citenamefont {Abe} \emph
  {et~al.}}]{T2K20}%
  \BibitemOpen
  \bibfield  {author} {\bibinfo {author} {\bibfnamefont {K.}~\bibnamefont
  {Abe}} \emph {et~al.} (\bibinfo {collaboration} {The T2K Collaboration}),\
  }\href {\doibase 10.1038/s41586-020-2177-0} {\bibfield  {journal} {\bibinfo
  {journal} {Nature}\ }\textbf {\bibinfo {volume} {580}},\ \bibinfo {pages}
  {339} (\bibinfo {year} {2020})}\BibitemShut {NoStop}%
\bibitem [{\citenamefont {{NuWro official repository}}()}]{NuWro-web}%
  \BibitemOpen
  \bibfield  {author} {\bibinfo {author} {\bibnamefont {{NuWro official
  repository}}},\ }\href {https://github.com/NuWro/nuwro} {\bibinfo  {journal}
  {https://github.com/NuWro/nuwro}\ }\BibitemShut {NoStop}%
\bibitem [{\citenamefont {Nikolakopoulos}\ \emph {et~al.}(2019)\citenamefont
  {Nikolakopoulos}, \citenamefont {Jachowicz}, \citenamefont {Van~Dessel},
  \citenamefont {Niewczas}, \citenamefont {Gonz\'alez-Jim\'enez}, \citenamefont
  {Ud\'{\i}as},\ and\ \citenamefont {Pandey}}]{Nikolakopoulos19}%
  \BibitemOpen
\bibfield  {journal} {  }\bibfield  {author} {\bibinfo {author} {\bibfnamefont
  {A.}~\bibnamefont {Nikolakopoulos}}, \bibinfo {author} {\bibfnamefont
  {N.}~\bibnamefont {Jachowicz}}, \bibinfo {author} {\bibfnamefont
  {N.}~\bibnamefont {Van~Dessel}}, \bibinfo {author} {\bibfnamefont
  {K.}~\bibnamefont {Niewczas}}, \bibinfo {author} {\bibfnamefont
  {R.}~\bibnamefont {Gonz\'alez-Jim\'enez}}, \bibinfo {author} {\bibfnamefont
  {J.~M.}\ \bibnamefont {Ud\'{\i}as}}, \ and\ \bibinfo {author} {\bibfnamefont
  {V.}~\bibnamefont {Pandey}},\ }\href {\doibase
  10.1103/PhysRevLett.123.052501} {\bibfield  {journal} {\bibinfo  {journal}
  {Phys. Rev. Lett.}\ }\textbf {\bibinfo {volume} {123}},\ \bibinfo {pages}
  {052501} (\bibinfo {year} {2019})}\BibitemShut {NoStop}%
\bibitem [{\citenamefont {Day}\ \emph {et~al.}(1990)\citenamefont {Day},
  \citenamefont {McCarthy}, \citenamefont {Donnelly},\ and\ \citenamefont
  {Sick}}]{Day90}%
  \BibitemOpen
  \bibfield  {author} {\bibinfo {author} {\bibfnamefont {D.~B.}\ \bibnamefont
  {Day}}, \bibinfo {author} {\bibfnamefont {J.~S.}\ \bibnamefont {McCarthy}},
  \bibinfo {author} {\bibfnamefont {T.~W.}\ \bibnamefont {Donnelly}}, \ and\
  \bibinfo {author} {\bibfnamefont {I.}~\bibnamefont {Sick}},\ }\href {\doibase
  10.1146/annurev.ns.40.120190.002041} {\bibfield  {journal} {\bibinfo
  {journal} {Ann. Rev. Nucl. Part. Sci.}\ }\textbf {\bibinfo {volume} {40}},\
  \bibinfo {pages} {357} (\bibinfo {year} {1990})}\BibitemShut {NoStop}%
\bibitem [{\citenamefont {Moreno}\ \emph {et~al.}(2014)\citenamefont {Moreno},
  \citenamefont {Donnelly}, \citenamefont {Van~Orden},\ and\ \citenamefont
  {Ford}}]{Moreno14}%
  \BibitemOpen
  \bibfield  {author} {\bibinfo {author} {\bibfnamefont {O.}~\bibnamefont
  {Moreno}}, \bibinfo {author} {\bibfnamefont {T.~W.}\ \bibnamefont
  {Donnelly}}, \bibinfo {author} {\bibfnamefont {J.~W.}\ \bibnamefont
  {Van~Orden}}, \ and\ \bibinfo {author} {\bibfnamefont {W.~P.}\ \bibnamefont
  {Ford}},\ }\href {\doibase 10.1103/PhysRevD.90.013014} {\bibfield  {journal}
  {\bibinfo  {journal} {Phys. Rev. D}\ }\textbf {\bibinfo {volume} {90}},\
  \bibinfo {pages} {013014} (\bibinfo {year} {2014})}\BibitemShut {NoStop}%
\bibitem [{\citenamefont {Amaro}\ \emph {et~al.}(2020)\citenamefont {Amaro},
  \citenamefont {Barbaro}, \citenamefont {Caballero}, \citenamefont
  {Gonz{\'{a}}lez-Jim{\'{e}}nez}, \citenamefont {Megias},\ and\ \citenamefont
  {Simo}}]{Amaro20}%
  \BibitemOpen
  \bibfield  {author} {\bibinfo {author} {\bibfnamefont {J.~E.}\ \bibnamefont
  {Amaro}}, \bibinfo {author} {\bibfnamefont {M.~B.}\ \bibnamefont {Barbaro}},
  \bibinfo {author} {\bibfnamefont {J.~A.}\ \bibnamefont {Caballero}}, \bibinfo
  {author} {\bibfnamefont {R.}~\bibnamefont {Gonz{\'{a}}lez-Jim{\'{e}}nez}},
  \bibinfo {author} {\bibfnamefont {G.~D.}\ \bibnamefont {Megias}}, \ and\
  \bibinfo {author} {\bibfnamefont {I.~R.}\ \bibnamefont {Simo}},\ }\href
  {\doibase 10.1088/1361-6471/abb128} {\bibfield  {journal} {\bibinfo
  {journal} {J. Phys. G: Nucl. Part. Phys.}\ }\textbf {\bibinfo {volume}
  {47}},\ \bibinfo {pages} {124001} (\bibinfo {year} {2020})}\BibitemShut
  {NoStop}%
\bibitem [{\citenamefont {Pandey}\ \emph {et~al.}(2016)\citenamefont {Pandey},
  \citenamefont {Jachowicz}, \citenamefont {Martini}, \citenamefont
  {Gonz\'alez-Jim\'enez}, \citenamefont {Ryckebusch}, \citenamefont
  {Van~Cuyck},\ and\ \citenamefont {Van~Dessel}}]{Pandey16}%
  \BibitemOpen
  \bibfield  {author} {\bibinfo {author} {\bibfnamefont {V.}~\bibnamefont
  {Pandey}}, \bibinfo {author} {\bibfnamefont {N.}~\bibnamefont {Jachowicz}},
  \bibinfo {author} {\bibfnamefont {M.}~\bibnamefont {Martini}}, \bibinfo
  {author} {\bibfnamefont {R.}~\bibnamefont {Gonz\'alez-Jim\'enez}}, \bibinfo
  {author} {\bibfnamefont {J.}~\bibnamefont {Ryckebusch}}, \bibinfo {author}
  {\bibfnamefont {T.}~\bibnamefont {Van~Cuyck}}, \ and\ \bibinfo {author}
  {\bibfnamefont {N.}~\bibnamefont {Van~Dessel}},\ }\href {\doibase
  10.1103/PhysRevC.94.054609} {\bibfield  {journal} {\bibinfo  {journal} {Phys.
  Rev. C}\ }\textbf {\bibinfo {volume} {94}},\ \bibinfo {pages} {054609}
  (\bibinfo {year} {2016})}\BibitemShut {NoStop}%
\bibitem [{\citenamefont {{Van Orden}}\ and\ \citenamefont
  {Donnelly}(1981)}]{VanOrden81}%
  \BibitemOpen
  \bibfield  {author} {\bibinfo {author} {\bibfnamefont {J.}~\bibnamefont {{Van
  Orden}}}\ and\ \bibinfo {author} {\bibfnamefont {T.}~\bibnamefont
  {Donnelly}},\ }\href {\doibase https://doi.org/10.1016/0003-4916(81)90038-5}
  {\bibfield  {journal} {\bibinfo  {journal} {Annals of Physics}\ }\textbf
  {\bibinfo {volume} {131}},\ \bibinfo {pages} {451} (\bibinfo {year}
  {1981})}\BibitemShut {NoStop}%
\bibitem [{\citenamefont {De Pace}\ \emph {et~al.}(2004)\citenamefont
  {De Pace}, \citenamefont {Nardi}, \citenamefont {Alberico}, \citenamefont
  {Donnelly},\ and\ \citenamefont {Molinari}}]{DePace04}%
  \BibitemOpen
  \bibfield  {author} {\bibinfo {author} {\bibfnamefont {A.}~\bibnamefont
  {De Pace}}, \bibinfo {author} {\bibfnamefont {M.}~\bibnamefont {Nardi}},
  \bibinfo {author} {\bibfnamefont {W.}~\bibnamefont {Alberico}}, \bibinfo
  {author} {\bibfnamefont {T.}~\bibnamefont {Donnelly}}, \ and\ \bibinfo
  {author} {\bibfnamefont {A.}~\bibnamefont {Molinari}},\ }\href {\doibase
  https://doi.org/10.1016/j.nuclphysa.2004.06.014} {\bibfield  {journal}
  {\bibinfo  {journal} {Nuclear Physics A}\ }\textbf {\bibinfo {volume}
  {741}},\ \bibinfo {pages} {249} (\bibinfo {year} {2004})}\BibitemShut
  {NoStop}%
\bibitem [{\citenamefont {Amaro}\ \emph {et~al.}(2002)\citenamefont {Amaro},
  \citenamefont {Barbaro}, \citenamefont {Caballero}, \citenamefont
  {Donnelly},\ and\ \citenamefont {Molinari}}]{Amaro02}%
  \BibitemOpen
  \bibfield  {author} {\bibinfo {author} {\bibfnamefont {J.}~\bibnamefont
  {Amaro}}, \bibinfo {author} {\bibfnamefont {M.}~\bibnamefont {Barbaro}},
  \bibinfo {author} {\bibfnamefont {J.}~\bibnamefont {Caballero}}, \bibinfo
  {author} {\bibfnamefont {T.}~\bibnamefont {Donnelly}}, \ and\ \bibinfo
  {author} {\bibfnamefont {A.}~\bibnamefont {Molinari}},\ }\href {\doibase
  https://doi.org/10.1016/S0370-1573(02)00195-3} {\bibfield  {journal}
  {\bibinfo  {journal} {Physics Reports}\ }\textbf {\bibinfo {volume} {368}},\
  \bibinfo {pages} {317} (\bibinfo {year} {2002})}\BibitemShut {NoStop}%
\bibitem [{\citenamefont {Martini}\ \emph {et~al.}(2009)\citenamefont
  {Martini}, \citenamefont {Ericson}, \citenamefont {Chanfray},\ and\
  \citenamefont {Marteau}}]{Martini09}%
  \BibitemOpen
  \bibfield  {author} {\bibinfo {author} {\bibfnamefont {M.}~\bibnamefont
  {Martini}}, \bibinfo {author} {\bibfnamefont {M.}~\bibnamefont {Ericson}},
  \bibinfo {author} {\bibfnamefont {G.}~\bibnamefont {Chanfray}}, \ and\
  \bibinfo {author} {\bibfnamefont {J.}~\bibnamefont {Marteau}},\ }\href
  {\doibase 10.1103/PhysRevC.80.065501} {\bibfield  {journal} {\bibinfo
  {journal} {Phys. Rev. C}\ }\textbf {\bibinfo {volume} {80}},\ \bibinfo
  {pages} {065501} (\bibinfo {year} {2009})}\BibitemShut {NoStop}%
\bibitem [{\citenamefont {Nieves}\ \emph {et~al.}(2011)\citenamefont {Nieves},
  \citenamefont {Simo},\ and\ \citenamefont {Vacas}}]{Nieves11}%
  \BibitemOpen
  \bibfield  {author} {\bibinfo {author} {\bibfnamefont {J.}~\bibnamefont
  {Nieves}}, \bibinfo {author} {\bibfnamefont {I.~R.}\ \bibnamefont {Simo}}, \
  and\ \bibinfo {author} {\bibfnamefont {M.~J.~V.}\ \bibnamefont {Vacas}},\
  }\href {\doibase 10.1103/PhysRevC.83.045501} {\bibfield  {journal} {\bibinfo
  {journal} {Phys. Rev. C}\ }\textbf {\bibinfo {volume} {83}},\ \bibinfo
  {pages} {045501} (\bibinfo {year} {2011})}\BibitemShut {NoStop}%
\bibitem [{\citenamefont {Van~Cuyck}\ \emph {et~al.}(2017)\citenamefont
  {Van~Cuyck}, \citenamefont {Jachowicz}, \citenamefont {Gonz\'alez-Jim\'enez},
  \citenamefont {Ryckebusch},\ and\ \citenamefont {Van~Dessel}}]{VanCuyck17}%
  \BibitemOpen
  \bibfield  {author} {\bibinfo {author} {\bibfnamefont {T.}~\bibnamefont
  {Van~Cuyck}}, \bibinfo {author} {\bibfnamefont {N.}~\bibnamefont
  {Jachowicz}}, \bibinfo {author} {\bibfnamefont {R.}~\bibnamefont
  {Gonz\'alez-Jim\'enez}}, \bibinfo {author} {\bibfnamefont {J.}~\bibnamefont
  {Ryckebusch}}, \ and\ \bibinfo {author} {\bibfnamefont {N.}~\bibnamefont
  {Van~Dessel}},\ }\href {\doibase 10.1103/PhysRevC.95.054611} {\bibfield
  {journal} {\bibinfo  {journal} {Phys. Rev. C}\ }\textbf {\bibinfo {volume}
  {95}},\ \bibinfo {pages} {054611} (\bibinfo {year} {2017})}\BibitemShut
  {NoStop}%
\end{thebibliography}%

\begin{figure*}[htbp]
\centering  
(a)\includegraphics[width=.3\textwidth,angle=270]{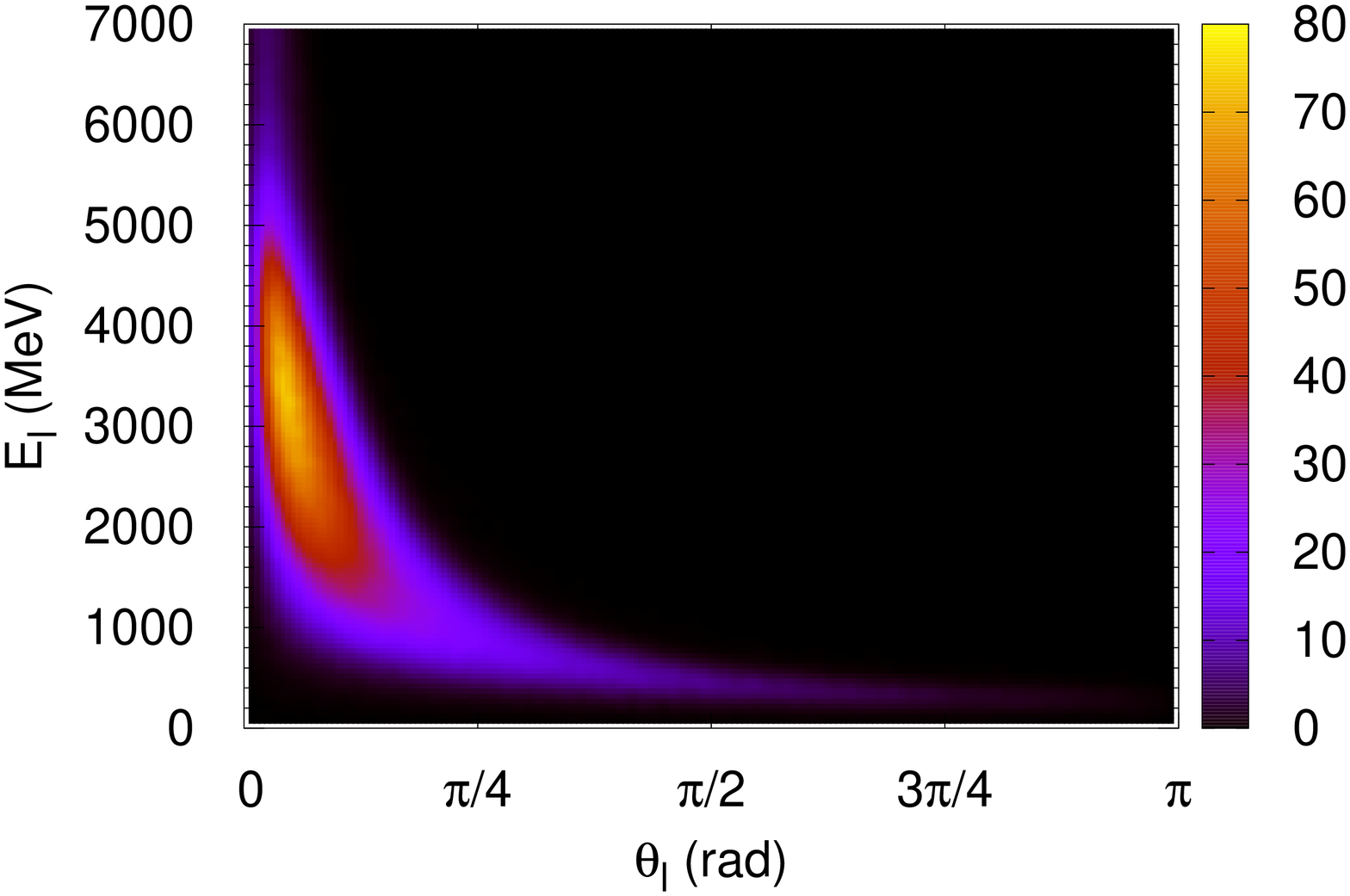}
(b)\includegraphics[width=.3\textwidth,angle=270]{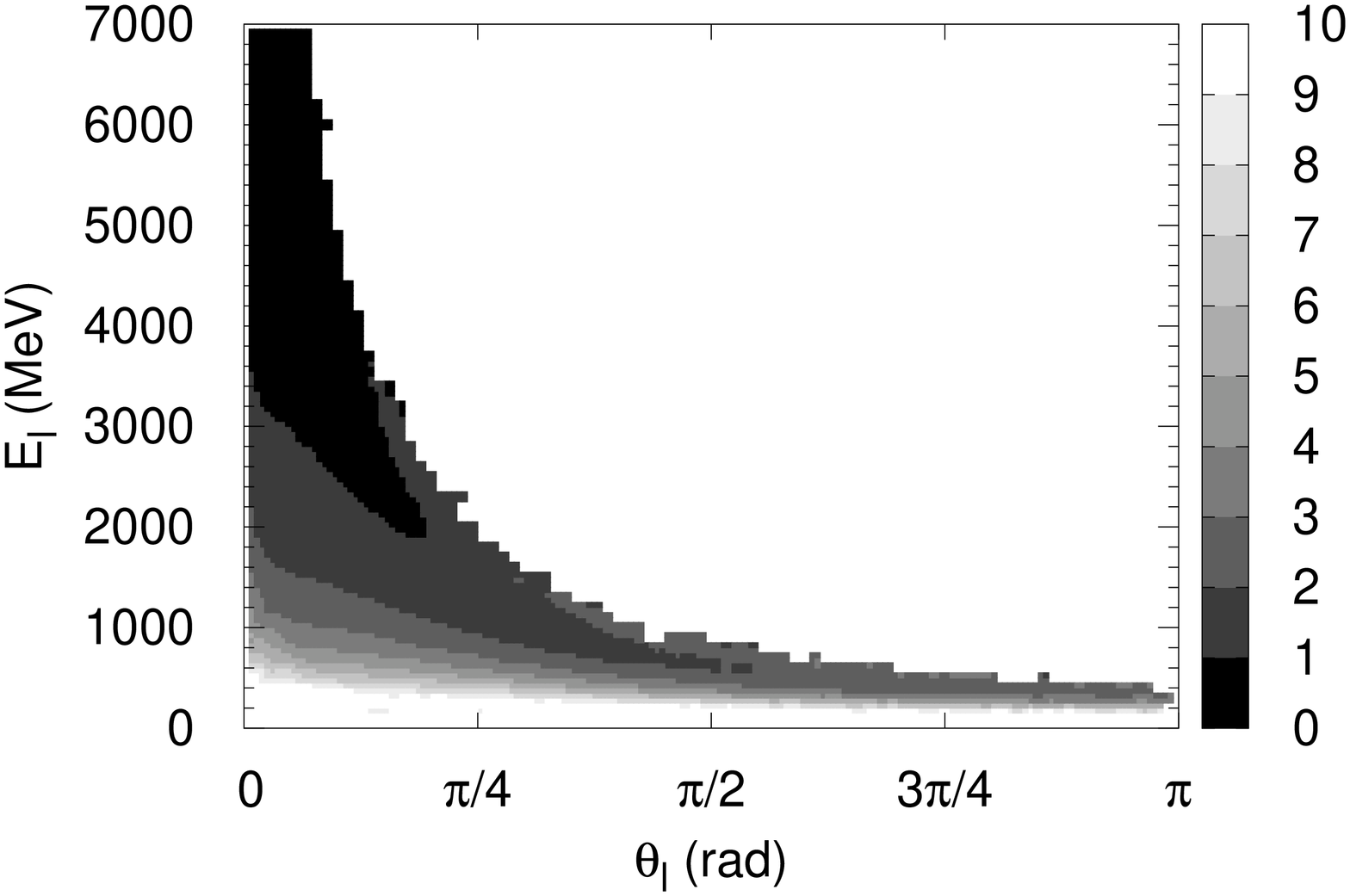}\\
\vspace{-0.5cm}
(c)\includegraphics[width=.3\textwidth,angle=270]{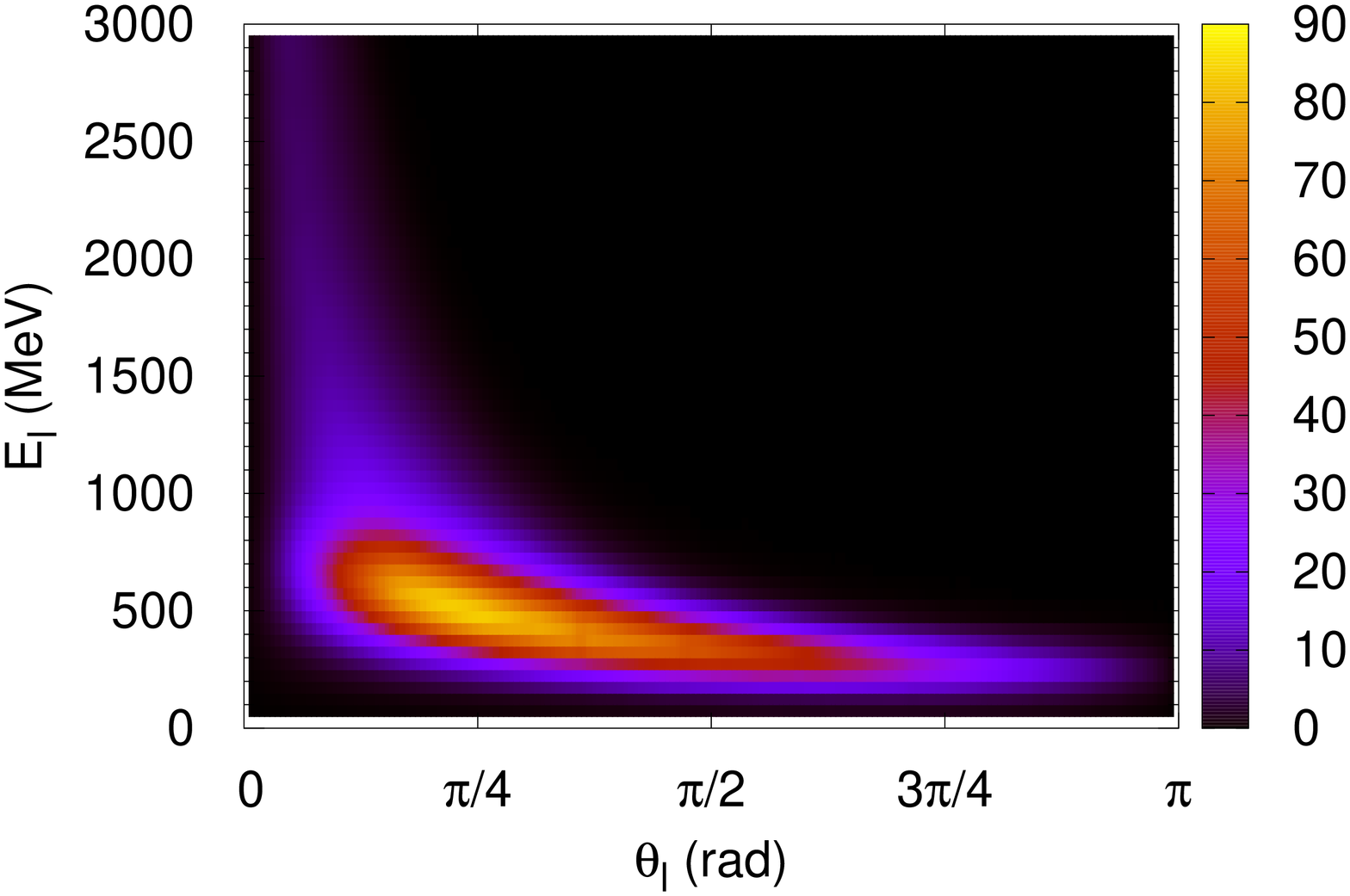}
(d)\includegraphics[width=.3\textwidth,angle=270]{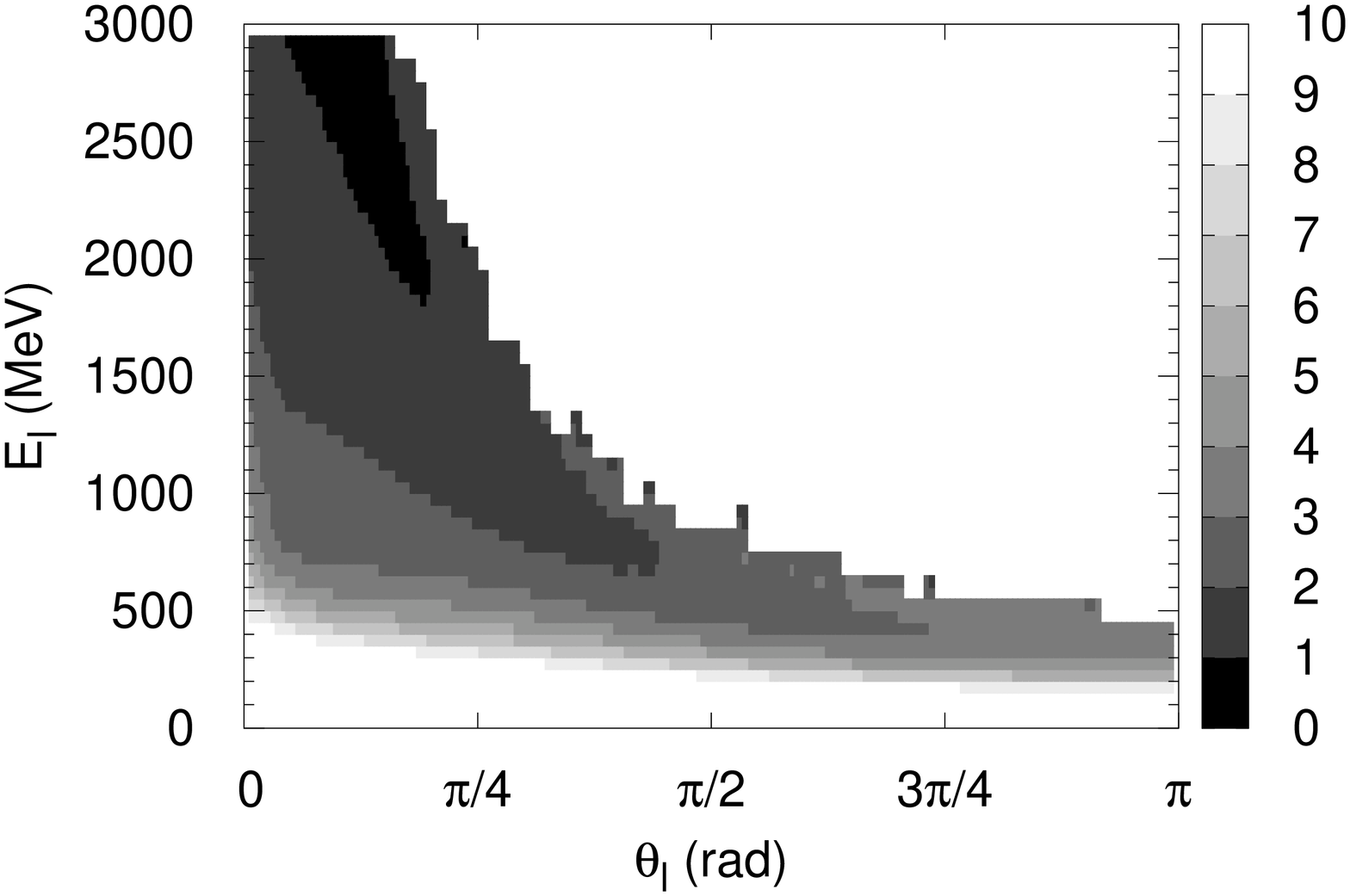}
\vspace{-0.3cm}
\caption{(Color online) (a) and (c) Double-differential cross section $d\sigma/(dE_l d\theta_l) [10^{-42}$cm$^2$/(MeV rad)] as a function of $E_l$ and $\theta_l$. 
(b) and (d) Average $\Delta E$ per bin in percentage. The darkest region corresponds to bins with $\Delta E<1\%$, while whitest are either bins with $\Delta E>10\%$ or with no events. The upper (lower) panels are the results for the DUNE (T2K) flux. We have used the rROP model for all calculations, although similar results are found with the other models. 
}\label{fig:El_thetal}
\end{figure*}
\begin{figure*}[htbp]
\centering  
(a)\includegraphics[width=.3\textwidth,angle=270]{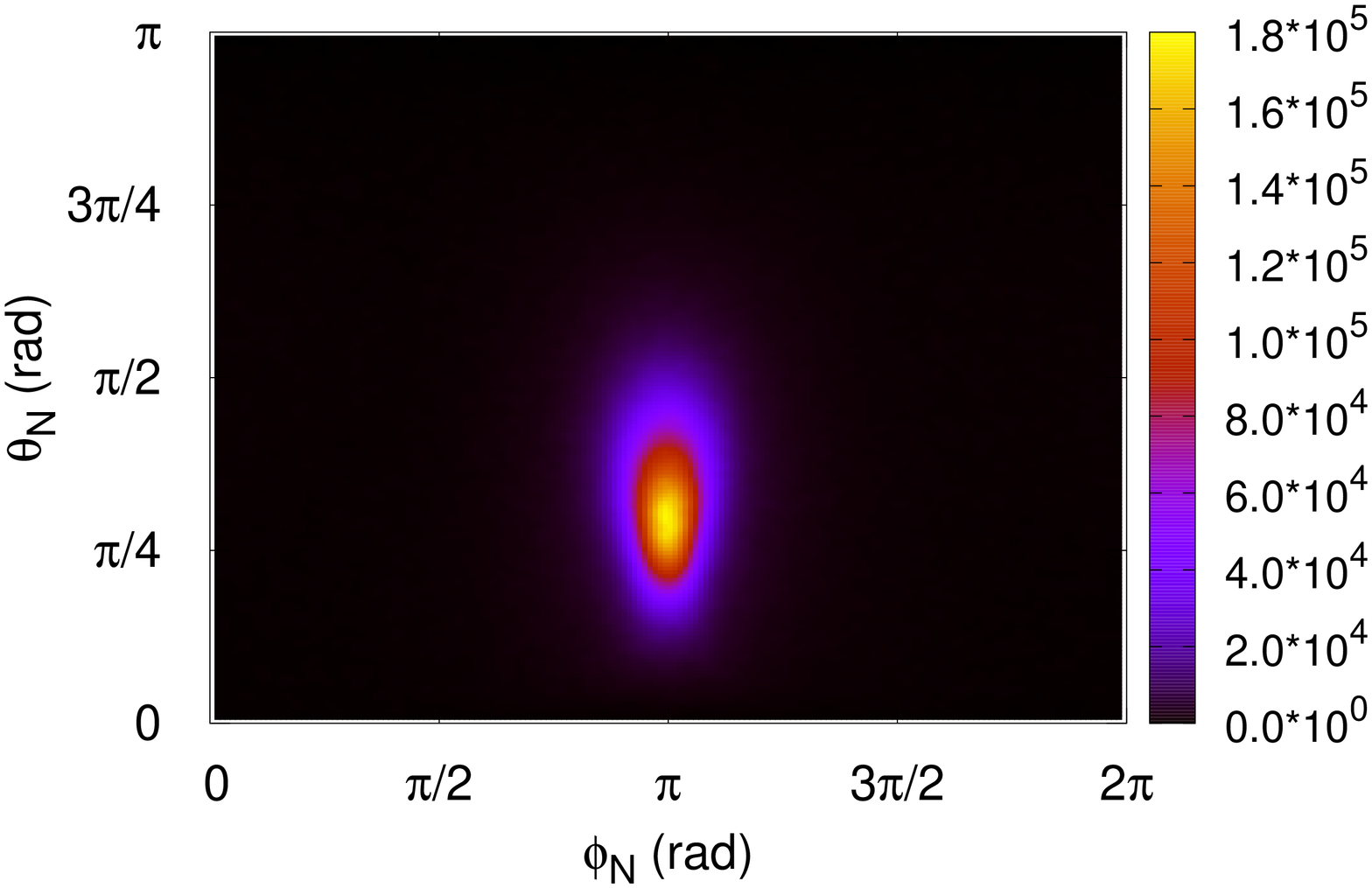}
(b)\includegraphics[width=.3\textwidth,angle=270]{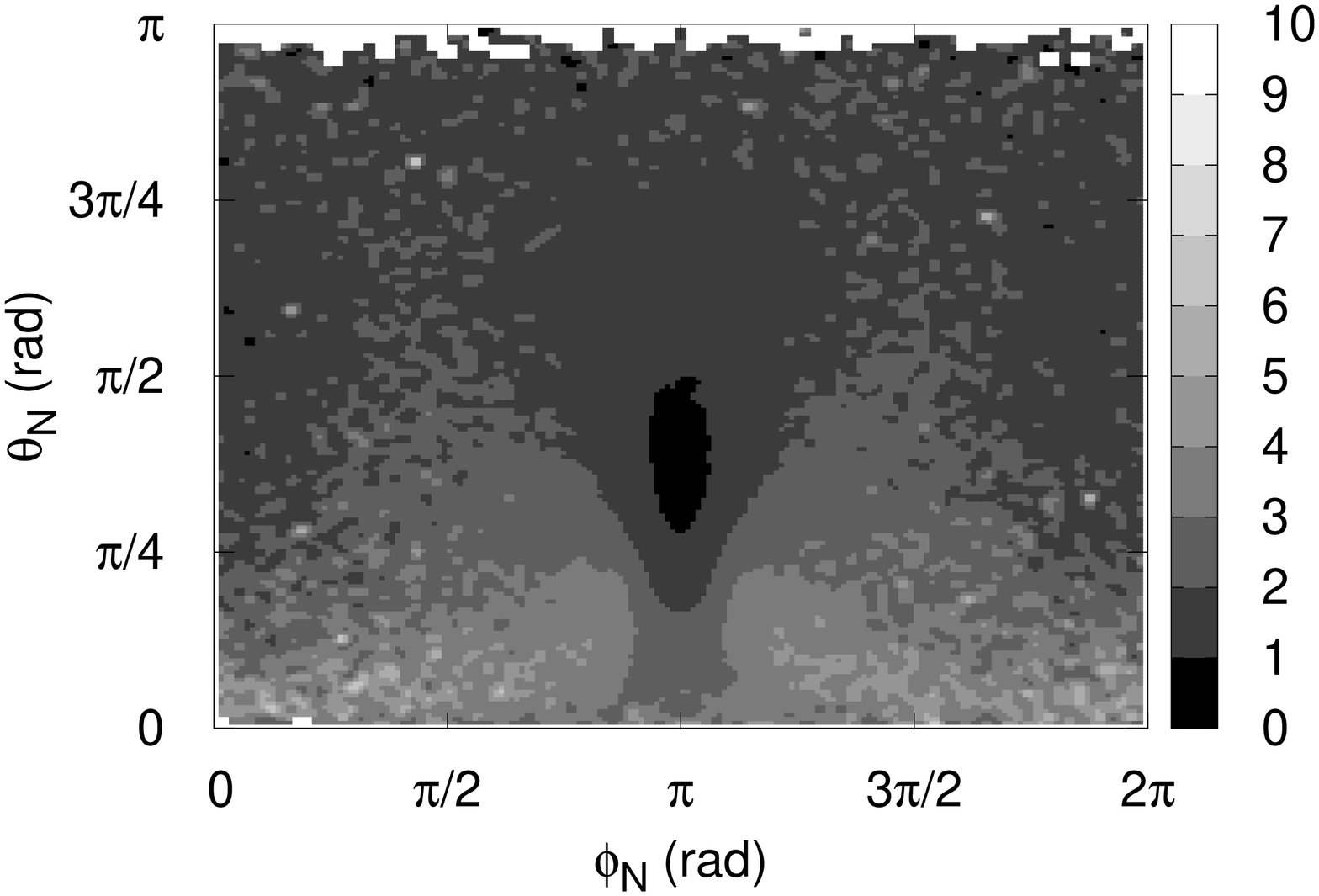}\\
\vspace{-0.5cm}
(c)\includegraphics[width=.3\textwidth,angle=270]{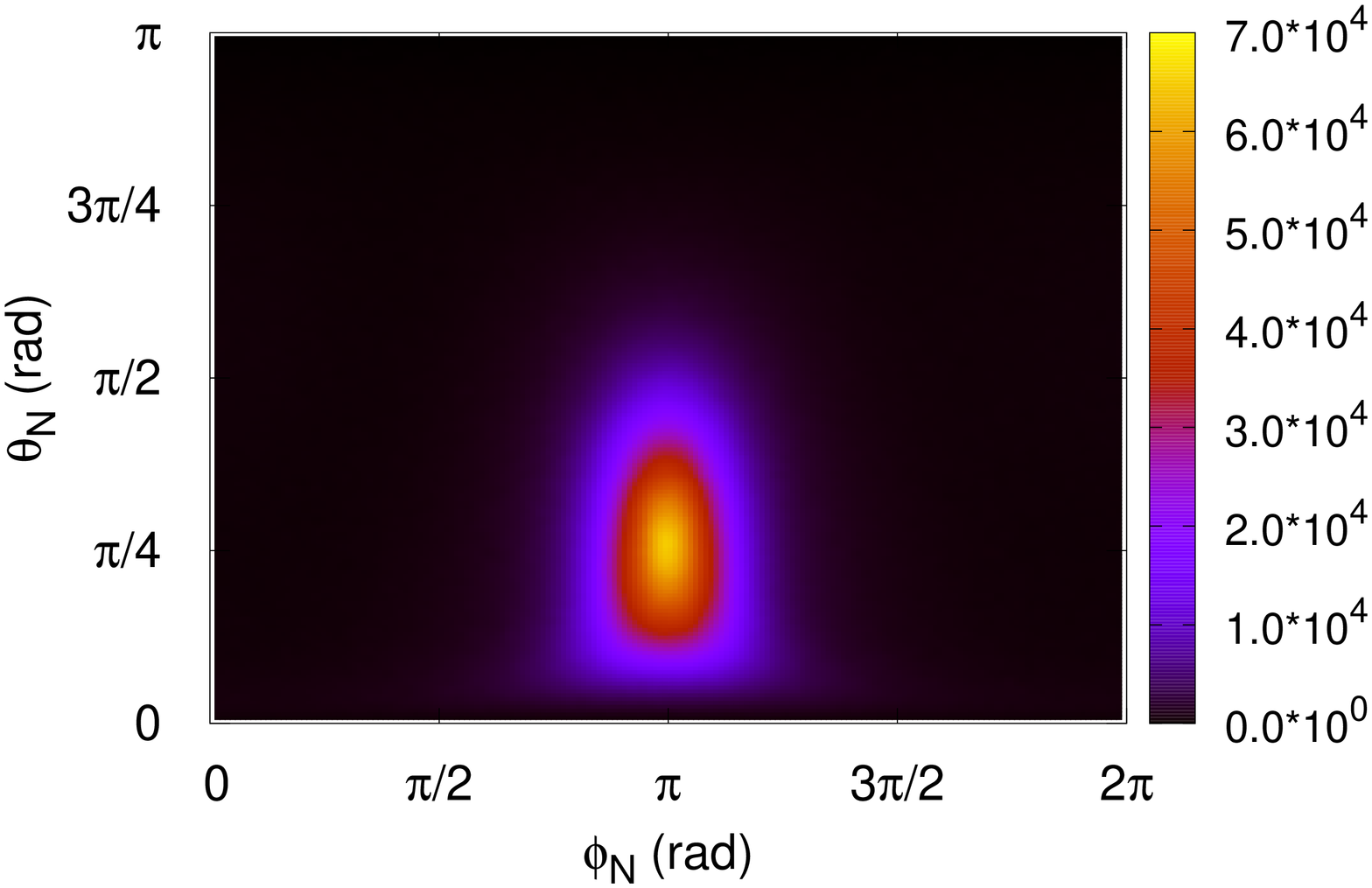}
(d)\includegraphics[width=.3\textwidth,angle=270]{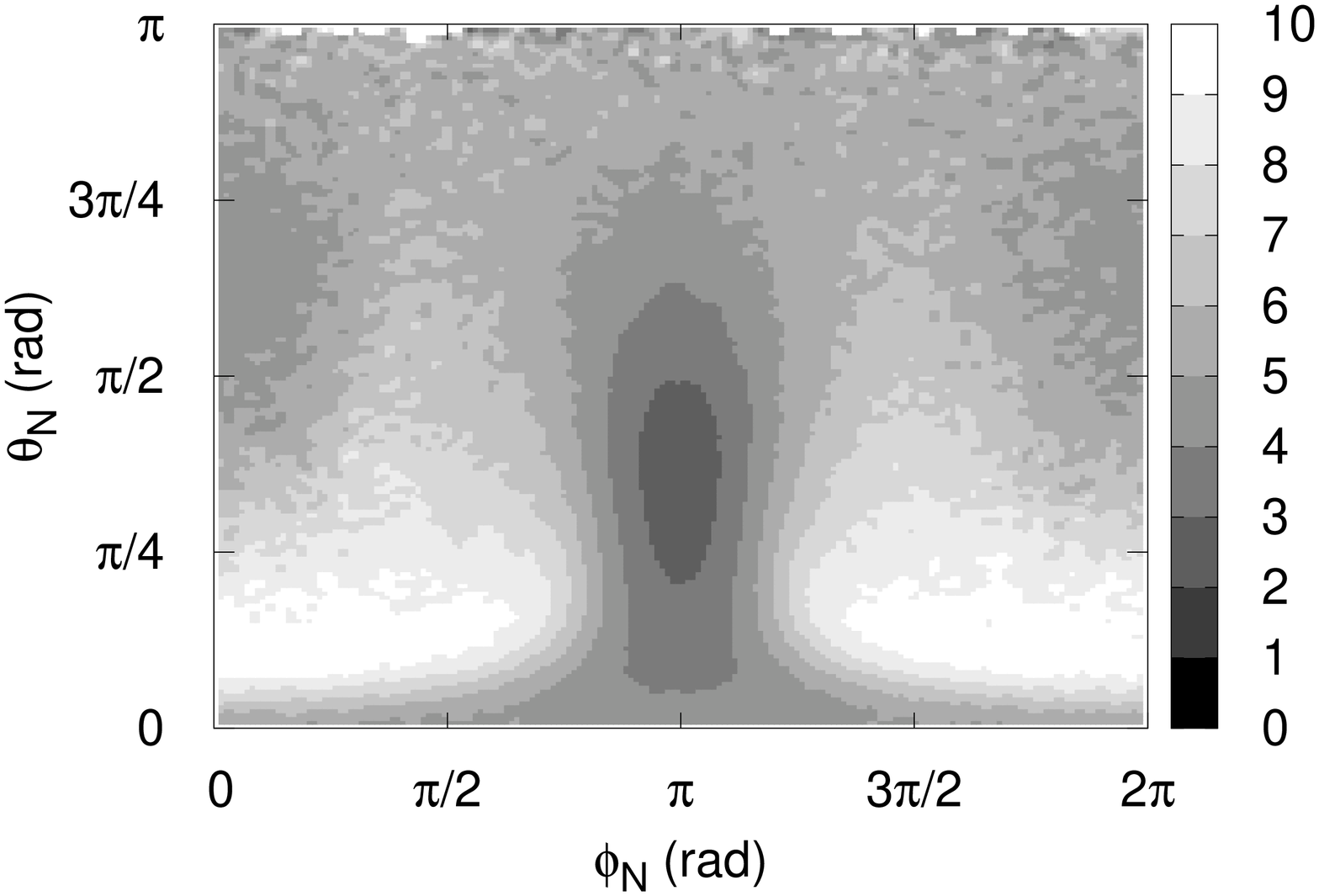}
\vspace{-0.3cm}
\caption{(Color online) As for Fig.~\ref{fig:El_thetal}, except here we represent the double-differential cross section $d\sigma/(d\theta_N d\phi_N) (10^{-42}$cm$^2$/rad$^2)$ as a function of $\theta_N$ and $\phi_N$. }\label{fig:thetaN_phiN}
\end{figure*}
\begin{figure*}[htbp]
\centering  
(a)\includegraphics[width=.3\textwidth,angle=270]{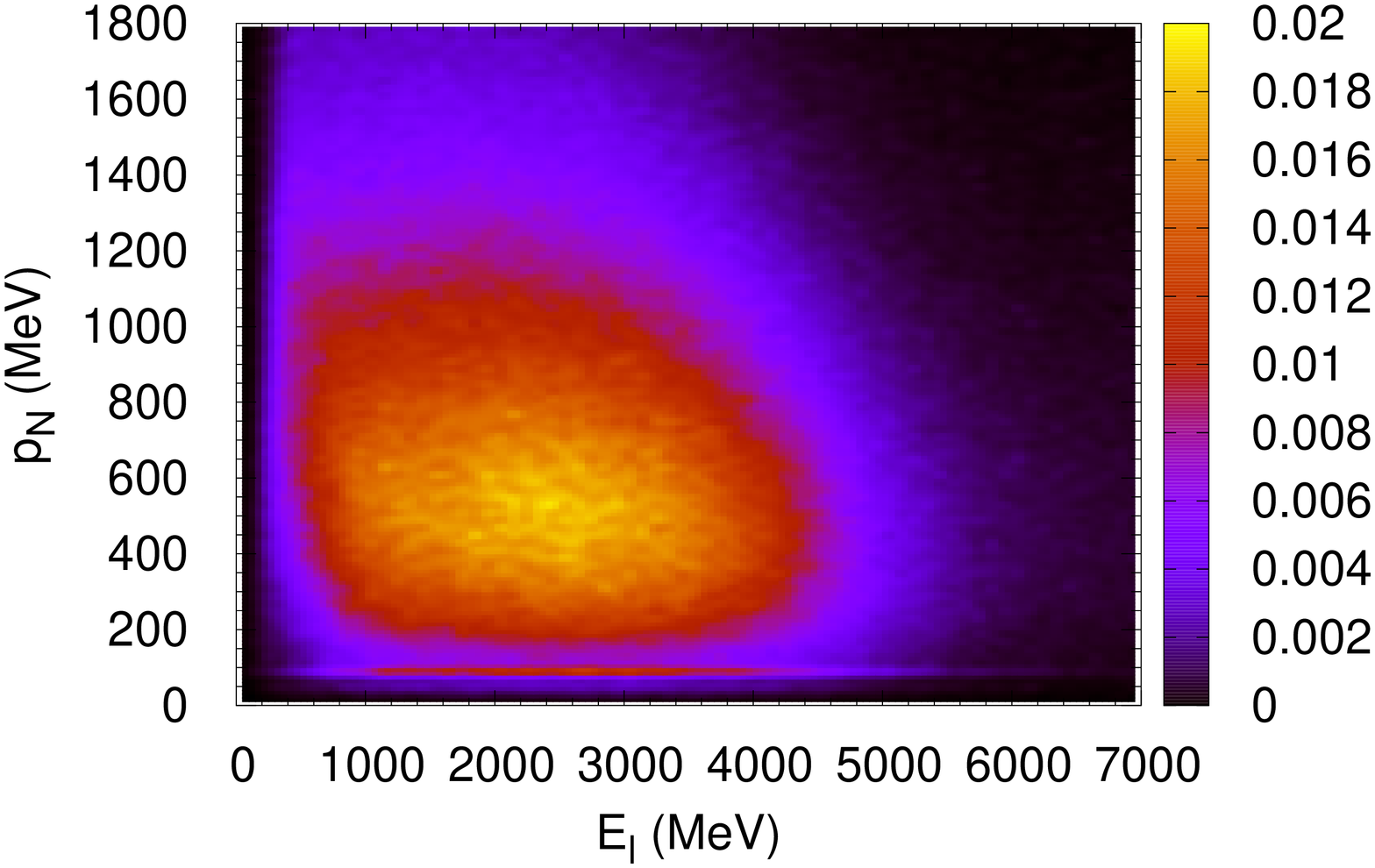}
(b)\includegraphics[width=.3\textwidth,angle=270]{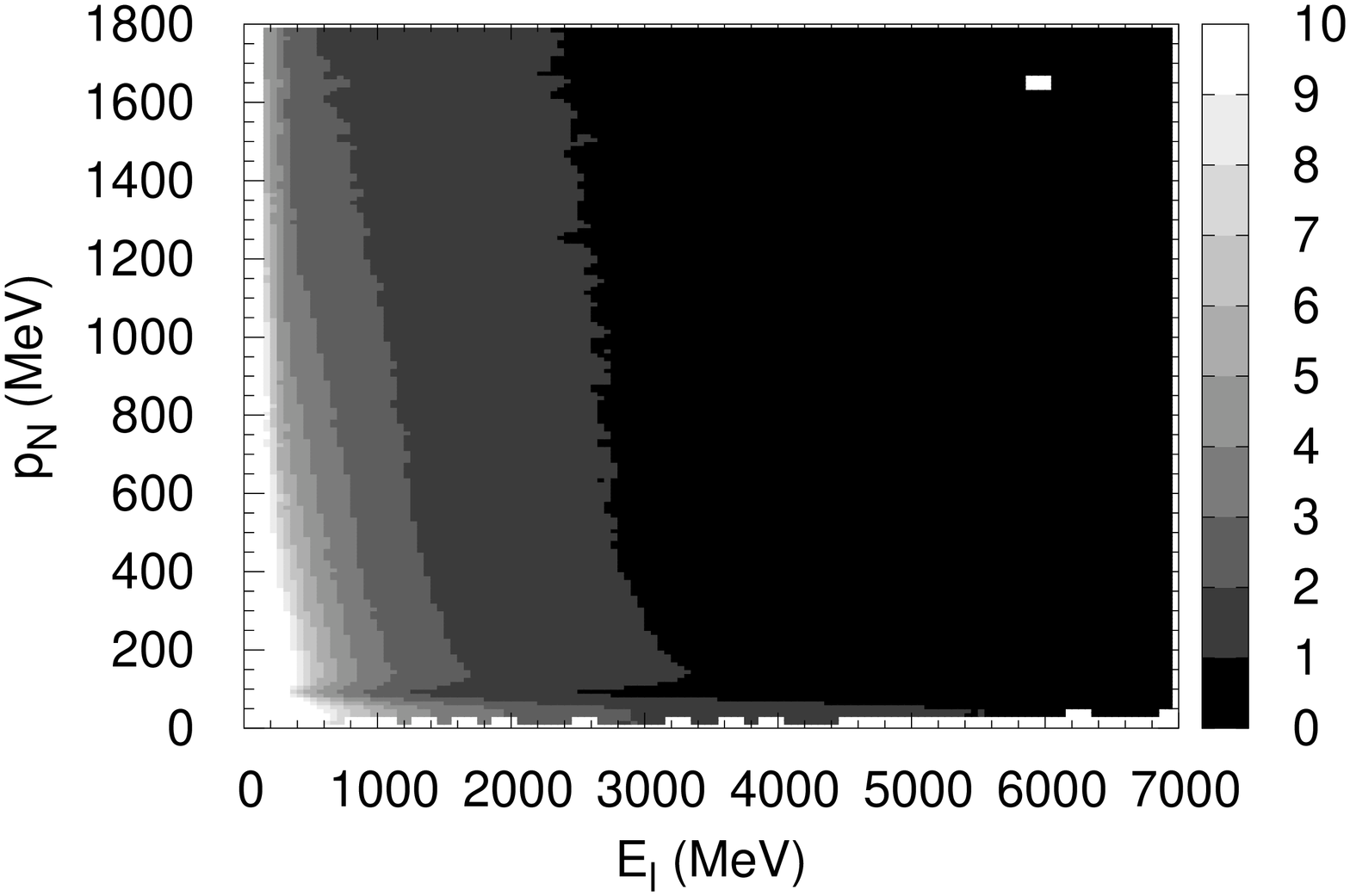}\\
\vspace{-0.5cm}
(c)\includegraphics[width=.3\textwidth,angle=270]{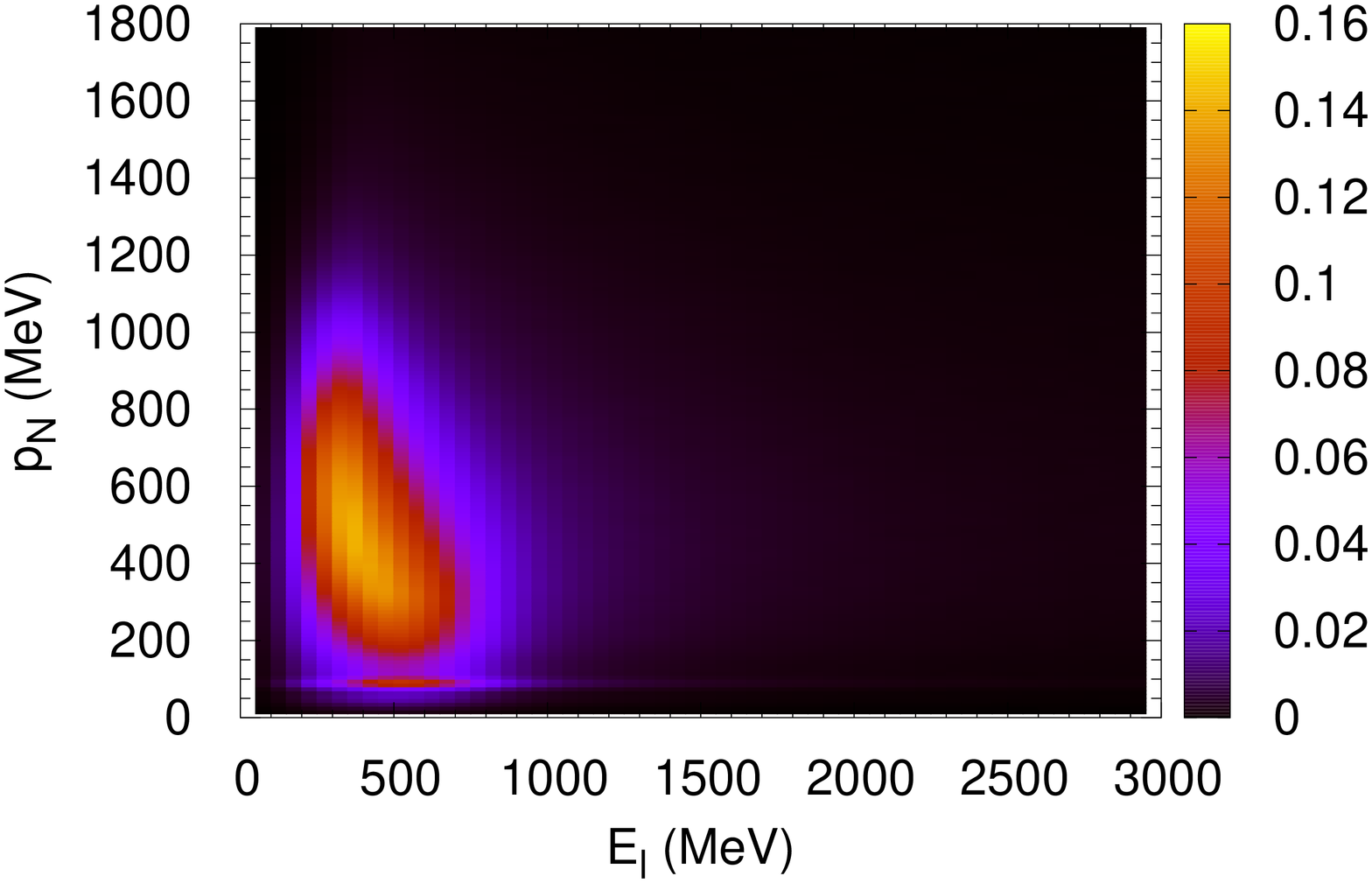}
(d)\includegraphics[width=.3\textwidth,angle=270]{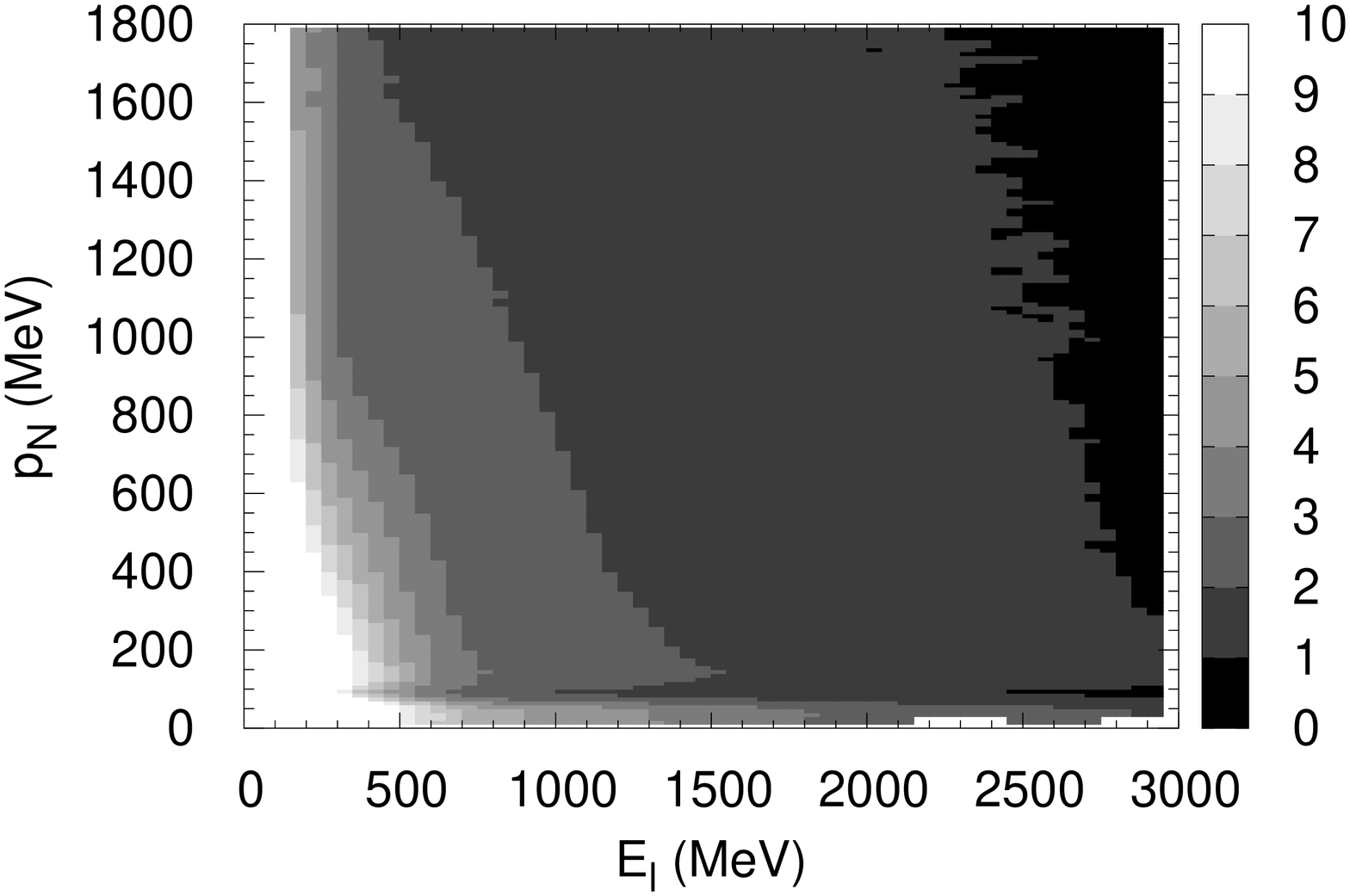}
\vspace{-0.3cm}
\caption{(Color online) As for Fig.~\ref{fig:El_thetal}, except here we represent the double-differential cross section $d\sigma/(dE_l dp_N) [10^{-42}$cm$^2$/MeV$^2]$ as a function of $E_l$ and $p_N$. 
}\label{fig:pN_El}
\end{figure*}
\begin{figure*}[htbp]
\centering  
(a)\includegraphics[width=.3\textwidth,angle=270]{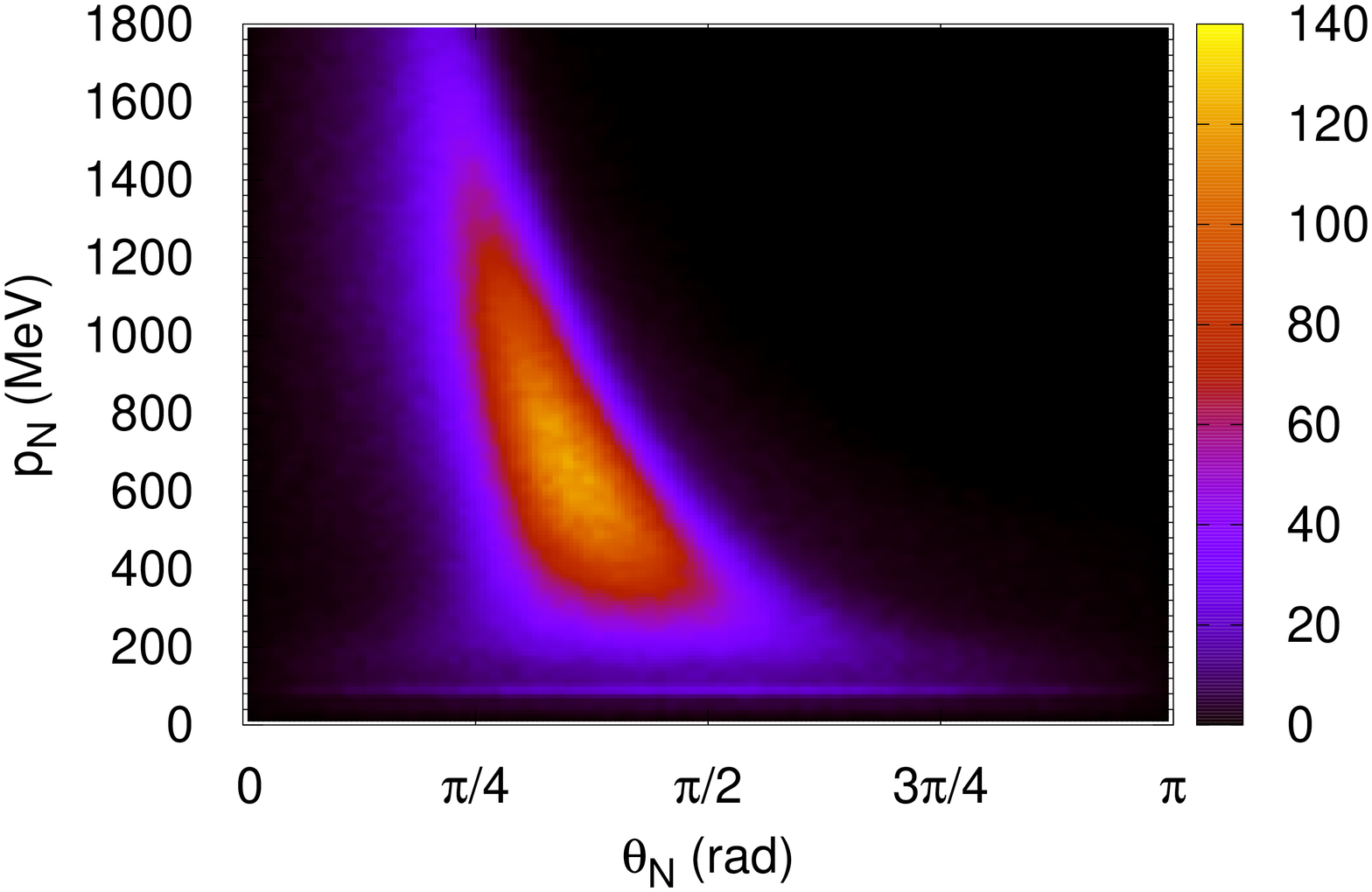}
(b)\includegraphics[width=.3\textwidth,angle=270]{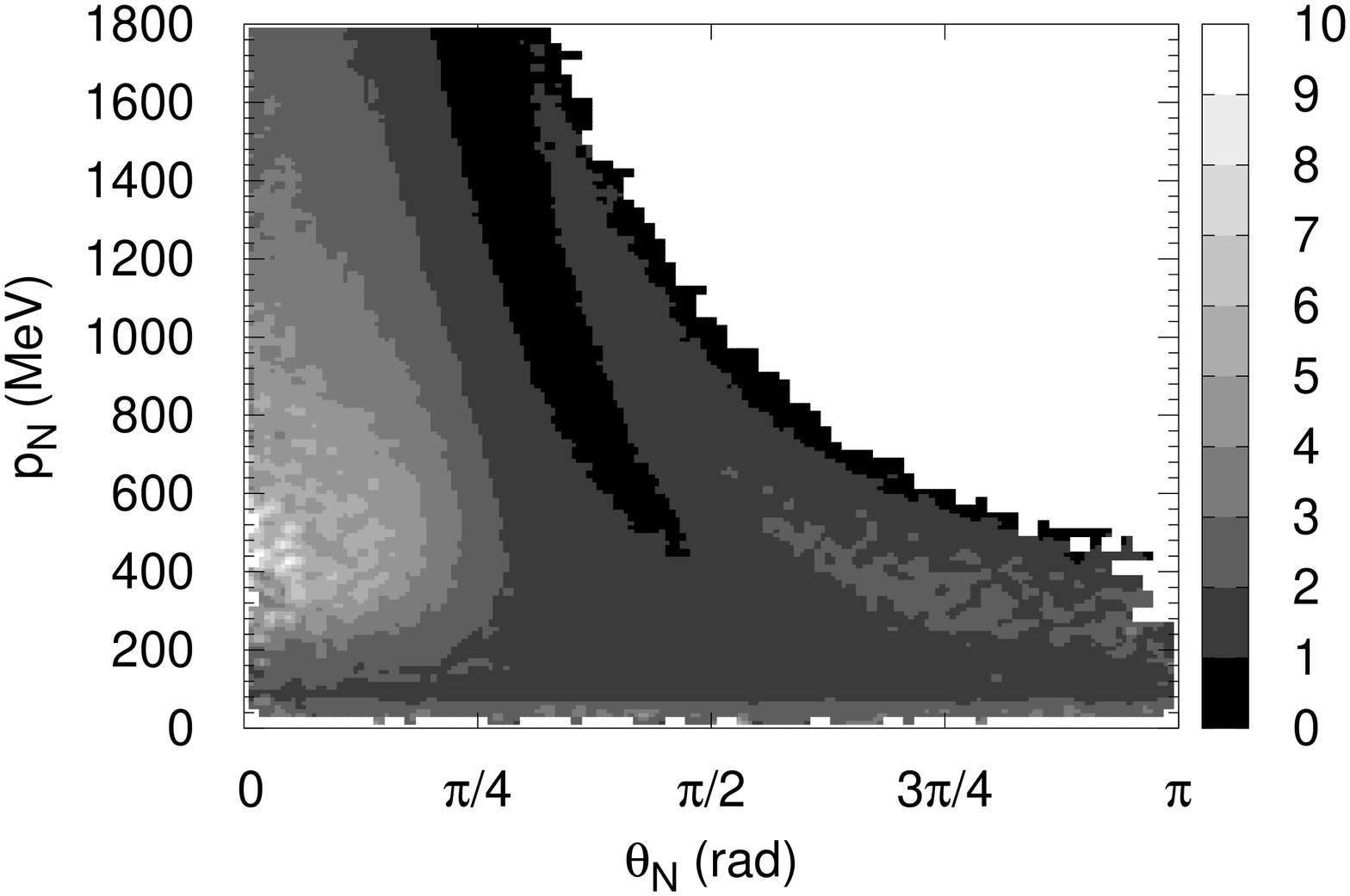}\\
\vspace{-0.5cm}
(c)\includegraphics[width=.3\textwidth,angle=270]{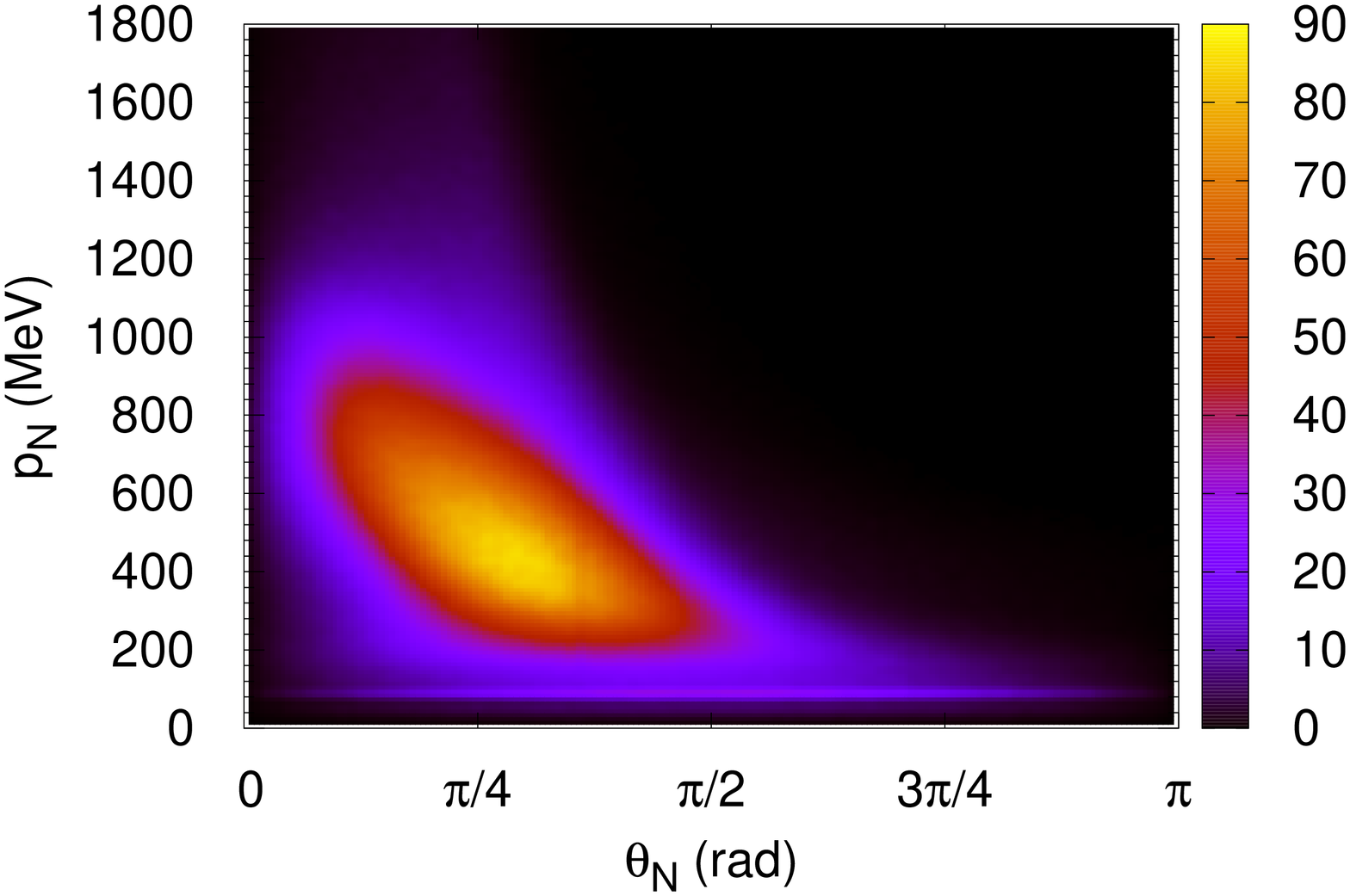}
(d)\includegraphics[width=.3\textwidth,angle=270]{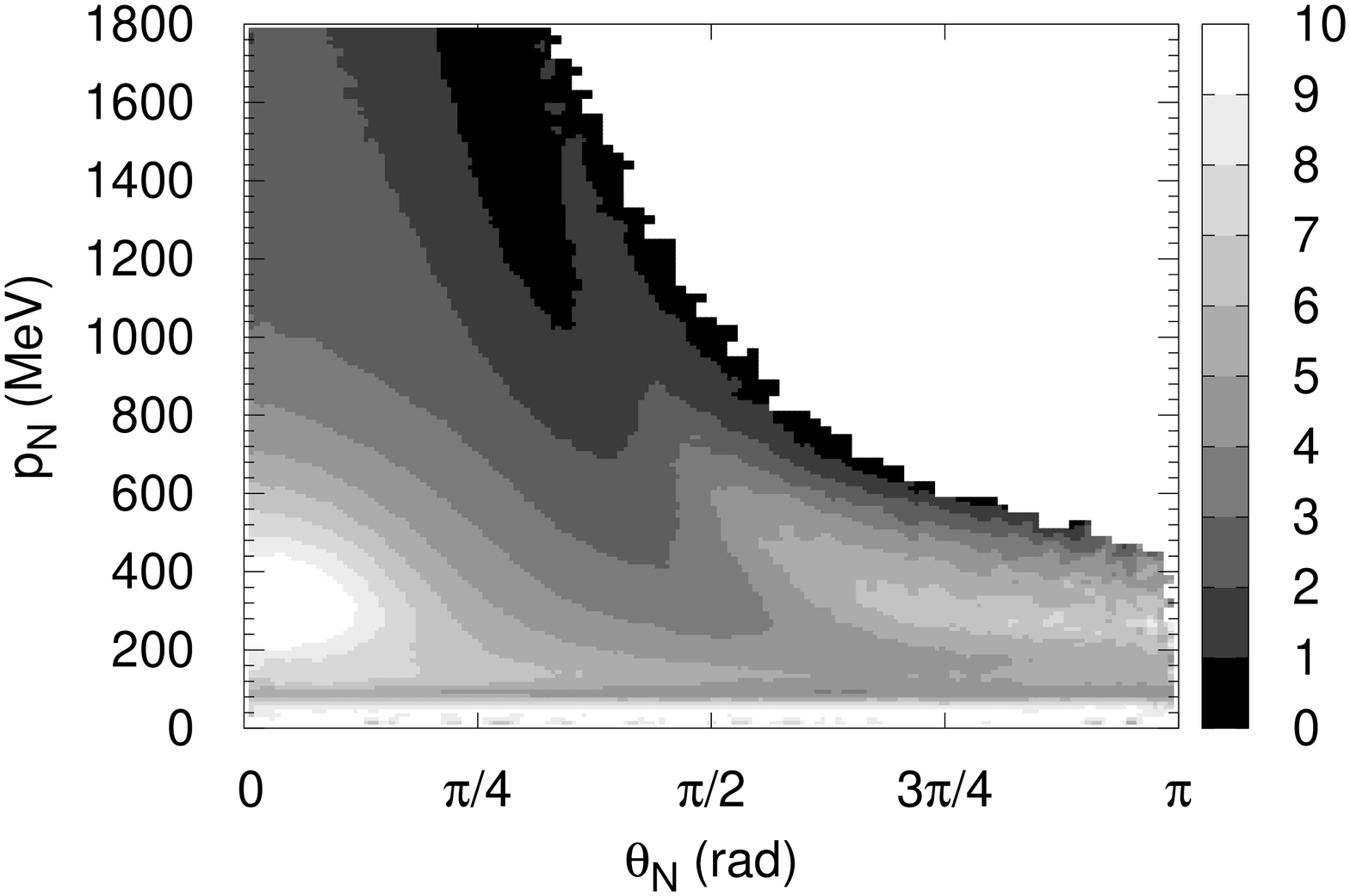}
\vspace{-0.3cm}
\caption{(Color online) As for Fig.~\ref{fig:El_thetal}, except here we represent the double-differential cross section $d\sigma/(dp_N d\theta_N) [10^{-42}$cm$^2$/(MeV rad)] as a function of $\theta_N$ and $p_N$. 
}\label{fig:pN_thetaN}
\end{figure*}

\end{document}